\documentclass[useAMS, usenatbib]{mnras}

\usepackage{graphicx}
\usepackage[all]{hypcap}
\usepackage{amsmath}

\title[Compaction and Quenching Inside-out]
{Evolution of Density Profiles in High-$z$ Galaxies: Compaction and Quenching Inside-Out}

\author[S. Tacchella, A. Dekel, C. M. Carollo, et al.]
{Sandro Tacchella$^{1}$\thanks{E-mail: \href{mailto:sandro.tacchella@phys.ethz.ch}{sandro.tacchella@phys.ethz.ch}}, 
Avishai Dekel$^{2}$, 
C. Marcella Carollo$^{1}$, 
Daniel Ceverino$^{3,4}$, \newauthor
Colin DeGraf$^{2}$,
Sharon Lapiner$^{2}$,
Nir Mandelker$^{2}$,
Joel R. Primack$^{5}$
\\
\\
\\
$^{1}$Department of Physics, Institute for Astronomy, ETH Zurich, CH-8093 Zurich, Switzerland \\
$^{2}$Center for Astrophysics and Planetary Science, Racah Institute of Physics, The Hebrew University, Jerusalem 91904, Israel \\
$^{3}$Centro de Astrobiologia (CSIC-INTA), Ctra de Torrejon a Ajalvir, km 4, E-28850 Torrejon de Ardoz, Madrid, Spain \\
$^{4}$Astro-UAM, Universidad Autonoma de Madrid, Unidad Asociada CSIC, E-28049 Madrid, Spain \\
$^{5}$Department of Physics, University of California, Santa Cruz, CA 95064, USA \\
}

\begin{document}

\date{\textit{Draft Version:} \today}

\pagerange{\pageref{firstpage}--\pageref{lastpage}} \pubyear{2015}

\maketitle

\label{firstpage}

\begin{abstract}

Using cosmological simulations, we address the interplay between structure and star formation in high-redshift galaxies via the evolution of surface density profiles. Our sample consists of 26 galaxies evolving in the redshift range $z=7-1$, spanning the stellar mass range $(0.2-6.4)\times 10^{10}M_\odot$ at $z=2$. We recover the main trends by stacking the profiles in accordance to their evolution phases. Following a wet compaction event that typically occurs when the stellar mass is $\sim10^{9.5}~M_{\odot}$ at $z\sim2-4$, the gas develops a cusp inside the effective radius, associated with a peak in star-formation rate (SFR). The SFR peak and the associated feedback, in the absence of further gas inflow to the centre, marks the onset of gas depletion from the central 1 kpc, leading to quenching of the central SFR. An extended, star-forming ring that forms by fresh gas during the central quenching process shows as a rising specific SFR (sSFR) profile, which is interpreted as inside-out quenching. Before quenching, the stellar density profile grows self-similarly, maintaining its log-log shape because the sSFR is similar at all radii. During the quenching process, the stellar density saturates to a constant value, especially in the inner 1 kpc. The stellar mass and SFR profiles deduced from observations show very similar shapes, consistent with the scenario of wet compaction leading to inside-out quenching and the subsequent saturation of a dense stellar core. We predict a cuspy gas profile during the blue nugget phase, and a gas-depleted core, sometimes surrounded by a ring, in the post-blue nugget phase.

\end{abstract}

\begin{keywords}
galaxies: evolution --- galaxies: formation --- galaxies: quenching --- galaxies: high-redshift --- galaxies: structure --- galaxies: fundamental parameters
\end{keywords}

\section{Introduction}\label{sec:Introduction}

At the present epoch, many early-type galaxies are massive systems, with stellar masses of $M_{\star}\ga10^{11}~M_{\odot}$, central stellar surface densities above $10^{10}~M_{\odot}~\mathrm{kpc}^{-2}$, and hosting only little ongoing star-formation. In the high-redshift universe ($z=1-4$), the progenitors of such massive galaxies had a significant gas-rich disc-like component and were forming stars at rates of up to a few hundred $M_{\odot}~\mathrm{yr}^{-1}$. In this paper, we investigate the physical mechanism responsible for the cessation of star formation, or ``quenching'', and shed light on the interplay between the transformation of morphology and the changes in star-formation activity in these galaxies. 

There is solid observational evidence that the shut-down of star formation in galaxies is correlated with both galaxy mass and environment \citep[e.g.,][]{dressler80, balogh04_bimodality, baldry06, kimm09}. The effects of galaxy stellar mass and environment in the quenching of galaxies appear to be separable up to redshift $z=1$, i.e., it is possible to write the fraction of quiescent galaxies as the product of two functions, one a function of stellar mass only, and the other a function of ``environment'' only \citep{peng10_Cont, knobel13, kovac14}. Furthermore, quenching has been observed to correlate strongly with morphology and galaxy structure in the local universe \citep{kauffmann03, franx08, robaina12, mendel13, cibinel13a, fang13, schawinski14, omand14, bluck14, woo15} and at high-$z$ \citep{wuyts11, wuyts12, bell12, cheung12, szomoru12, barro13, lang14, tacchella15}. On the other hand, quenched disc galaxies have also been found \citep{mcgrath08, van-dokkum08, bundy10, salim12, bruce12, carollo14}. An important question is whether the separability of the quenched fraction with mass and environment, highlighted by \citet{peng10_Cont}, does imply two different quenching mechanisms, or is merely a reflection of the dependence on halo mass of the quenched fraction of satellite galaxies, as pointed out in \citet{carollo14}. 

In this paper, we focus on quenching of galaxies at $z\sim1-4$. Even in the picture of two separate quenching mechanisms, at high redshifts the environmental quenching channel would be less relevant than the quenching channel that affects all galaxies independent of environment \citep{peng10_Cont, ilbert13, muzzin13}. \citet[][hereafter T15b]{tacchella15_sci} mapped out the $M_{\star}$ and star-formation rate (SFR) distribution on scales of 1 kpc in $z\sim2$ star-forming galaxies (SFGs). They found that $\sim10^{10}~M_{\odot}$ galaxies on the star-forming main sequence have flat sSFR profiles, whereas $\sim10^{11}~M_{\odot}$ galaxies have sSFR profiles that rise toward the outskirts. Consistent results have been found at lower redshifts \citep[e.g.][]{nelson15_insideout}. Furthermore, T15b used an empirical model to constrain the timescale quenching, finding that the massive $z\sim2$ galaxies quench inside-out, where the star formation in the centre ceases within $\la200$ Myr, whereas the outskirts still form stars for $1-3$ Gyr. However, the physical nature of quenching, and its link to morphology, is yet to be understood. Several quenching mechanisms have been proposed. Galaxies can run out of gas by rapid gas consumption into stars, in combination with the associated outflows driven by stellar feedback \citep[e.g., ][]{dekel86, murray05} or super-massive black hole feedback \citep[e.g.,][]{di-matteo05, croton06, ciotti07, cattaneo09}, and/or by a slowdown of gas supply into the galaxy \citep{rees77, dekel06, feldmann10,hearin13, feldmann15_quench}. Morphological / gravitational quenching suggests that the growth of a central mass concentration, i.e., a massive bulge, stabilizes a gas disc against fragmentation \citep{martig09, martig13, genzel14a, forbes14}. Long-term suppression of external gas supply happens once the halo mass grows above a threshold mass of $\sim10^{12} M_{\odot}$, either via virial shock heating \citep{birnboim03, keres05, dekel06, keres09}, or by gravitational infall heating \citep{dekel08, birnboim11, khochfar08}, which can be aided by AGN feedback coupled to the hot halo gas \citep{dekel06, cattaneo09, fabian12}. 

Regarding the morphology, the massive galaxies have transformed from a highly perturbed disc-like appearance in the rest-frame optical light at $z=1-4$ to spheroidal-like appearance today. To understand how and when the galaxies assembled their stellar masses and developed structural components, the size-mass relation has been extensively studied at different redshifts \citep[e.g., ][]{daddi05, trujillo07, cimatti08, mcgrath08, van-dokkum08, szomoru11, newman12a, barro13, dullo13, poggianti13, shankar13, carollo13, cassata13}. Among other things, these studies found that the average half-light radii of massive quiescent galaxies increase by a factor of $3-5$ (at a constant mass, i.e., they were $30-100$-times denser) from redshift of $z\approx2$ to the present. The origin of this evolution is debated. Often, this trend has been entirely ascribed to the physical growth of individual galaxies, where massive quiescent galaxies grow through accretion of smaller galaxies in minor gas-poor mergers \citep[e.g.][]{naab09, hopkins09b, nipoti09, feldmann10, oser12}. On the other hand, several studies \citep{carollo13, cassata13, poggianti13} showed that the size growth of quiescent galaxies can be explained by quenching of SFGs that keeps producing, at later epochs, new quenched galaxies with larger size than those of galaxies quenched at earlier epochs. The main challenge addressed here is to study the origin of a massive compact stellar core, with surface densities reaching $\sim10^{10}~M_{\odot}~\mathrm{kpc}^{-2}$. Observations of $z\sim0-2.5$ galaxies inside radii of $\sim1$ kpc indicate that these high central densities have formed early, before $z\sim2$ \citep[e.g.,][]{wuyts12, saracco12, barro13, patel13, van-dokkum10, van-dokkum13, barro14, van-dokkum14_dense_cores, morishita15, tacchella15_sci, van-dokkum15}. In this paper, we show that these densities must have originated from a dissipative compaction of gas into a massive dense gas core that has largely converted into stars \citep{dekel14_nugget}. 

We know from cosmological simulations that high-$z$ galaxies are continuously fed by streams from the cosmic web, consisting of smooth gas and a whole spectrum of merging clumps and galaxies \citep{birnboim03, keres05, dekel06, ocvirk08, keres09, dekel09, danovich12, danovich15}. The high gas fraction and the high density of the universe at these high redshifts, combined with constant triggering by the intense instreaming of minor mergers, induce and maintain violent disc instabilities (VDIs), which are characterized by turbulence and perturbations in the form of large transient features and giant clumps \citep{noguchi98, immeli04a, immeli04b, genzel06, bournaud07, genzel08, agertz09, dekel09b, ceverino10, ceverino12, mandelker14}. To achieve the high gas mass in the centre, the gas has to flow to the centre at a rate that is faster than the SFR, such that the inflow is wet and the compaction is complete, not leaving behind an extended disc.

\citet{dekel14_nugget} have introduced a toy model for describing the main aspects of this dissipative shrinkage of gaseous discs into compact star-forming systems. \citet[][hereafter Z15]{zolotov15} investigated the wet gas compaction phase, and the way it leads to quenching, and identified several physical mechanisms that can be responsible for the onset of compaction. They found that the wet compaction is sometimes associated with VDI, and it is likely triggered by an intense episode of gas inflow through gas-rich (mostly minor) mergers \citep[e.g.,][]{barnes91, mihos96, hopkins06a}. The compaction could alternatively be triggered by counter-rotating streams \citep{danovich15}, by low-angular-momentum recycled gas, and by tidal compression. Z15 put forward a picture in which galaxies evolve through compaction (to compact, star-forming galaxies, ``blue nuggets'') and quenching phases. Following their definition of nuggets, we refer to a massive compact galaxy as a nugget, i.e., a galaxy that has a massive $\sim1$ kpc core of high central density in stellar mass and possibly in gas density. Star-forming nuggets are called blue nuggets, but recall that the blue nuggets could actually be red due to dust.

In a companion paper \citep[][hereafter T15a]{tacchella15_MS}, we have taken the analysis of Z15 one step further by investigating how the galaxies of these simulations evolve along the star-forming main sequence, and how they cease their star formation activity. We found that galaxies oscillate about an equilibrium at the main-sequence ridge on timescales of $\sim0.4~t_{\rm Hubble}$, ranging from 0.5 Gyr to 1 Gyr from $z=4$ to 2. Galaxies on the upper part of the main sequence are compact, star-forming galaxies (blue nuggets), with higher gas fractions and shorter depletion times, consistent with observations \citep[e.g.,][]{magdis12, sargent14, huang14, genzel15, silverman15, scoville15}. We find that the blue nugget phase comes to an end when the balance of flow rates in the central 1 kpc tips to inflow rate $<$ SFR $+$ outflow rate. The immediate reason is that while the SFR $+$ outflow rate are at their maximum, the inflow is suppressed because the disc has largely disappeared (shrunk) and whatever is left in the disc is low density. In addition, the bulge has grown, so the remaining disc is stable gravitationally (morphological quenching, see \citealt{martig09, genzel14a, tacchella15_sci}) and does not drive more gas inward. If there is fresh gas supply from the halo on a sufficiently short timescale, it allows the disc to revive and a new compaction may be triggered to cause a new oscillation about the MS ridge. In order to maintain long-term quenching, one needs to prevent fresh gas supply from the halo. This is more likely at lower redshifts, where the cosmological accretion timescale is longer than the depletion time, and especially so in a hot halo above a critical threshold halo mass \citep{birnboim03, keres05, dekel06}.

In this paper, we study the compaction and quenching of simulated galaxies using the time evolution of the density profiles of stellar mass, gas mass, and SFR. This is part of the effort to understand the interplay between the transformations of morphology and star-formation activity in massive, high-$z$ galaxies. We use high-resolution, zoom-in, hydro-cosmological simulations of 26 galaxies in the redshift range $z=7$ to $z=1$. These are the same simulations as studied by Z15 and T15a. The suite of galaxies analyzed here were simulated at a maximum resolution of $\sim25$ pc and include supernova and radiative stellar feedback. At $z\sim2$, the halo masses are in the range $M_{\rm vir}\sim10^{11}-10^{12}~M_{\odot}$ and the stellar masses are in the range $(0.2-5.7)\times10^{10}~\mathrm{M_{\odot}}$. 

This paper is organized as follows. In Section~\ref{sec:Simulations} we give an overview of the simulations, including a summary of their limitations. In Section~\ref{sec:Growth_Galaxies}, we investigate surface density profiles of individual galaxies and all galaxies based on average profies. Furthermore, we look into stellar mass and sSFR evolution in the centres as well as the outskirts of the simulated galaxies. Section~\ref{sec:phases} presents the emerging picture with the key figure showing average density profiles stacked by evolutionary stages of the galaxies. In Section~\ref{sec:OBS} we present a comparison of the simulations with observational data, focusing on the $M_{\star}$ and SFR profile. Finally, we summarize our results in Section~\ref{sec:Conclusion}.

\section{Simulations}\label{sec:Simulations}

To investigate the evolution of density profiles in high-$z$ galaxies, we use zoom-in hydro-cosmological simulations of 26 moderately massive galaxies, a subset of the 35-galaxy \texttt{VELA} simulation suite. Two papers (\citealt{ceverino14_radfeed}; Z15) have presented the full details of these \texttt{VELA} simulations. Z15 and T15a used the same sample of 26 simulations and investigated similar questions concerning compaction and quenching. Z15 focused on the signatures of compaction in the simulation and T15a addressed the connection between compaction, quenching and the confinement of the main sequence of SFGs. Additional analysis of the same suite of simulations are discussed in \citet{moody14}, \citet{snyder15_morph} and \citet{ceverino15b}. In this section, we give an overview of the key aspects of the simulations and their limitations.

\subsection{Cosmological Simulations}

The \texttt{VELA} cosmological simulations make use of the Adaptive Refinement Tree (ART) code \citep{kravtsov97, kravtsov03, ceverino09}, which accurately follows the evolution of a gravitating N-body system and the Eulerian gas dynamics using an adaptive mesh refinement approach. The adaptive mesh refinement maximum resolution is $17-35$ pc at all times, which is achieved at densities of $\sim10^{-4}-10^3~\mathrm{cm}^{-3}$. In the circumgalactic medium (at the virial radius of the dark-matter halo), the median resolution amounts to $\sim500$ pc. Beside gravity and hydrodynamics, the code includes many physical process relevant for galaxy formation: gas cooling by atomic hydrogen and helium, metal and molecular hydrogen cooling, photoionization heating by the UV background with partial self-shielding, star formation, stellar mass loss, metal enrichment of the ISM and stellar feedback. Supernovae and stellar winds are implemented by local injection of thermal energy as in \citet{ceverino09, ceverino10} and \citet{ceverino12}. Radiative stellar feedback is implemented at a moderate level, following \citet{dekel13a}, as described in \citet{ceverino14_radfeed}.

Cooling and heating rates are tabulated for a given gas density, temperature, metallicity and UV background based on the CLOUDY code \citep{ferland98}, assuming a slab of thickness 1 kpc. A uniform UV background based on the redshift-dependent \citet{haardt96} model is assumed, except at gas densities higher than $0.1~\mathrm{cm}^{-3}$, where a substantially suppressed UV background is used ($5.9\times10^ 6~\mathrm{erg}~\mathrm{s}^{-1}~\mathrm{cm}^{-2}~\mathrm{Hz}^{-1}$) in order to mimic the partial self-shielding of dense gas, allowing dense gas to cool down to temperatures of $\sim300~\mathrm{K}$. The assumed equation of state is that of an ideal mono-atomic gas. Artificial fragmentation on the cell size is prevented by introducing a pressure floor, which ensures that the Jeans scale is resolved by at least 7 cells (see \citealt{ceverino10}).

Star formation is assumed to occur at densities above a threshold of $1~\mathrm{cm}^{-3}$ and at temperatures below $10^4~\mathrm{K}$. Most stars ($>90~\%$) form at temperatures well below $10^3~\mathrm{K}$, and more than half of the stars form at $300~\mathrm{K}$ in cells where the gas density is higher than $10~\mathrm{cm}^{-3}$. The code implements a stochastic star-formation model that yields a star-formation efficiency per free-fall time of $\sim2~\%$, using the stellar initial mass function of \citet{chabrier03}. At the given resolution, this efficiency roughly mimics the empirical Kennicutt-Schmidt law \citep{kennicutt98}. 

The thermal stellar feedback model incorporated in the code releases energy from stellar winds and supernova explosions as a constant heating rate over $40~\mathrm{Myr}$ following star formation. The heating rate due to feedback may or may not overcome the cooling rate, depending on the gas conditions in the star-forming regions \citep{dekel86, ceverino09}. There is no artificial shutdown of cooling implemented in these simulations. Runaway stars are included by applying a velocity kick of $\sim10~\mathrm{km}~\mathrm{s}^{-1}$ to $30~\%$ of the newly formed stellar particles. The code also includes the later effects of Type Ia supernova and stellar mass loss, and it follows the metal enrichment of the ISM. We incorporate radiation pressure through the addition of a non-thermal pressure term to the total gas pressure in regions where ionizing photons from massive stars are produced and may be trapped. This ionizing radiation injects momentum in the cells neighbouring massive star particles younger than $5~\mathrm{Myr}$, and whose column density exceeds $10^{21}~\mathrm{cm}^{-2}$, isotropically pressurizing the star-forming regions. \citet{ceverino14_radfeed} provides more details.

The initial conditions for the simulations are based on dark-matter haloes that were drawn from dissipationless N-body simulations at lower resolution in three comoving cosmological boxes (box-sizes of 10, 20, and 40 Mpc/h). We assume the standard $\Lambda$CDM cosmological model with the WMAP5 values of the cosmological parameters, namely $\Omega_m=0.27$, $\Omega_{\Lambda}=0.73$, $\Omega_b=0.045$, $h=0.7$ and $\sigma_8=0.82$ \citep{komatsu09}. We selected each halo to have a given virial mass at $z = 1$ and no ongoing major merger at $z=1$. This latter criterion eliminates less than $10~\%$ of the haloes which tend to be in a dense, proto-cluster environment at $z\sim1$. The virial masses at $z=1$ were chosen to be in the range $M_{\rm vir}=2\times10^{11}-2\times10^{12}~M_{\odot}$, about a median of $4.6\times10^{11}~M_{\odot}$. If left in isolation, the median mass at $z=0$ was intended to be $\sim10^{12}~M_{\odot}$. Realistically, the actual mass range is broader, with some of the haloes merging into more massive haloes that eventually host groups at $z=0$.

\subsection{Limitations of the Current Simulations}\label{subsec:limitations}

The cosmological simulations used in this paper are state-of-the-art in terms of high-resolution adaptive mesh refinement hydrodynamics and the treatment of key physical processes at the subgrid level. In particular, they trace the cosmological streams that feed galaxies at high redshift, including mergers and smooth flows, and they resolve the VDI that governs high-$z$ disc evolution and bulge formation \citep{ceverino10, ceverino12, ceverino15, mandelker14}. 

Like in other simulations, the treatment of star formation and feedback processes can still be improved. Currently, the code assumes a SFR efficiency per free fall time that is rather realistic, but it does not yet follow in detail the formation of molecules and the effect of metallicity on SFR \citep{krumholz12b}. Furthermore, the resolution does not allow the capture of the Sedov-Taylor adiabatic phase of supernova feedback. The radiative stellar feedback assumed no infrared trapping, in the spirit of low trapping advocate by \citet{dekel13a} based on \citet{krumholz12}. Other works assume more significant trapping \citep{murray10, krumholz10, hopkins12}, which makes the assumed strength of the radiative stellar feedback lower than in other simulations. Finally, AGN feedback, and feedback associated with cosmic rays and magnetic fields, are not yet implemented. Nevertheless, as shown in \citet{ceverino14_radfeed}, the star formation rates, gas fractions, and stellar-to-halo mass ratio are all in the ballpark of the estimates deduced from observations, providing a better match to observations than earlier simulations.

The uncertainties and any possible remaining mismatches by a factor of order 2 are comparable to the observational uncertainties. For example, the stellar-to-halo mass ratio is not well constrained observationally at $z\ga1$. Recent estimates by \citet{burkert15} (see their Fig. 5), based on the observed kinematics of $z\sim0.6-2.8$ SFGs, reveal significantly larger ratios than the estimates based on abundance matching \citep{conroy09, moster10, moster13, behroozi10, behroozi13b} at $M_{\rm vir}<10^{11.8}~M_{\odot}$. We compare in Figure~\ref{FigApp:stellarhalo} the stellar-to-halo mass ratio of our simulations with the ones from observations. At $z=1$, the simulated galaxies have a median halo mass of $\log_{10}~M_{\rm vir}=11.7\pm0.3$ and a stellar to halo mass ratio of $\log_{10}~M_{\star}/M_{\rm vir}=-1.3\pm0.2$. \citet{burkert15} found $\log_{10}~M_{\star}/M_{\rm vir}=-1.5\pm0.3$ at $\log_{10}~M_{\rm vir}=11.7$, which is consistent with our simulations, while \citet{behroozi13b} found $\log_{10}~M_{\star}/M_{\rm vir}=-1.8\pm0.2$ via abundance matching, which is a factor of 3 lower than our simulated estimate. At $z=2$, we find only little evolution in the halo-to-stellar mass ratio: our simulated galaxies have a median halo mass of $\log_{10}~M_{\rm vir}=11.5\pm0.3$ and a stellar to halo mass ratio of $\log_{10}~M_{\star}/M_{\rm vir}=-1.4\pm0.2$. Via abundance matching, \citet{behroozi13b} found $\log_{10}~M_{\star}/M_{\rm vir}=-2.3\pm0.2$ at $\log_{10}~M_{\rm vir}=11.5$, but \citet{burkert15} found $\log_{10}~M_{\star}/M_{\rm vir}=-1.6\pm0.3$, which is again consistent with our simulations. Altogether, we conclude that our simulations produce stellar-to-halo mass ratios that are in the ballpark of the values estimated from observations, and within the observational uncertainties, which are also comparable to the uncertainties associated with the feedback recipes in the simulations. Therefore, it is sensible and adequate to use our simulations with the adopted feedback prescription,  while bearing in mind the factor-of-two uncertainties.

It seems that in the current simulations, the compaction and the subsequent onset of quenching occur at cosmological times that are consistent with observations (see Fig. 12 of \citealt{zolotov15} and Fig. 2 of \citealt{barro13}). However, with some of the feedback mechanisms not yet incorporated (e.g., fully resolved supernova feedback and AGN feedback), full quenching to very low sSFR values may not be entirely reached in many galaxies by the end of the simulations at $z\sim1$. One should be aware of this limitation of the current simulations. Nevertheless, for all the above reasons, we adopt in this work the hypothesis that the simulations grasp the qualitative features of the main physical mechanisms that govern galaxy evolution through the processes of compaction and subsequent quenching.

\subsection{Simulated Galaxy Sample and Measurements}\label{subsec:sample}

\begin{table*}
\centering
\begin{tabular}{@{}lccccccccc}
\multicolumn{10}{c}{{\bf The suite of 26 simulated galaxies.}} \\
\hline
Galaxy & $M_{\rm vir}$ & $M_{\star}$ & $M_{\rm g}$ & SFR & sSFR & $R_{\rm vir}$ & $R_{\rm e}$ & $a_{\mathrm{fin}}$ & $z_{\mathrm{fin}}$ \\
  & $10^{12}~M_{\odot} $& $10^{10}~M_{\odot}$ & $10^{10}~M_{\odot}$ & $M_{\odot}/$yr & Gyr$^{-1}$ & kpc & kpc & & \\
  & ($z=2$) & ($z=2$) & ($z=2$) & ($z=2$) & ($z=2$) & ($z=2$) & ($z=2$) & & \\
\hline
\hline
V01 & 0.16 & 0.22 & 0.12 & 2.65 & 1.20 & 58.25 & 1.08 & 0.50 & 1.00 \\
V02 & 0.13 & 0.19 & 0.16 & 1.84 & 0.94 & 54.50 & 2.23 & 0.50 & 1.00 \\
V03 & 0.14 & 0.43 & 0.10 & 3.76 & 0.87 & 55.50 & 1.73 & 0.50 & 1.00 \\
V06 & 0.55 & 2.22 & 0.33 & 20.72 & 0.93 & 88.25 & 1.11 & 0.37 & 1.70 \\
V07 & 0.90 & 6.37 & 1.42 & 26.75 & 0.42 & 104.25 & 3.41 & 0.50 & 1.00 \\
V08 & 0.28 & 0.36 & 0.19 & 5.76 & 1.58 & 70.50 & 0.81 & 0.50 & 1.00 \\
V09 & 0.27 & 1.07 & 0.31 & 3.97 & 0.37 & 70.50 & 1.84 & 0.39 & 1.56 \\
V10 & 0.13 & 0.64 & 0.11 & 3.27 & 0.51 & 55.25 & 0.54 & 0.50 & 1.00 \\
V11 & 0.27 & 1.02 & 0.58 & 17.33 & 1.69 & 69.50 & 3.76 & 0.46 & 1.17 \\
V12 & 0.27 & 2.06 & 0.19 & 2.91 & 0.14 & 69.50 & 1.27 & 0.39 & 1.56 \\
V13 & 0.31 & 0.96 & 0.98 & 21.23 & 2.21 & 72.50 & 5.36 & 0.39 & 1.56 \\
V14 & 0.36 & 1.40 & 0.59 & 27.61 & 1.97 & 76.50 & 0.41 & 0.42 & 1.38 \\
V15 & 0.12 & 0.56 & 0.14 & 1.71 & 0.30 & 53.25 & 1.33 & 0.50 & 1.00 \\
V20 & 0.53 & 3.92 & 0.48 & 7.27 & 0.19 & 87.50 & 1.99 & 0.44 & 1.27 \\
V21 & 0.62 & 4.28 & 0.57 & 9.76 & 0.23 & 92.25 & 1.89 & 0.50 & 1.00 \\
V22 & 0.49 & 4.57 & 0.21 & 12.05 & 0.26 & 85.50 & 1.36 & 0.50 & 1.00 \\
V23 & 0.15 & 0.84 & 0.19 & 3.32 & 0.39 & 57.00 & 1.40 & 0.50 & 1.00 \\
V24 & 0.28 & 0.95 & 0.28 & 4.39 & 0.46 & 70.25 & 1.87 & 0.48 & 1.08 \\
V25 & 0.22 & 0.76 & 0.08 & 2.35 & 0.31 & 65.00 & 0.88 & 0.50 & 1.00 \\
V26 & 0.36 & 1.63 & 0.25 & 9.76 & 0.60 & 76.75 & 0.80 & 0.50 & 1.00 \\
V27 & 0.33 & 0.90 & 0.52 & 8.75 & 0.97 & 75.50 & 2.98 & 0.50 & 1.00 \\
V29 & 0.52 & 2.67 & 0.39 & 18.74 & 0.70 & 89.25 & 2.47 & 0.50 & 1.00 \\
V30 & 0.31 & 1.71 & 0.41 & 3.84 & 0.22 & 73.25 & 1.64 & 0.34 & 1.94 \\
V32 & 0.59 & 2.74 & 0.37 & 15.04 & 0.55 & 90.50 & 2.66 & 0.33 & 2.03 \\
V33 & 0.83 & 5.17 & 0.45 & 33.01 & 0.64 & 101.25 & 1.42 & 0.39 & 1.56 \\
V34 & 0.52 & 1.73 & 0.42 & 14.79 & 0.85 & 86.50 & 2.10 & 0.35 & 1.86 \\
\hline
\end{tabular}
 \caption{Quoted are the total virial mass, $M_{\rm vir}$, the stellar mass, $M_{\star}$, the gas mass, $M_{\rm g}$, the star formation rate, SFR, the specific star formation rate, sSFR, the virial radius, $R_{\rm vir}$, the effective stellar (half-mass) radius, $R_{\rm e}$ (all at $z=2$) and the final simulation snapshot, $a_{\mathrm{fin}}$, and redshift, $z_{\mathrm{fin}}$. The $M_{\star}$, $M_{\rm g}$, SFR, and sSFR are measured within a radius of $0.2\times R_{\rm vir}$.}
\label{tab:sample}
\end{table*}

From the suite of 35 galaxies, we have excluded three low-mass galaxies as well as six galaxies which have not been simulated down to $z=2.0$. Therefore, our final sample consists of 26 galaxies. The virial and stellar properties of these galaxies are listed in Table~\ref{tab:sample}. The virial mass $M_{\rm vir}$ is the total mass within a sphere of radius $R_{\rm vir}$ that encompasses an overdensity of $\Delta(z)=(18\pi^2-82\Omega_{\Lambda}(z)-39\Omega_{\Lambda}(z)^2)/\Omega_{m}(z)$, where $\Omega_{\Lambda}(z)$ and $\Omega_{m}(z)$ are the cosmological parameters at $z$ \citep{bryan98, dekel06}. The stellar mass $M_{\star}$ is measured within a radius of $0.2\times R_{\rm vir}$..

We start the analysis at the cosmological time corresponding to expansion factor $a=0.125$ (redshift $z=7$). At earlier times, the fixed resolution scale typically corresponds to a non-negligible fraction of the galaxy size. As can be seen in Table~\ref{tab:sample}, most galaxies reach $a=0.50$ ($z=1$). Each galaxy is analysed at output times separated by a constant interval in $a$, $\Delta a=0.01$, corresponding at $z=2$ to $\sim100~\mathrm{Myr}$ (roughly half an orbital time at the disc edge). Our sample consists of totally $\sim900$ snapshots in the redshift range $z=6-1$ from 26 galaxies that at $z = 2$ span the stellar mass range $(0.2-6.4)\times10^{11}~\mathrm{M_{\odot}}$. The half-mass sizes $R_{\rm e}$ are determined from the $M_{\star}$ that are measured within a radius of $0.2~R_{\rm vir}$ and they range $R_{\rm e}\simeq0.4-3.2~\mathrm{kpc}$ at $z=2$.

The SFR for a simulated galaxy is obtained by $\mathrm{SFR}=\langle M_{\star}(t_{\rm age}<t_{\rm max})/t_{\rm max} \rangle_{t_{\rm max}}$, where $M_{\star}(t_{\rm age}<t_{\rm max})$ is the mass at birth in stars younger than $t_{\rm max}$. The average $\langle\cdot\rangle_{t_{\rm max}}$ is obtained by averaging over all $t_{\rm max}$ in the interval $[40,80]~\mathrm{Myr}$ in steps of 0.2 Myr. The $t_{\rm max}$ in this range are long enough to ensure good statistics. Our 26 galaxies have SFR ranging from $\sim1$ to $33~\mathrm{M_{\odot}}~\mathrm{yr}^{-1}$ at $z\sim2$.

The instantaneous mass of each star particle is derived from its initial mass at birth and its age
using a fitting formula for the mass loss from the stellar population represented by the star particle,
according to which 10\%, 20\% and 30\% of the mass is lost after 30 Myr, 260 Myr, and 2 Gyr from birth, respectively. 
We consistently use here the instantaneous stellar mass, $M{_\star}$, and define the specific SFR by $\mathrm{sSFR}=\mathrm{SFR}/M_{\star}$\footnote{Note that this sSFR is not strictly the inverse of the timescale for the formation of the stellar system, for which the instantaneous masses of stars have to be replaced by their initial masses. This makes only a small difference for our purpose here, given that in our galaxies the mass-loss rate is small compared to the SFR.}.

We split the simulated galaxies in a low- and high-mass sample based on the total stellar mass at $z=2$. We divide the sample in about half at $\log_{10}~M_{\star}/M_{\odot}=10.2$. Galaxies below and above this mass scale have a mean mass of $\log_{10}~M_{\star}/M_{\odot}=9.9\pm0.2$ and $\log_{10}~M_{\star}/M_{\odot}=10.5\pm0.2$, respectively. This shows that our galaxies are relatively low- and high-mass galaxies, and we do not study the most massive or the least massive galaxies at $z\sim2-3$. We caution that this is a rough division and that the galaxies which are in a certain mass bin at $z=2$ do not necessarily belong to the same mass bin at all times. Generally, we find that the high-mass galaxies evolve through similar phases as the low-mass galaxies, but at earlier times and in a more pronounced way. 

The density profiles presented here are two dimensional surface density profiles $\Sigma(r)$. We have chosen to present the surface density profiles instead of the three dimensional volume density profiles $\rho(r)$ because the surface density profiles are straight-forwardly accessible observationally. The determination of the centre of the galaxy is outlined in detail in Appendix B of \citet{mandelker14}. Briefly, starting form the most bound star, the centre is refined iteratively by calculating the centre of mass of stellar particles in spheres of decreasing radii, updating the centre and decreasing the radius at each iteration. We begin with an initial radius of 600 pc, and decrease the radius by a factor of 1.1 at each iteration. The iteration terminates when the radius reaches 130 pc or when the number of stellar particles in the sphere drops below 20. The three dimensional density profiles $\rho(r)$ are then constructed from concentric spheres around this centre. We have used the simple relation $\Sigma(r)=2 \cdot r \cdot \rho(r)$ to convert the three dimensional density profiles to two dimensional surface density profiles, which is a good approximation for the face-on projection. The differences in surface density at a given radius for the face-on projection and our approach are $\la20~\%$, i.e. well within our systematic uncertainty of the simulation.

\section{Growth of Galaxies }\label{sec:Growth_Galaxies}

\begin{figure*}
\includegraphics[width=\textwidth]{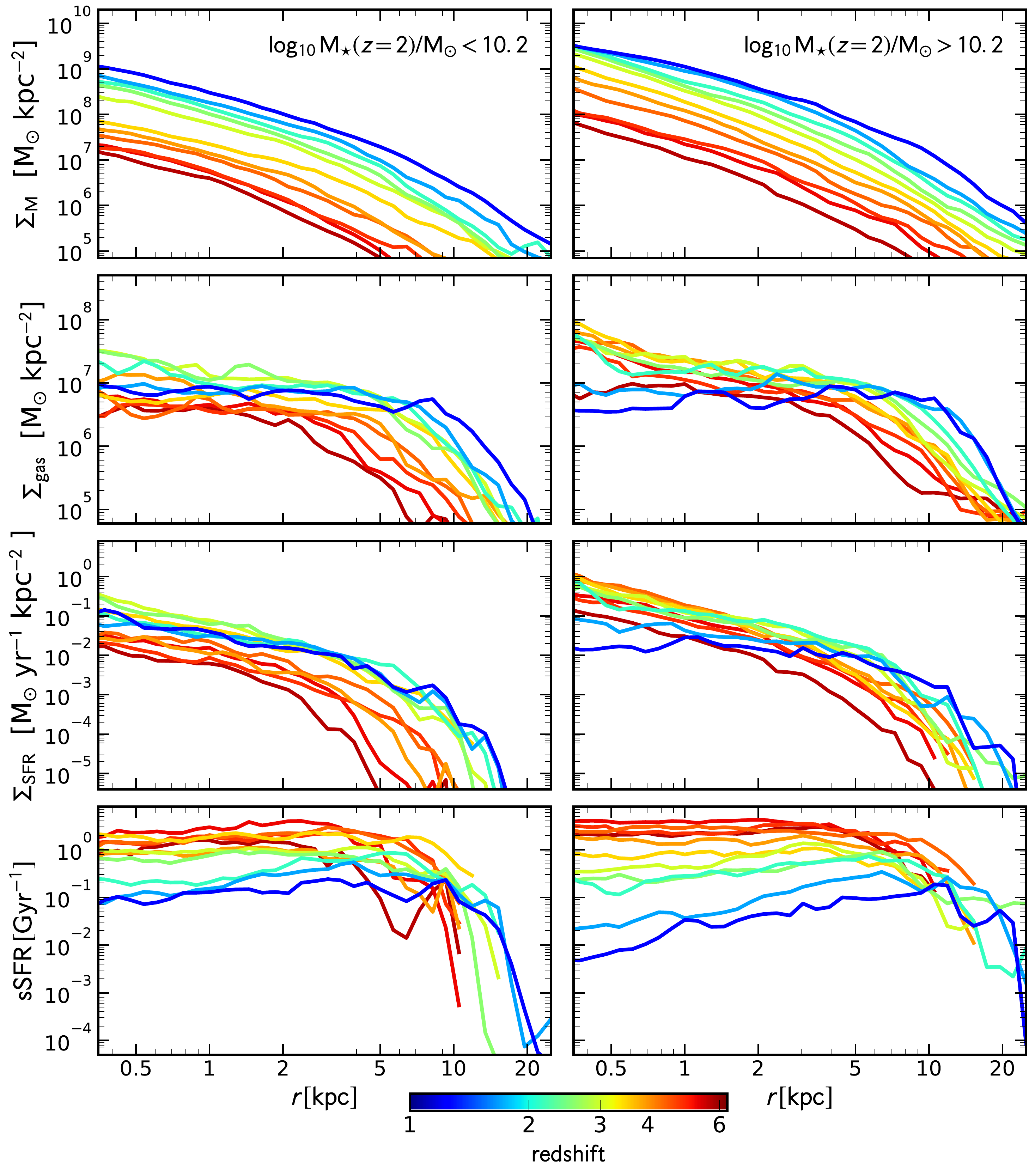} 
\caption{Evolution of the stacked density profiles. From top to bottom, the surface density profiles of stellar mass ($\Sigma_{\rm M}$), cold ($T<10^4~\mathrm{K}$) gas mass ($\Sigma_{\rm gas}$), SFR ($\Sigma_{\rm SFR}$), and sSFR are shown. The left and right panels show the stacks for the low- and high-mass galaxies, which are selected based on the stellar mass at $z=2$. The colour coding corresponds to redshift, according to the colour bar at the bottom. The key feature is the development of a gas cusp in the massive galaxies that is associated with the wet compaction. Subsequently, the gas density in the centre decreases. The sSFR profile is flat which leads to a self-similar growth of the stellar mass profiles. Towards lower redshifts, the sSFR decreases from inside out. The compaction with the succeeding central depletion occurs in different galaxies at different times, which smears out the evolutionary features in these stacks, especially for low-mass galaxies. The profiles for all galaxies are shown individually in Figures~\ref{FigA1_Profile}-\ref{FigA4_Profile}.}
\label{Fig_All_Density}
\end{figure*}

We investigate how the surface density profiles of stellar mass, gas mass, SFR, and sSFR are evolving with time for individual simulated galaxies as well as for the whole sample of galaxies. In particular, we use the evolution of the gas profile to study the wet compaction process and the subsequent phase of gas depletion from the core. We study the emergence of a dense stellar system through the evolution of the stellar density in the central region in comparison to the growth of the total mass. We then analyse the SFR density profile and the progression of quenching within the galaxies. Finally, we study the evolution of the half-mass sizes as a function of stellar mass and cosmic time.

\subsection{Evolution of the Surface Density Profiles}\label{subsec:surface_profiles}

Figure~\ref{Fig_All_Density} shows the cosmic evolution of the stacked surface density profiles of stellar mass, gas mass, SFR, and sSFR for low-mass (left) and high-mass (right) galaxies, respectively. We split the simulated galaxies in a low- and high-mass sample based on the total stellar mass at $z=2$, as described in Section~\ref{subsec:sample}. The individual profiles of the simulations are scaled before stacking to have the same effective density, $\Sigma_{\rm eff}=Q_{\rm tot}/(2\pi R_{\rm e}^2)$, where $Q_{\rm tot}$ is the total relevant quantity and $R_{\rm e}$ is the half-mass radius in 3D. The final, stacked profile has, by construction, a $\Sigma_{\rm eff}$ value that is equal to the median $\Sigma_{\rm eff}$ of all the galaxies in the sample. A complete list of all the surface density profiles for all galaxies individually, ordered by total stellar mass at $z=2$, can be found in Appendix~\ref{App:Profiles} (Figures~\ref{FigA1_Profile}-\ref{FigA4_Profile}). 

The evolution of the high-mass galaxies is quiet synchronized in cosmic time. Therefore, the key evolutionary phases of compaction and quenching are visible in the stacked profiles (Figure~\ref{Fig_All_Density}, right panel). The gas density evolves from a flat profile at $z>4$, through a cuspy profile at $z\sim3-4$, to a ring-like gas distribution surrounding a gas-poor core at $z<2-3$. These evolutionary stages are more noticeable in the individual profiles as these characteristic phases occur in different galaxies at different times. The gas in the inner region is mostly cold. The cusp forms in a wet compaction event that occurs preferentially at intermediate redshifts ($z\sim2-4$) within the central 1 kpc. As mentioned in the Introduction, minor mergers, counter-rotating streams, recycling and tidal compression drive the gas compaction, possibly through VDI, into a compact, star-forming blue nugget \citep{dekel14_nugget, zolotov15}. The SFR profile follows the gas profile, in accordance with the Kennicutt relation between gas mass and SFR that was built-in into the simulations, leading to a central cusp in the SFR profile, which characterizes the blue nugget phase. In the phase of gas shrinkage and blue nugget formation, the extended disc gas mass and SFR is reduced or stays constant, while the gas density and SFR in the core goes up, indicating an early outside-in quenching phase. The stellar mass density within 1 kpc grows to a maximum in this stage, and it remains constant thereafter, i.e., the central stellar mass density saturates. As highlighted by Z15 (see their Figure 1), based on the same sample of simulations used here, the fraction of in-situ star formation in the bulge is high, gradually declining from above $60\%$ at $z\sim5$ to $\sim50\%$ at $z\sim1$. This is clear evidence for wet compaction at high redshifts.

Following compaction and the blue nugget phase, the centres of the galaxies get depleted from gas, a central hole in the gas density develops, and most of the gas is found in an outer ring of radius $\sim3-10$ kpc, and in one case (V07) even 15 kpc. Therefore, in this stage, the star-formation also takes place in the outskirts, leading to stellar mass growth in the outer regions. The galaxies decrease the sSFR within the central 1 kpc by $2-3$ orders of magnitude, while the sSFR in the outskirts stays at a similar level as at earlier times. We interpret this as inside-out quenching because the star-formation activity reduces from inside-out. We will analyse the inside-out quenching and its connection to the characteristic evolutionary pattern of our high-$z$ galaxies in more detail in Sections~\ref{subsec:insideout_quenching} and \ref{sec:phases}, respectively. 

For the low-mass galaxies (Figure~\ref{Fig_All_Density}, left panel), the features described before are milder. The average surface density profiles of gas mass show only little evolution between $z=1-4$. However, the individual low-mass galaxies show substantial evolution of the gas profiles with several phases of high central gas density. The cuspy phases are washed out in the stacked profile because these phases happen at different epochs for different galaxies. The low-mass galaxies tend to grow their stellar mass in a self-similar way, increasing the stellar mass at all radii with the same multiplicative pace that keeps the profile shape the same in log-log space. This is consistent with the SFR profile roughly following the shape of the stellar mass profile, which leads to a rather flat sSFR profile out to $\sim7$ kpc, especially so at $z>2$. Since the sSFR is the inverse of the mass $e$-folding time, the flat sSFR profile implies that at all radii, the galaxy roughly doubles its mass within the same time interval. Only at late cosmic times ($z\sim1$) do the low-mass galaxies begin to show a rising sSFR from the central $\sim0.5$ kpc to the outskirts ($\sim4$ kpc).  

Summarizing, we find in each galaxy and in the stacked profiles at least one phase of wet gas compaction with the formation of a compact, gas-rich, star-forming nugget characterized by a cuspy profile of gas and SFR. The blue nugget phase is the onset of central gas depletion. Thus, the outside-in quenching phase is followed by an inside-out quenching phase, associated with stellar mass growth in the outskirts at late times. This evolutionary pattern is more noticeable in the stacked profiles of high-mass galaxies, since they tend to go through just one major compaction phase during the same cosmological epoch. Lower-mass galaxies tend to go through several compaction events, not all at the same time, which washes out the signature. In Section~\ref{sec:phases}, we will stack all galaxies according to this evolutionary pattern of compaction and quenching, and find that the low-mass galaxies also follow a similar evolutionary pattern.

\subsection{Evolution of the S\'{e}rsic Index $n$}\label{subsec:sersic}

\begin{figure}
\includegraphics[width=\linewidth]{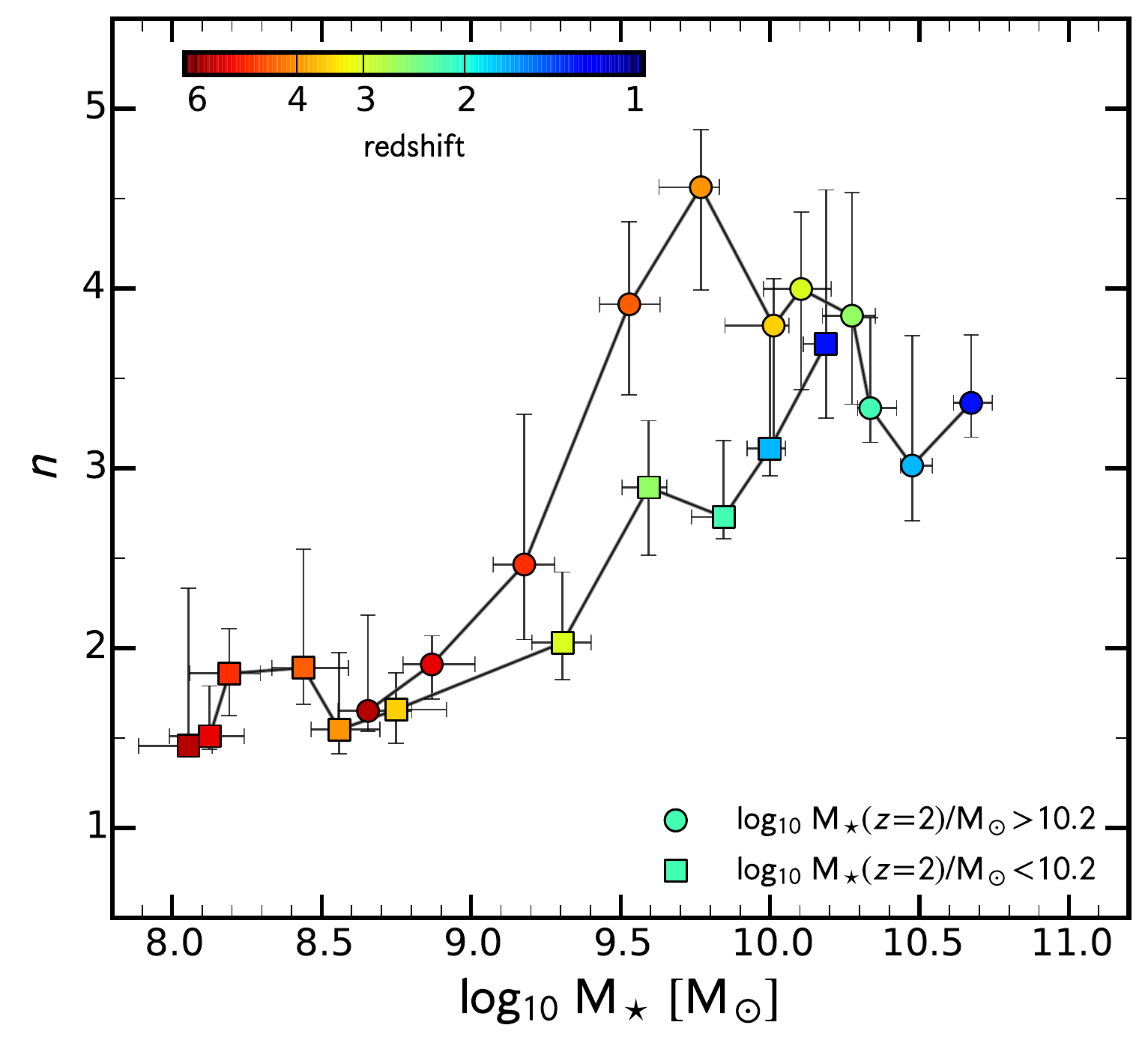} 
\caption{S\'{e}rsic index $n$ versus stellar mass $M_{\star}$. Squares and circles show the median S\'{e}rsic index for the low- and high-mass galaxies, respectively. The colours correspond to redshift according to the colour bar. The errorbars indicate the $1~\sigma$ scatter. The median S\'{e}rsic index increases with stellar mass towards lower redshifts from $n\approx1.5-2.0$ to $n\approx3-4$. For the high mass galaxies, the median S\'{e}rsic index peaks ($n\approx4.5$) at $z\sim4.0$. }
\label{Fig_sersic-mass}
\end{figure}

We fit the stellar mass surface density profiles with \citet{sersic68} profiles to characterize their shapes. The evolution with cosmic time and stellar mass of the median S\'{e}rsic index $n$ is shown in Figure~\ref{Fig_sersic-mass}. Overall, galaxies at later cosmic time (lower redshifts) and higher masses have a higher S\'{e}rsic index: $n$ increases from $1.5-2.0$ at $M_{\star}<10^9~M_{\odot}$ and $z>4$ to $n=3-4$ at $M_{\star}>10^{10}~M_{\odot}$ and $z<2$. This shows that the simulated galaxies transform from a disc-like $n\sim1$ profile to a spheroidal-like $n\sim4$ profile. The galaxies of the high-mass subsample evolve ahead of the galaxies of low-mass subsample and have on average a higher median S\'{e}rsic index. Furthermore, the median S\'{e}rsic index of the high-mass galaxies shows a clear peak of $n=4.6^{+0.3}_{-0.6}$ at $z\sim4$, indicating that these galaxies are very compact. After this compaction, the S\'{e}rsic index decreases to $n=3.4^{+0.4}_{-0.2}$ at $z\sim1$. These classical, de-Vaucouleurs-like profiles are consistent with the results from previous simulations with weaker feedback models \citep{ceverino15}. We also fitted a two-component S\'{e}rsic model to the mass profile, but the two-component fits do not represent the data better, i.e., we do not find a stellar surface-density excess towards the centres beyond the already rather steep S\'{e}rsic profile.

\subsection{Inner versus Outer Evolution}\label{subsec:innerVSouter}

\begin{figure}
\includegraphics[width=\linewidth]{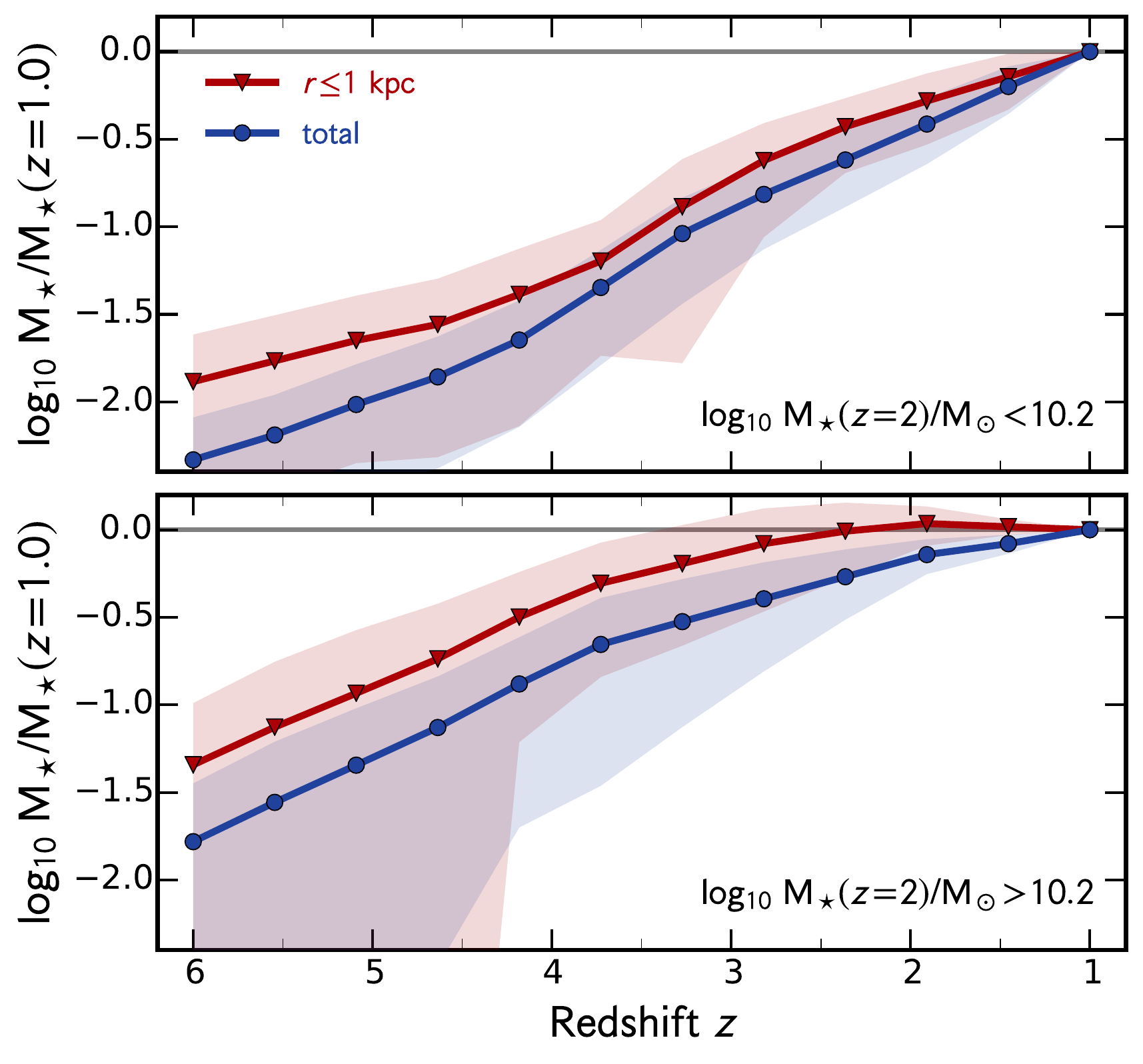} 
\caption{Growth curve of stellar mass for the whole galaxy (blue line) and within 1 kpc (red line). The top and bottom panels refer to the low-mass ($M_{\star}<10^{10.2}~M_{\odot}$ at $z=2$) and high-mass ($M_{\star}>10^{10.2}~M_{\odot}$ at $z=2$) subsamples, respectively. The masses are normalized to their value at $z=1$. The shaded region indicates the $1~\sigma$ scatter. The growth at high redshift is exponential in $z$, as predicted by Equation~\ref{eq:fitting_growth}. For high-mass galaxies, the growth of the central stellar mass saturates starting at $z\approx3$.}
\label{Fig_Mass_Growth}
\end{figure}

In Figure~\ref{Fig_Mass_Growth}, we investigate the stellar mass growth, either within a sphere of 1 kpc or for the whole galaxy within a radius of $0.2\times R_{\rm vir}$, for low-mass and high-mass galaxies. The growth curves are normalized by the corresponding stellar mass at $z=1$. We find that low-mass galaxies (top panel) increase their mass inside both radii at roughly the same pace: the central mass grows concurrently with the total stellar mass, though there is a tendency for the mass in the central 1 kpc to grow somewhat earlier, e.g., by $\sim0.1$ dex at $z=2-3$. The mass increases by a factor of $\sim8$ from $z=3$ to $z=1$. For the high-mass galaxies (bottom panel), the mass in the inner 1 kpc is in place significantly earlier than the total mass, such that at any redshift in the range $6-2.5$, the core mass is a higher fraction of its final mass at $z=1$, by a factor of $\sim3$. The central and total stellar mass grow at a similar pace only at $z\ga3$. After $z\simeq3$, the central stellar mass saturates and it remains roughly constant after $z\sim2.5$, while the total stellar mass continues to grow, by a factor of $\sim2$ from $z=2.5$ to $z=1$.

We quantify the rate of growth where it resembles a straight line in this semi-log plot by fitting the curve of growth with the theory-motivated functional form

\begin{equation}
M(z)=M_0\times\exp[-\alpha (z-z_0)],
\label{eq:fitting_growth}
\end{equation}
\noindent
where the mass at some fiducial redshift $z_0$ is given to be $M_0$. For a given choice of $z_0$, we determine the best-fit value of the slope $\alpha$ within a redshift range between $z_0$ and a certain higher $z$. We set $z_0=1$. 

The functional form in Equation~\ref{eq:fitting_growth} is motivated by theory, see \citet{dekel13} for a simple derivation. Briefly, it can be shown that the average specific mass accretion rate (sMAR) onto haloes can be approximated in the Einstein-deSitter regime ($z>1$) by $\mathrm{sMAR}=\dot{M}/M\simeq s\cdot (1+z)^{\mu}$ with $\mu\rightarrow5/2$. This can be simply integrated, resulting in the growth of halo mass as a function of $z$ as in Equation~\ref{eq:fitting_growth} with $\alpha=(3/2)st_1$ and $t_1=2/3\Omega_m^{-1/2}H_{0}^{-1}\simeq17.5~\mathrm{Gyr}$ for standard $\Lambda$CDM. This functional form was indeed found to be a good fit to the growth in earlier cosmological $N$-body simulations of lower resolution \citep{wechsler02, neistein08}. Then, one can assume that, even if the accretion rate into the central galaxy may be lower than the sMAR into the halo, the specific accretion rates are comparable. This has been confirmed in simulations \citep{dekel13}. Finally, conservation of mass through the ``bathtub'' model \citep{bouche10, lilly13_bathtube, dekel14_bathtube} yields a quasi-stable state with $\mathrm{sSFR}\sim\mathrm{sMAR}$, so Equation~\ref{eq:fitting_growth} applies also for the growth of stellar mass. 
 
In T15a we have investigated the evolution of the main sequence of SFGs in these simulations and we have found that $s\simeq0.046~\mathrm{Gyr}^{-1}$ at $M_{\star}=10^{10}~M_{\odot}$, resulting in $\alpha\simeq1.21$. The parameter $s$ has a weak mass dependence $\propto M^{0.14}$, associated with the slope of the $\Lambda$CDM power spectrum on the relevant scales. This crudely translates to a similar mass dependence in $\alpha$. \citet{wechsler02} found for the halo mass growth $\alpha\sim1.2$ at $M_{\rm vir}=10^{12}$, and a mass dependence of $\alpha\propto M^{0.13}$. As mentioned in Section~\ref{sec:Simulations}, our galaxies have halo masses of $M_{\rm vir}=2\times10^{11}-2\times10^{12}$, i.e., we expect $\alpha$ in the range of 0.97 to 1.31.

For the low-mass galaxies, we fit $\alpha$ in the redshift range $z=1-6$ while fixing $z_0=1.0$. For the total stellar mass, we find $\alpha=0.99\pm0.01$, i.e., low-mass galaxies increase their stellar mass by an $e$-fold over one unit in redshift. For the stellar mass within 1 kpc, we find a slightly slower growth with $\alpha=0.82\pm0.03$. For the high-mass sample, we can already see in Figure~\ref{Fig_Mass_Growth} that the growth slows down with cosmic time, i.e., $\alpha$ decreases. Fitting the data in two redshift bins ($z=1.0-3.5$ and $z=3.5-6.0$), we find for the total mass $\alpha=0.98\pm0.03$ with $z_0=3.5$ and $\alpha=0.41\pm0.03$ with $z_0=1.0$ for the high-$z$ and low-$z$ bin, respectively. For the stellar mass within 1 kpc, we find again only a slightly smaller $\alpha$ with $\alpha=0.92\pm0.04$ at high redshifts, whereas $\alpha=0.12\pm0.03$ at low redshifts. 

Summarizing, we see that low-mass galaxies throughout cosmic time and high-mass galaxies at early times ($z\ga3.5$) grow stellar mass with the same pace in the inner ($<1$ kpc) region and overall (with a semi-log slope $\alpha\simeq0.8-1.0$). This is in good agreement with the halo growth rate of $N$-body simulations and shows that the sSFR is indeed tightly coupled to the specific halo mass increase, as seen for an earlier suite of the simulations in \citet{dekel13}. The most interesting result for our purpose here is that, for massive galaxies, after $z\sim 3$, the growth rate of stellar mass slows down compared to the halo growth rate. This is especially pronounced inside the inner 1 kpc, where the stellar mass saturates after $z \sim 3$, establishing a compact stellar core already then, while the stellar mass continues to build up outside this core at later times.

\subsection{Inside-Out Quenching}\label{subsec:insideout_quenching}

\begin{figure}
\includegraphics[width=\linewidth]{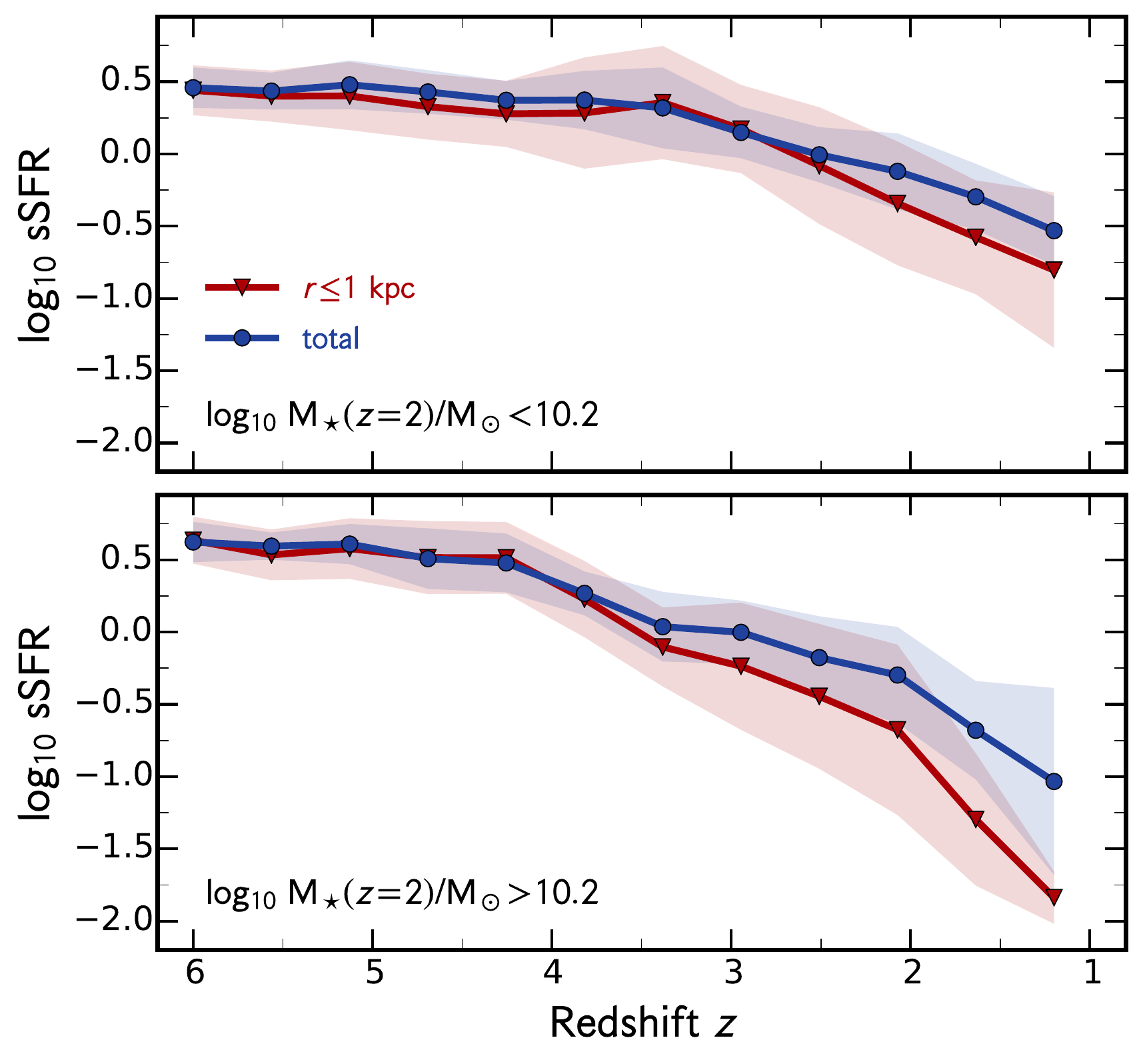} 
\caption{Evolution of the sSFR for the whole galaxy (blue line) and within 1 kpc (red line). The top and bottom panels refer to the low-mass ($M_{\star}<10^{10.2}~M_{\odot}$ at $z=2$) and high-mass ($M_{\star}>10^{10.2}~M_{\odot}$ at $z=2$) subsamples, respectively. The shaded region indicates the $1~\sigma$ scatter. The sSFR is rather constant down to $z=3.5$ (4.2) for the low-mass (high-mass) sample, after which it declines gradually, rather steeply for the massive galaxies at $z\leq2$. For low-mass galaxies the sSFR decreases with the same pace at all radii, while for high-mass galaxies the sSFR decreases faster in the inner regions.}
\label{Fig_sSFR_Evolution}
\end{figure}

Figure~\ref{Fig_sSFR_Evolution} shows the sSFR in the central 1-kpc region and for the whole galaxy as a function of redshift. Low-mass galaxies have a constant sSFR down to $z=3.4$, after which it declines gradually. The sSFR is the same within the two apertures at $z>2.5$. At lower redshifts we see a deviation: the sSFR is declining, slightly faster in the central 1 kpc than over the whole galaxy. The average total sSFR is decreasing from $3~\mathrm{Gyr}^{-1}$ at $z\sim6$ to $0.3~\mathrm{Gyr}^{-1}$ at $z\sim1$, while the sSFR within 1 kpc drops from $3~\mathrm{Gyr}^{-1}$ to below $0.2~\mathrm{Gyr}^{-1}$. 

High-mass galaxies have a constant sSFR till $z=4.2$, before the sSFR declines. Below $z=3.5$, the sSFR declines faster in the inner 1 kpc region. The sSFR in the central region is $3.2~\mathrm{Gyr}^{-1}$ at $z\sim6$ and it decreases to $0.01~\mathrm{Gyr}^{-1}$ at $z\sim1$, while the total sSFR of the whole galaxy only decreases to $0.07~\mathrm{Gyr}^{-1}$ at $z\sim1$. The sSFR reduces in the centre by more than two orders of magnitudes, while the sSFR in the outskirts stays high and declines later. This is consistent with a picture of inside-out quenching in the post-compaction phase and it is in line with what we have seen in Figure~\ref{Fig_Mass_Growth}, where the stellar mass within the central 1 kpc saturates after $z\sim3$, while the mass in the outskirts keeps growing.

\subsection{Size-Mass Evolution}\label{subsec:size-mass}

\begin{figure}
\includegraphics[width=\linewidth]{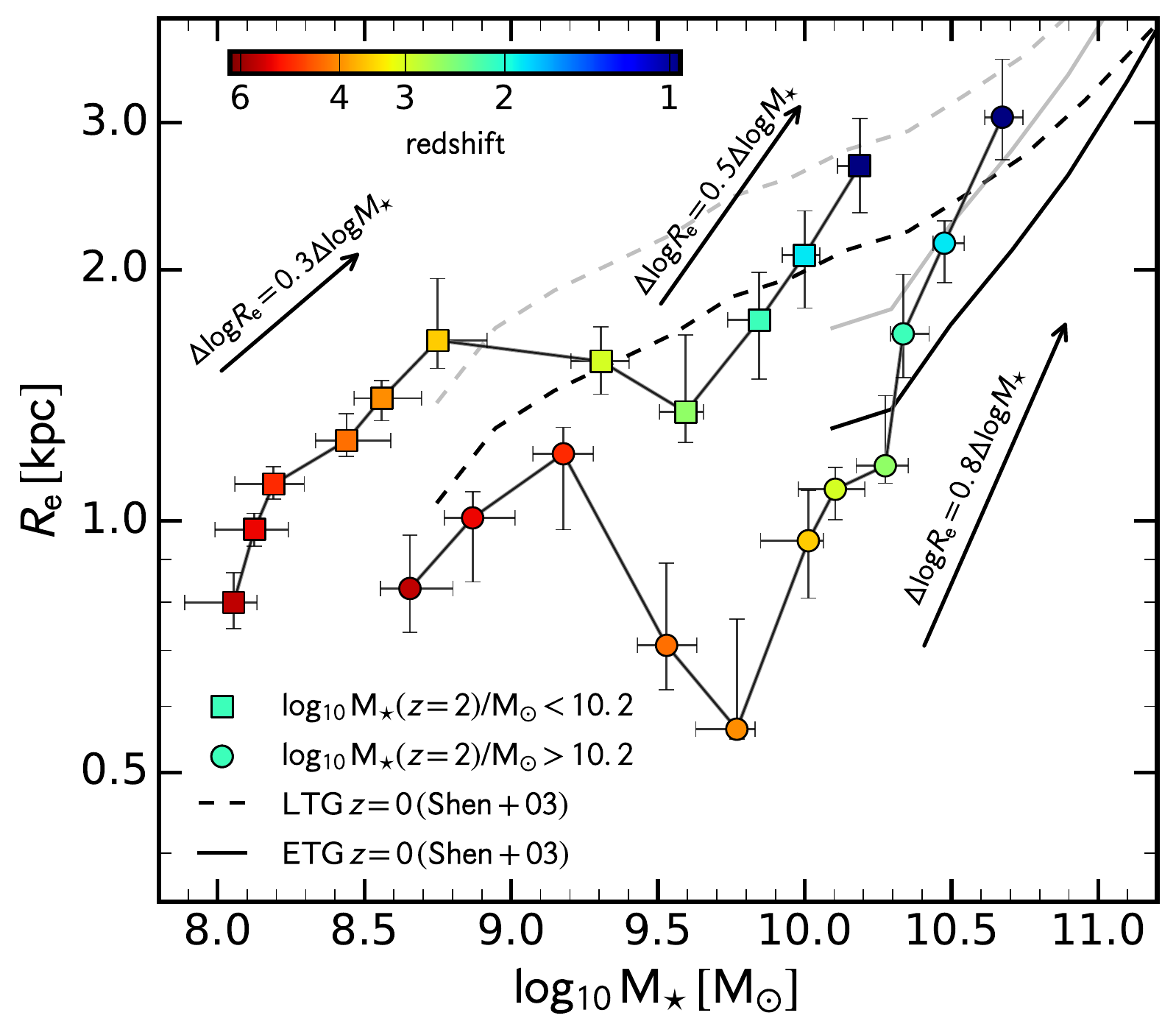} 
\caption{Half-mass size $R_{\rm e}$ versus stellar mass $M_{\star}$. Squares and circles show the median size for the low- and high-mass galaxies, respectively. The colours correspond to redshift according to the colour bar. The errorbars indicate the $1~\sigma$ scatter. The gray solid and dashed lines indicate the half-light sizes for early type (mostly quiescent) galaxies and late-type (mostly star-forming) galaxies from observations of \citet{shen03} at $z=0$. The corresponding black lines show the half-mass sizes by correcting the half-light sizes by $25\%$. The median size of low-mass galaxies stays within $1-2$ kpc, whereas the one of high-mass galaxies decreases from $z=4.9$ to 4.0 from 1.5 kpc to 0.6 kpc and increases from $z=4.0$ to 1.0 from 0.6 kpc to 2.2 kpc.}
\label{Fig_size-mass}
\end{figure}

Z15 studied the size evolution of individual galaxies in the simulations presented here (see their Figure 9). They found that the galaxies indeed go through a compact, nugget phase, with a compactness that is comparable to observations. We take this discussion one step further and look at the median effective size of all galaxies in our evolving sample, as a function of stellar mass and redshift, and compare it with observations. An important point is that most sizes quoted in observations refer to the half-light radius, whereas we determine the stellar half-mass radius in the simulations (and do it in three dimensions). \citet{szomoru13} (see also \citealt{hernquist90}) found that the average (rest-frame $g$-band) light-mass size difference of $\sim25\%$ is roughly the same at all redshifts ($0.5<z<2.5$), and does not strongly correlate with stellar mass, specific star formation rate, effective surface density, S\'ersic index, or galaxy size. 

In Figure~\ref{Fig_size-mass}, we plot the median half-mass size $R_{\rm e}$ versus stellar mass $M_{\star}$ for our low-mass (squares) and high-mass (circles) samples in the redshift range $z=1-6$. At high redshifts ($z>4$) and low stellar masses ($M_{\star}<10^9~M_{\odot}$), both low-mass and high-mass galaxies increase their sizes according to
\begin{equation}
\Delta\log R_{\rm e} = 0.3 \Delta \log M_{\star},
\label{eq:size_growth}
\end{equation}
\noindent
i.e., galaxies increase their size by a factor of 2 for a factor of 10 increase in their stellar mass. At $z\sim3-4$, high-mass galaxies show a significant contraction in a relatively narrow mass range around $M_{\star}\sim10^{9.5}~M_{\odot}$. The median size decrease from $z=4.9$ to 4.0 from $1.2^{+0.1}_{-0.2}$ kpc to $0.56^{+0.20}_{-0.02}$ kpc, and increase from $z=4.0$ to 1.0 from $0.56^{+0.20}_{-0.02}$ kpc to $3.0^{+0.5}_{-0.3}$ kpc. The lower mass galaxies also show the signature of compaction, but to a lesser extent: they compact by a factor of $\sim~20\%$ near $M_{\star}\sim10^{9.5}~M_{\odot}$ and $z\sim2-3$. The more mild compaction seen on average for the low-mass subsample indicates that each compaction event is less pronounced (e.g., because it happens later when the density is lower and the gas fraction is lower), and that the compaction events are spread over a larger range of redshifts and masses, so their signal is smeared in the stacked evolution track.

In the post compaction phase, the galaxies increase their size more significantly. Their growth in the mass-size plane is well approximated by $\Delta\log R_{\rm e} = 0.5 \Delta \log M_{\star}$ and $\Delta\log R_{\rm e} = 0.8 \Delta \log M_{\star}$ for the low-mass and high-mass galaxies, respectively. Both slopes are clearly steeper than before compaction. Due to the pronounced compaction phase at intermediate redshifts and masses, there is only a mild overall growth of the median size from $z=6$ to 1, where $R_{\rm e}$ grows by only a factor $\sim4$ as the mass grows by a factor $\sim100$. The overall growth of size is minor compared to the mass growth for all galaxies over the same period. This is consistent with the picture discussed before in which the stellar mass increases at all radii with a similar pace. Furthermore, $R_{\rm e}$ for the more massive galaxies tends to be smaller at every given mass, indicating that galaxies that formed earlier in a denser universe have also on average a higher stellar (effective) density \citep[e.g.,][]{carollo13a}. 

Interestingly, the maximum compaction tends to occur in a relatively small range of stellar masses around $M_{\star}\sim10^{9.5}~M_{\odot}$. This is apparently the stellar mass range where the core switches from being dark-matter-dominated to baryon-dominated, which largely occurs near the point of maximum compaction (\citealt{zolotov15, ceverino15b}; Tomassetti et al., in prep.). 

In Figure~\ref{Fig_size-mass}, the solid and dashed lines indicate the crude observed size evolution of early-type and late-type galaxies at $z=0$ of \citet{shen03}. The gray lines show the measured half-light sizes, whereas the black lines show the half-mass sizes (assuming a constant correction factor of 25 \%). We see that the sizes of our low and high-mass galaxies at $z=1$ are consistent with the sizes of early-type galaxies at $z=0$, which are mostly quiescent galaxies. As shown before, most of our simulated galaxies are in process of being quenched at $z=1$. Therefore, the simple shut down of star-formation in our galaxies with no additional size (and mass) growth would roughly produce the observed galaxies at $z=0$. 

\begin{figure}
\includegraphics[width=\linewidth]{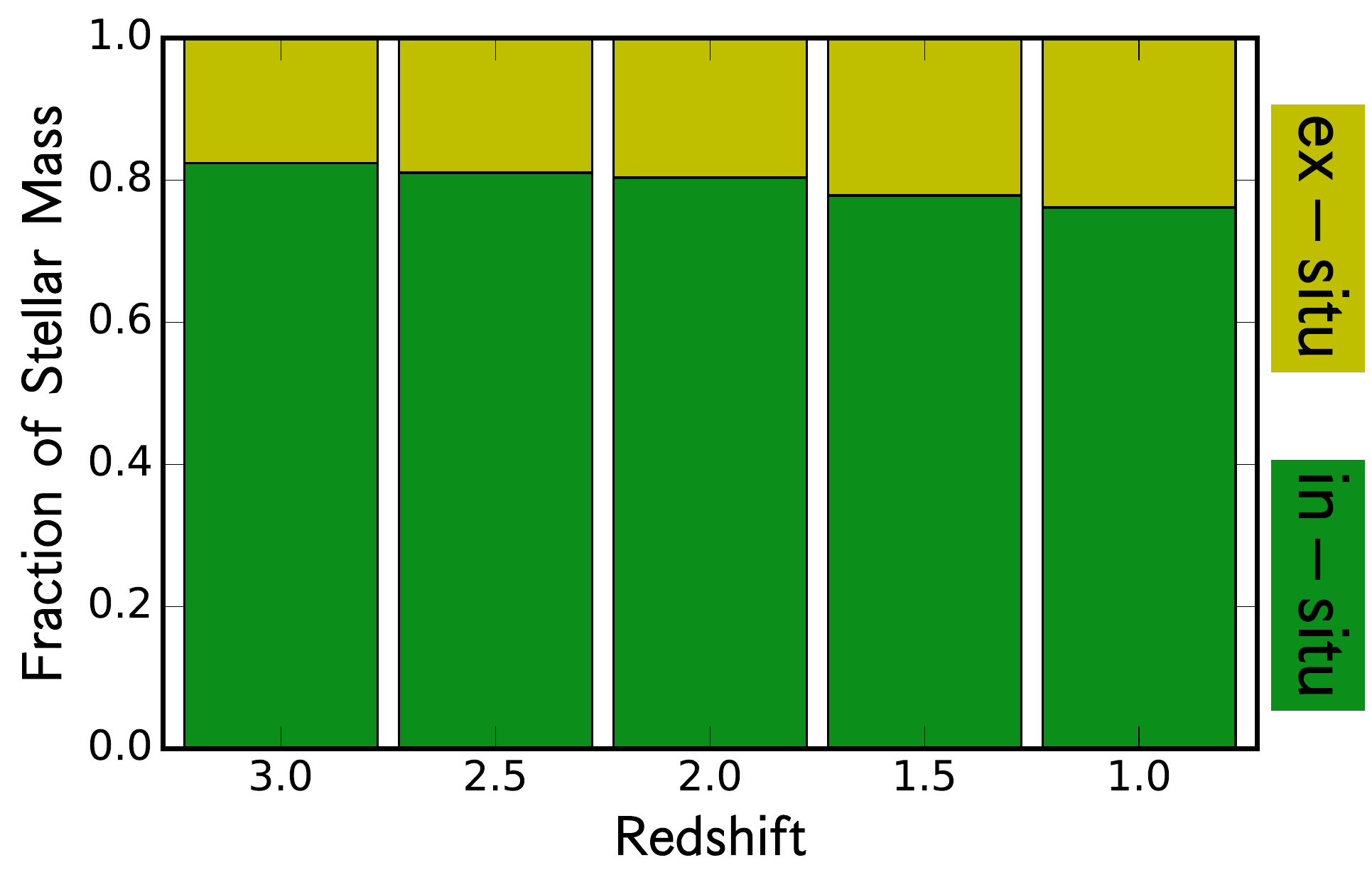} 
\caption{Average fraction of the stellar mass in the outskirts ($1-5~R_{\rm e}$) formed in-situ and ex-situ. A large fraction of $\sim80\%$ of the stellar mass in the disc has formed in-situ, i.e., in the galaxy, and only about 20\% has been accreted. This fraction stays roughly constant with redshift. }
\label{Fig_Where_Stars_Form}
\end{figure}

The late ($z<3$) size growth of our simulated galaxies can be understood by the addition of stellar mass in the outskirts, as shown in Section~\ref{sec:Growth_Galaxies} by analysing the stellar density profiles. To understand if this outer stellar mass growth happens due to accretion of stellar mass or in-situ star formation in an extended ring, we plot in Figure~\ref{Fig_Where_Stars_Form} the fraction of stellar mass of the stars in the outskirts that formed in-situ versus ex-situ. At a certain redshift, stars in the outskirts were selected by choosing all stars within $1-5~R_{\rm e}$. If a star was formed outside of 10\% of the virial radius ($>0.1~R_{\rm vir}$), then the star was classified as ex-situ. On the other hand, all stars that formed within $0.1~R_{\rm vir}$ were classified as in-situ. We find that most stellar mass in the outskirts formed in-situ: $\sim80\%$ of the mass was formed in-situ at all redshifts down to $z\approx1$.

\section{Evolutionary Phases}\label{sec:phases}

\begin{figure*}
\includegraphics[width=\textwidth]{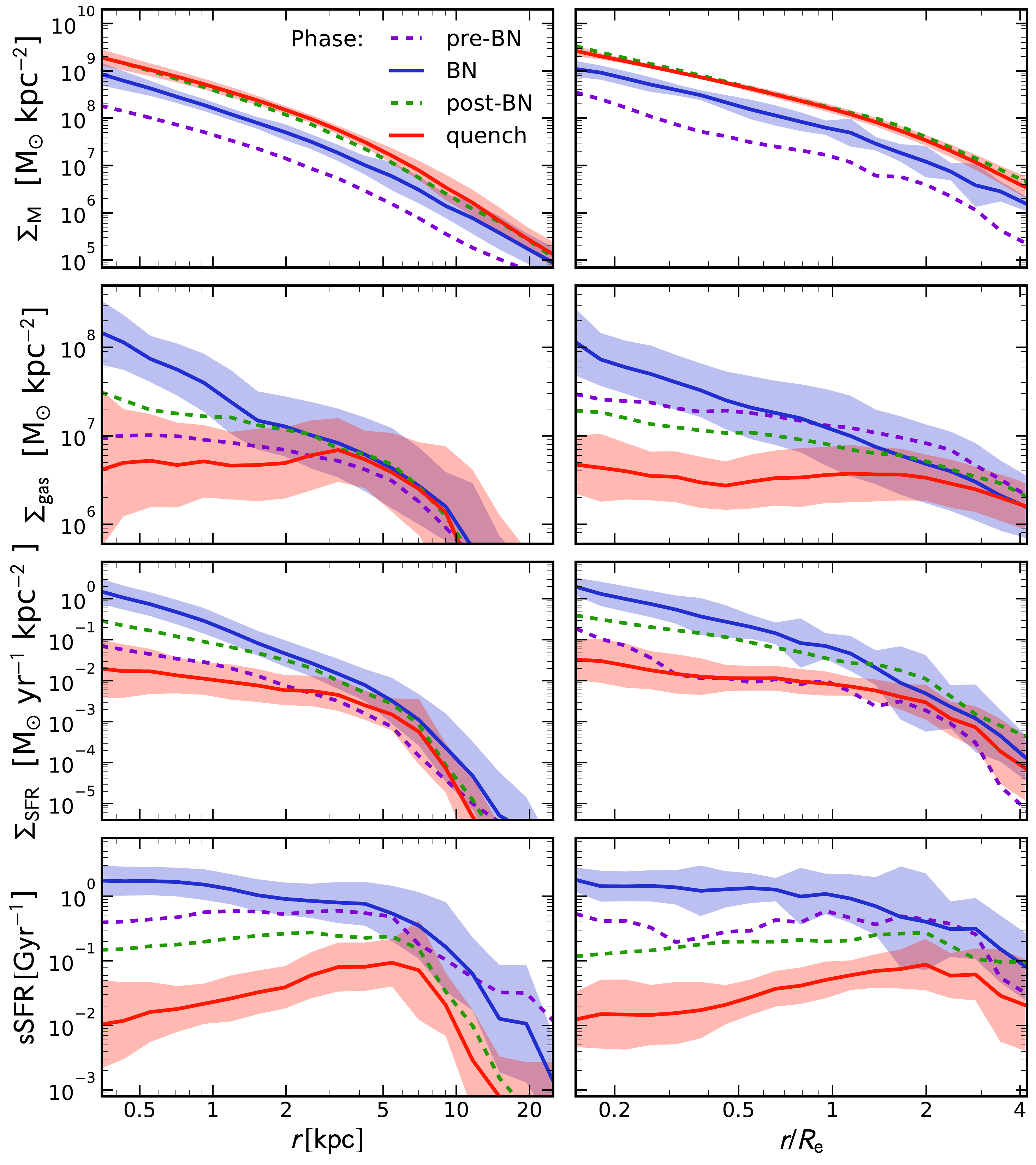} 
\caption{Stacked profiles during the four galaxy evolution phases. Shown from top to bottom are the stellar mass density, gas mass density, SFR density, and sSFR. On the left, the radius is in kpc, and on the right it is scaled by the effective radius $R_{\rm e}$. The individual profiles of the simulations are stacked to have the same median effective density. The four phases are pre-blue nugget (pre-BN; dashed purple), blue nugget (BN; solid blue), post-blue nugget (post-BN; dashed green), and quenching phase (quench; solid red), where the red nugget is surrounded by a star-forming ring. The shaded areas show $1\sigma$ scatter.}
\label{Fig_Season_Profile}
\end{figure*}

We are now going to interpret the evolution of the surface density profiles in the context of the characteristic evolutionary phases of high-$z$ galaxies. As mentioned in the Introduction, Z15 put forward a picture in which galaxies evolve through gas compaction to compact, star-forming blue nuggets and then quench inside-out. Instead of stacking the profiles at a certain redshift as carried out in Section~\ref{sec:Growth_Galaxies}, we now stack the profiles of the galaxies at four evolutionary stages: the pre-blue nugget phase, i.e., when the galaxy was an SFG close to the MS ridge, the blue nugget phase of peak central gas density and SFR, the post-blue nugget phase when central quenching has started, and the quenching phase, where the core has quenched to a red nugget. The individual profiles of the simulations are stacked to have the same effective density, following the procedure outlined in Section~\ref{subsec:surface_profiles}. Figure~\ref{Fig_Season_Profile} shows from top to bottom the surface density profiles of stellar mass, gas mass, SFR and sSFR. In the left panels, we plot the profiles as a function of the physical radius in kpc, while in the right, the profiles are shown as a function of the radius with respect to $R_{\rm e}$. The different lines indicate the mean profile at the pre-blue nugget phase (``pre-BN''), the blue nugget phase (``BN''), the post-blue nugget phase (``post-BN''), and the quenching phase (``quench''). 

These four phases have been identified for each galaxy individually based on Z15 and T15a. Briefly, in T15a we studied in detail the galaxy properties with respect to the star-forming main sequence ridge. The blue nugget phase is the snapshot in which the galaxy reaches its global maximum in the central 1 kpc gas mass. This happens for all galaxies at $z\sim1-4$. The pre-blue nugget phase is identified to be the snapshot where the galaxy is still on the main sequence ridge (distance from the main sequence $<0.1$ dex), before leaving it towards higher sSFR and entering the blue nugget phase. This is typically prior or during the compaction process. Usually, this is $3-5$ snapshots ($300-500$ Myr) before the blue nugget phase. The post-blue nugget phase is defined to be the snapshot after the blue nugget phase, where the galaxy is arriving at the lower edge of the main sequence, which usually is $2-4$ snapshots ($200-400$ Myr) after the blue nugget phase. Finally, the quenching phase is the point in time where the galaxy has the lowest sSFR with respect to the main-sequence ridge, which usually is one of the final snapshots ($z\sim1-2$). 

In Figure~\ref{Fig_Season_Profile}, there is a clear gas cusp in the inner 1 kpc in the blue nugget phase, where the gas surface density reaches $>10^8~M_{\odot}~\mathrm{kpc}^{-2}$ in the centre. Before and after the blue nugget phase (pre-BN and post-BN), the gas surface density in the inner region amounts to $\sim10^7~M_{\odot}~\mathrm{kpc}^{-2}$. Therefore, a gas compaction event takes place where the inflow rate $>$ SFR $+$ outflow rate, i.e., the contraction is wet. The SFR profile follows closely the gas mass profile, following the SFR recipe incorporated in the simulations. During the blue nugget phase, the SFR density in the centre increased by an order of magnitude and the SFR profile is cuspy. A weak, but interesting feature is that the sSFR in the disc before compaction is high ($>0.01~\mathrm{Gyr}^{-1}$) out to 20 kpc, whereas after the blue nugget phase, this high sSFR regime in the disc has shrunk to about 10 kpc. This is explained by the fact that in the compaction phase, the SFR slows down in the extended disc, because its gas has contracted and not replenished yet, while in the centre the SFR goes up to a peak at the blue nugget phase. We interpret this as \textit{outside-in} quenching in the early, pre-blue nugget phase. After the blue nugget phase, the flow rate in the central 1 kpc tips to inflow rate $<$ SFR $+$ outflow rate, reducing the central gas mass and SFR density. In the pre-BN, BN, and post-BN phases, the sSFR profile is flat out to $\sim7$ kpc, and therefore, the surface stellar mass density profiles are nearly identical in log-log shape, showing again self-similar growth. 

During the quenching phase, the gas in the central region gets depleted and a gas hole, usually associated with an outer ring, develops. This is consistent with the picture of Z15 (see their Figure 6), where a post-compaction extended gas ring forms, in which the SFR resumes. This is probably from fresh gas that comes with high angular momentum and cannot get rid of it. At later times, the SFR density varies only a little in the outskirts, whereas in the centre, the SFR density declines by about two orders of magnitude. The sSFR shows the typical signature of \textit{inside-out} quenching, where the sSFR in the central part is substantially lower (by about 2 orders of magnitude) than in the outskirts. The absolute difference in the stellar mass surface density profile is small, highlighting the convergence of the mass profile at later evolutionary time, after the compaction and throughout the quenching process. 

Comparing the profiles as a function of radius in kpc with the radius scaled by $R_{\rm e}$, there are no qualitative differences. We see that the cusp of gas density and star-formation rate develops inside a radius of $r<2$ kpc, which corresponds to $\la2~R_{\rm e}$. The sSFR profiles rises in the red-nugget phase out to $\sim5$ kpc, which corresponds to about 2 $R_{\rm e}$. 

Summarizing, a wet compaction causes a development of a gas central cusp in the blue nugget phase. There is an associated peak of SFR in the centre, while the SFR at larger radii decreases (outside-in quenching). The stellar mass profile grows self-similarly, consistent with the flat sSFR profile. After the blue nugget phase, central gas depletion leads to central quenching of the SFR, leading to the saturation of the stellar core. The sSFR profile is rising in the post-blue nugget phases, which can be termed inside-out quenching.

\section{Comparisons to Observations}\label{sec:OBS}

Thanks to the Hubble Space Telescope (\textit{HST}), accurate measurements of the stellar mass density on scales of $\sim1$ kpc have become possible. Several studies have investigated the stellar mass density distribution within galaxies at high redshifts \citep[e.g.,][]{elmegreen09, forster-schreiber11a, wuyts12, van-dokkum13, patel13, lang14, morishita15}. In addition to stellar mass density profiles, T15b presented resolved star-formation rate density distributions in $z\sim2.2$ galaxies. In this section, we compare our simulated stellar mass and SFR density profiles to these observations.

\subsection{Central Stellar Mass Density}\label{subsec:obs_central}

\begin{figure*}
\includegraphics[width=\textwidth]{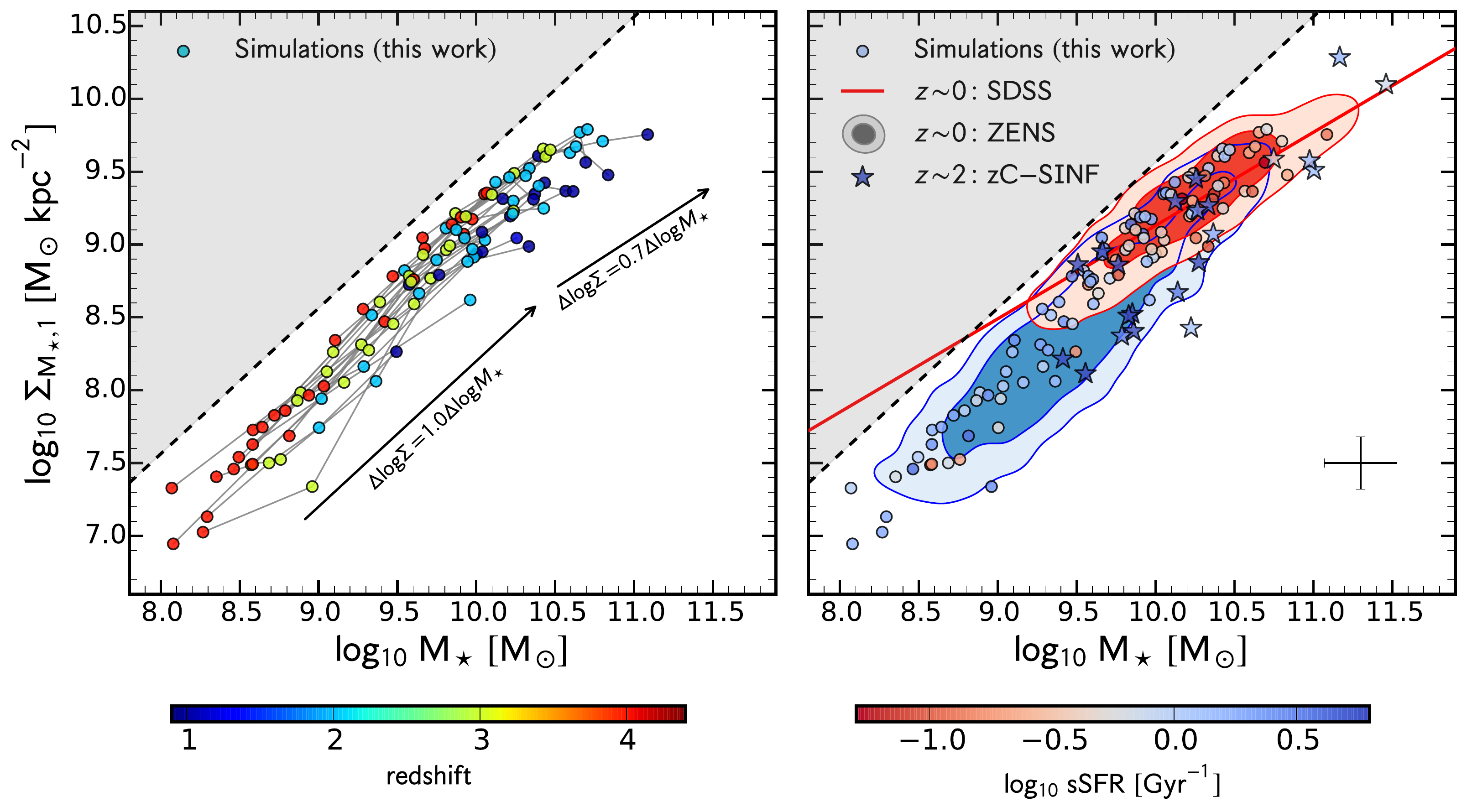} 
\caption{Central stellar mass density within 1 kpc ($\Sigma_{M_{\star},1}$) as a function of stellar mass ($M_{\star}$). The grey shaded area above the dashed line marks the forbidden parameter space where $\Sigma_{M_{\star},1}\geq M_{\star}/\pi$. The colour coding in the left panel corresponds to redshift, while the colour coding in the right panel indicates the total sSFR, highlighting the difference between SFGs and quiescent galaxies. All galaxies grow their central stellar mass density hand-in-hand with their total stellar mass. At $M_{\star}>10^{10.2}~M_{\odot}$, there is a flattening in the $\Sigma_{M_{\star},1}-M_{\star}$ relation, where the total stellar mas still increases while the central stellar mass density has saturated. We compare our simulated galaxies with observations at $z\sim0$ and $z\sim2$ in the right panel. In particular, the blue and red contours show the distribution of $z\sim0$ star-forming and quiescent galaxies of ZENS \citep{carollo13}. The red solid line marks the best fit to the $z\sim0$ red sequence galaxies of SDSS \citep{fang13}. As the $z\sim2$ comparison sample, we use the zC-SINF sample of T15b, plotted with star symbols. There is a good agreement between observations and simulations, and between $z\sim2$ and $z\sim0$. }
\label{Fig_Sigma_Mass_Compare}
\end{figure*}

In Figure~\ref{Fig_Sigma_Mass_Compare}, we plot the central stellar mass density within 1 kpc, $\Sigma_{M_{\star},1}$, as a function of stellar mass, $M_{\star}$, in our simulations compared with observations. The colour-coding of the symbols in the left panel corresponds to redshift. We find a rather tight, nearly linear relation between $\Sigma_{M_{\star},1}$ and $M_{\star}$. All galaxies grow along a well-defined sequence during their SFG phase, i.e., they are increasing the central stellar mass density concurrently with the total stellar mass, following the relation $\Sigma_{ M_{\star},1} \propto M_{\star}^{1.02\pm0.05}$. \citet{ceverino15} used different simulations with much weaker feedback, but finding the same linear relation between $\Sigma_{M_{\star},1}$ and $M_{\star}$, i.e., this suggests that the overall trend is independent of feedback. 

At the high-mass end, around $\sim10^{10.2}~M_{\odot}$, the evolution tracks tend to flatten: the central stellar mass density saturates, while the total mass continues to increase (see also Fig. 16 of Z15). We find the relation $\Sigma_{ M_{\star},1} \propto M_{\star}^{0.68\pm0.01}$ for galaxies with $M_{\star}>10^{10.2}~M_{\odot}$. This flattening starts at the end of the compaction phase and the tracks remain flat during the quenching process. This implies that any further stellar mass growth happens in the outskirts, avoiding the cores, consistent with the aforementioned picture. 

The colour-coding of the symbols in the right panel of Figure~\ref{Fig_Sigma_Mass_Compare} corresponds to the total sSFR. We compare our simulations at $z=1-4$ with observations at $z\sim2$ and at $z\sim0$. As the $z\sim2$ comparison sample, we use the zC-SINF sample of \citet{tacchella15}, indicated by the star symbols. At $z\sim0$, we show the best-fit line to the quiescent sample of SDSS from \citet{fang13} as red solid line ($\Sigma_{ M_{\star},1} \propto M_{\star}^{0.64}$). The contours show the $z\sim0$ ZENS sample \citep{carollo13, cibinel13a, cibinel13b} of star-forming (blue) and quiescent (red) galaxies. The zC-SINF sample as well as our simulations are not mass complete, and therefore, the comparison has to be take with care.

We find excellent agreement between the overall shape of the $\Sigma_{M_{\star},1}-M_{\star}$ sequence in the simulations and in the observations, both for SFGs and for quenched galaxies. T15b already pointed out that galaxies at high and low redshifts practically lie on top of each other in the $\Sigma_{M_{\star},1}-M_{\star}$ diagram. This may indicate that galaxies grow along this sequence, which is indeed confirmed in the simulations (Figure~\ref{Fig_Sigma_Mass_Compare} and also Z15). In both the simulations and observations, there is a flattening of the $\Sigma_{M_{\star},1}-M_{\star}$ relation in the high-$M_{\star}$/high-$\Sigma_{M_{\star},1}$ end, associated with the central quenching process.

\subsection{Density Profiles}\label{subsec:obs_density}

\begin{figure}
\includegraphics[width=\linewidth]{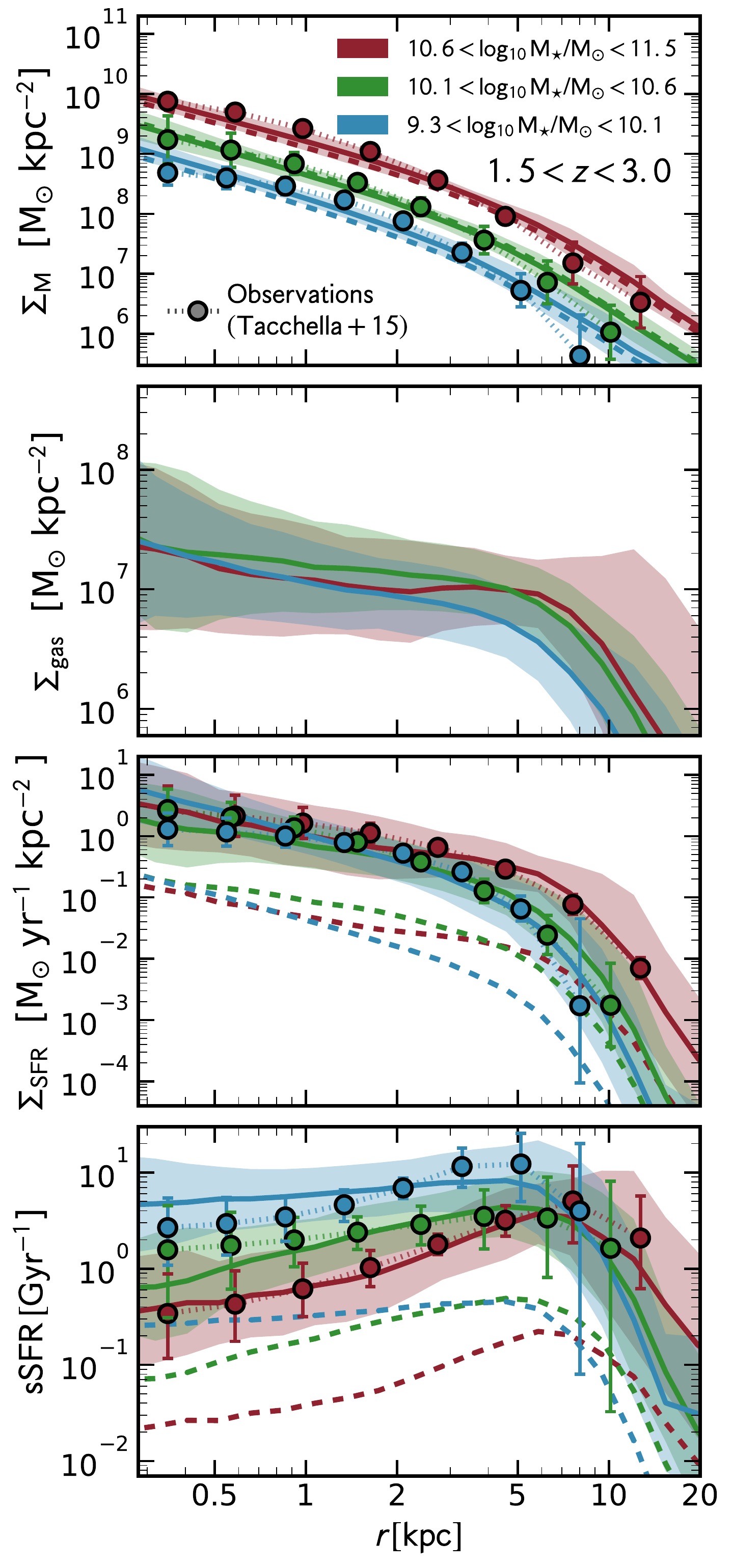} 
\caption{Comparison of the shapes of the stacked profiles from the simulations at $z=1.5-3.0$ with  the estimates from observations of star-forming galaxies \citep{tacchella15_sci}. The stacked surface density profiles of stellar mass, gas mass, SFR, and sSFR are shown from top to bottom. The color coding corresponds to three different mass bins as indicated in the legend. The circles, connected by dotted lines, show the observational data. The thick solid lines show the renormalized simulated galaxies, whereas the dashed lines show the measured profiles of the simulated galaxies. Overall, we find good agreement in the profile's shapes between simulations and observations. However, the normalization, especially of the SFR profiles, differ up to $\sim1$ dex between observations and simulations. }
\label{Fig_Profiles_Comparison_Data}
\end{figure}

The total stellar masses and SFR are two fundamental global parameters of each galaxy and therefore have been measured in many observational works. However, only very recently, it became possible to resolve galaxies with broad-band photometry from the \textit{HST} at high redshifts ($z\sim2-3$) and determine stellar masses on spacial scales of $1-2~\mathrm{kpc}$ \citep[e.g.,][]{wuyts12, van-dokkum13, patel13, lang14, morishita15}. \citet{morishita15} used $HST$/WFC3 and ACS multi-band imaging data taken in CANDELS and 3D-HST to study the stellar mass growth of the $z=0.5-3$ progenitors of local galaxies with mass comparable to the Milky-Way and more massive galaxies, based on a constant cumulative number density selection of galaxies \footnote{Connecting progenitors and descendants using a fixed comoving number density arguments is problematic, especially over long time-scales such as form $z=2$ and $z=0$. More careful analyses using an evolving number density approach or numerical simulations yield $2-3$ times lower masses for the progenitors \citep[e.g.,][]{behroozi13c, lu15, torrey15}.}. They conducted a radially resolved pixel SED fitting to obtain the radial distributions of the stellar mass and rest-frame colours by stacking several tens of galaxies at a given redshift and mass bin. They found that the stellar mass profiles of Milky-Way-like progenitors increase in a self-similar way, irrespective of radial distance. However, the profiles of massive galaxies grow inside-out; about 75\% of their current total mass has grown since $z\sim2.5$ outside the inner 2.5 kpc. This is consistent with previous studies conducted by \citet{van-dokkum13} and \citet{patel13}. 

As shown in Section~\ref{subsec:surface_profiles}, we find exactly the same qualitative behaviour in our simulations: the low-mass galaxies preserve the log-log shape of the stellar mass density profile at all redshifts, whereas the more massive galaxies grow self-similarly only at $z>3$. Toward lower redshifts, the stellar mass growth saturates first in the centre. The stellar mass in the outskirts, and thus the total stellar mass, still increases at lower redshifts; $\sim50\%$ of the total stellar mass is assembled between $z\sim2$ and $z\sim1$ in the outskirts ($r>2.5$ kpc). More quantitatively, we compare the exponential growth rate $\alpha$ from Equation~\ref{eq:fitting_growth} as derived from the observations and the simulations. For the lower mass galaxies, we fit the data at $z=1.0-2.5$ with $z_0=1.0$. We find $\alpha=1.16$ and $\alpha=1.17$ for the data of \citet{van-dokkum13} and \citet{morishita15}, respectively. In the simulations (Section~\ref{subsec:innerVSouter}), we obtained $\alpha=0.99$, which is in the same ball park, perhaps slightly smaller, i.e., the lower mass galaxies in the simulations tend to form stars a little earlier then the ones in the observations. For the more massive galaxies, we fit the data of \citet{patel13} at $z=1.0-2.5$ with $z_0=1.0$, finding $\alpha=0.52$. The data of \citet{morishita15} shows a break at $z=2.0$, so we fit two redshift bins independently from each other. At $z=2-3$ with $z_0=2$ we find $\alpha=1.89$, and for $z=1-2$ with $z_0=1$ we find $\alpha=0.46$. In the simulations, recall that we find $\alpha=0.41$ with $z_0=1.0$ at $z=1.0-3.5$, which is also in reasonable agreement with the observations.

\citet{morishita15} used the rest-frame $U-V$ colour to infer that the quenching of the central dense region is ahead of the outer disc, which is also qualitatively consistent with the inside-out quenching found in the simulations. Unfortunately, the rest-frame $U-V$ colour is an uncertain tracer of SFR, since it depends on the assumed star-formation history as well as dust properties. Instead of using broad-band photometry as a tracer for star-formation, T15b used the H$\alpha$ emission line from \textit{VLT}/SINFONI integral field unit spectroscopy as part of the SINS/zC-SINF effort \citep{newman13a, forster-schreiber14, genzel14a, tacchella15, tacchella15_sci}. The H$\alpha$ emission line traces the recent SFR (on timescales of $10^7$ yr), independent of the previous star-formation history. Thanks to adaptive optics, the same \textit{HST}-like, 1 kpc resolution is obtained for the H$\alpha$ emission line maps. T15b presented a sample of 22 $z\sim2.2$ galaxies with both \textit{HST} stellar masses and SINFONI SFR maps, which spans a wide range in stellar mass ($4\times10^9-5\times10^{11}~M_{\odot}$) and SFR ($20-300~M_{\odot}~\mathrm{yr}^{-1}$), probing the star-forming main sequence at $z\sim2$. 

In Figure~\ref{Fig_Profiles_Comparison_Data} we compare the stacked stellar mass, gas mass, SFR and sSFR profiles of the simulations with the observations at $z~1.5-3.0$. We stack the simulated galaxies in the same mass bins as the observations of T15b (with a median redshift of $z\sim2.2$), namely, in a low- ($9.3<\log_{10}~M_{\star}/M_{\odot}<10.1$), an intermediate- ($10.1<\log_{10}~M_{\star}/M_{\odot}<10.6$), and a high-mass bin ($10.6<\log_{10}~M_{\star}/M_{\odot}<11.5$). The individual profiles of the simulations and observations are each stacked to have the same effective stellar mass (and SFR) density, following the description outlined in Section~\ref{subsec:surface_profiles}. The dashed lines show the stacked profiles of the simulations. In order to compare the profiles' shapes, the main focus of this comparison presented here, we may need to re-normalize the profiles of the simulations by requiring them to have the same $\Sigma_{\rm eff}$ as the one from the observations (shown as solid lines).

Important to mention are the following two caveats in the comparison between simulations and observations in Figure~\ref{Fig_Profiles_Comparison_Data}. First, the SFRs are estimated in different ways, both associated with significant uncertainties. While the observed SFR is estimated from the dust-corrected H$\alpha$ emission-line flux, in the simulations it is estimated by counting the newly born stars over a given period. Second, both the observed and simulated samples are not complete samples in terms of mass or SFR, incomplete in different ways, which could lead to a bias. In particular, the observed sample consists mainly of SFGs on the main sequence, while no explicit selection on SFR is applied in the simulations. This can explain at least part of the difference between the average SFR at a given mass, which is lower in the simulations compared to the observations. The motivation for the comparison of the simulations to observed SFGs is our finding in T15a (where we modeled the star-forming main sequence self-consistently within the simulations) that most of our simulated galaxies are on the main sequence at $z > 2$. However, they do tend to fall bellow the Main Sequence toward lower redshifts and when they have higher masses, thus introducing a bias in the comparison. What matters to us here is that the shapes of the profiles in the different phases of evolution, including the SFR profiles, are consistent between the simulations and observations. This indicates that the simulations capture the important physical processes, at least qualitatively, and allow us to understand the evolution through compaction and quenching phases.

The top panel of Figure~\ref{Fig_Profiles_Comparison_Data} shows the simulated and observed stacked stellar mass density profiles. We obtain as a re-normalisation factor for the stellar mass profiles of the low-, intermediate-, and high-mass sample small offsets of $+0.13$, $-0.05$ and $+0.10$ dex, respectively, namely, the required re-normalization is negligible. In the intermediate- and high-mass bin, we find good agreement in the log-log profile shape. In the lower mass bin, there is a certain difference in the profile shapes: the profile from observations shows a more disc-like profile (with $n\sim1$), whereas the one from simulations is more spheroidal-like (with $n\sim3$, see Fig.~\ref{Fig_sersic-mass}).

The low-mass profile of the simulations does resemble the high-mass profile of the observations. The second panel from the top shows the gas mass profile of the simulations only, since there is no available data for the resolved gas mass profile. We see a rather flat profile between 1 and 5 kpc, and a steeper decline at larger radii. 

In the bottom two panels of Figure~\ref{Fig_Profiles_Comparison_Data}, we show the SFR and sSFR profiles. We find a substantial re-normalisation factor for the SFR profiles: we have to re-normalize the simulated SFR profiles of the low-, intermediate-, and high-mass sample by a factor of $+1.39$ dex, $+0.91$ dex, and $+1.33$ dex, respectively. This offset between observations and simulations could partly be due to the caveats mentioned above, and possibly also due to too low gas fractions and total SFRs in the simulations at $z\la2$ as a result of the SFR being overpredicted at much higher redshifts (see also T15a), but possibly also due to overestimation in the observations. 

Focusing on the shape of the profiles, we find overall good agreement between the profiles' shapes in the observations and in the simulations. For the low-mass galaxies, the simulated SFR profile is slightly flatter in the inner 3 kpc than the observed one. In the intermediate-mass bin it is the opposite. However, the shapes of the SFR profiles of the simulated galaxies agree within the errors with the observed ones. The high-mass profile of the simulations and observations are in very good agreement. As highlighted in Section~\ref{sec:Growth_Galaxies}, the sSFR is increasing towards the outskirts. This effect of reduced star-formation activity toward the centre is stronger for intermediate-mass galaxies than for low-mass galaxies. As discussed in Sections~\ref{sec:Growth_Galaxies} and \ref{sec:phases}, the rise in the sSFR profile can be interpreted as a central gas depletion after the blue nugget phase, which is followed by an extended star-forming ring. This is consistent with the inside-out quenching scenario proposed by T15b and Z15. Overall, the agreement between the shapes of the observed and simulated profiles at $z\sim2$ is an encouraging sanity check. It indicates that the observed profile shapes can be interpreted as reflecting the build-up of a compact bulge by wet compaction and the subsequent inside-out quenching, as seen in the evolution of the simulated profiles in Figure~\ref{Fig_Season_Profile}.

\subsection{Inside-Out Quenching}\label{subsec:obs_quenching}

\begin{figure}
\includegraphics[width=\linewidth]{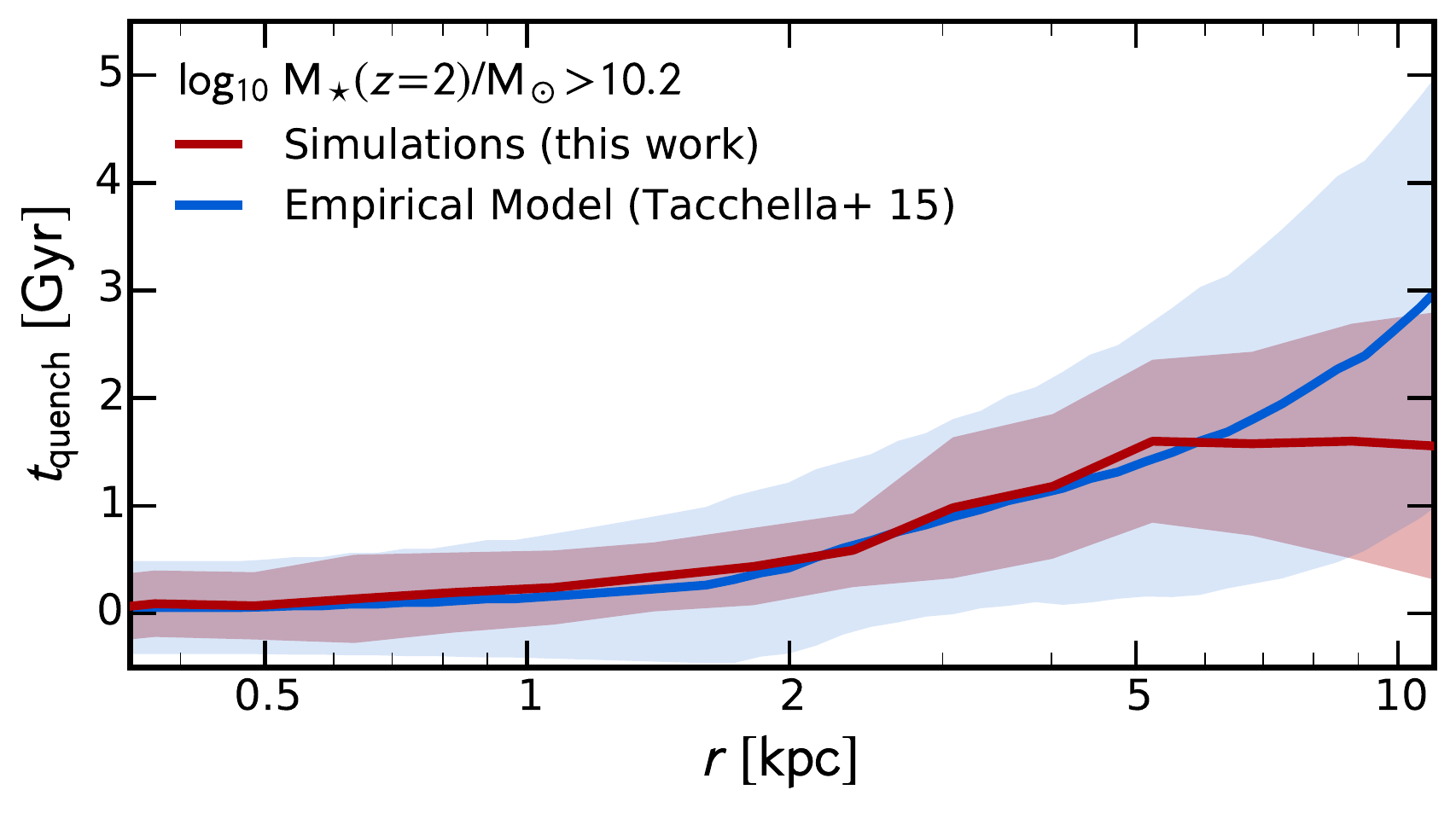} 
\caption{Progression of the ``quenching wave''. The red line shows progression of quenching within the massive galaxies ($M_{\star}>10^{10.2}~M_{\odot}$) in our simulation (shaded regions indicates the $1~\sigma$ scatter). We find that the galaxies quench their outskirts ($>5~\mathrm{kpc}$) $1-2~\mathrm{Gyrs}$ after they started quenching in their centre. The blue line represents the empirical model based on observations of \citet{tacchella15_sci}, showing excellent agreement. }
\label{Fig_Quenching_Time_Profile}
\end{figure}

To quantify the inside-out quenching phase more quantitatively in the simulations and allow a more direct comparison to observations, we measure the time when the sSFR falls below a critical sSFR threshold at every given radius. We set this threshold to be $\mathrm{sSFR_{thresh}}=0.1~\mathrm{Gyr^{-1}}$. The motivation for this number is that a sSFR of $0.1~\mathrm{Gyr^{-1}}$ roughly corresponds to a mass doubling time of $10~\mathrm{Gyr}$, i.e., several times the Hubble time at $z\ga2$. This means that the mass increase from star-formation below $\mathrm{sSFR_{thresh}}$ is negligible. Changing $\mathrm{sSFR_{thresh}}$ in the range $0.05-0.2~\mathrm{Gyr^{-1}}$ does not change our results significantly. 

Figure~\ref{Fig_Quenching_Time_Profile} shows the progression of the inside-out ``quenching wave''. For all massive galaxies ($M_{\star}>10^{10.2}~M_{\odot}$), we measure at which cosmic epoch and radius the sSFR drops below $\mathrm{sSFR_{thresh}}$. The ``quenching'' time $t_{\rm quench}$ of a galaxy starts ticking when the first radial bin quenches. The massive galaxies in the simulations start quenching ($t_{\rm quench}=0$) at $z=2.4^{+0.7}_{-0.5}$. By $z=1$, seven out of 12 massive galaxies are fully quiescent, i.e., they are $>0.3$ dex below the main sequence (T15a), while the other 5 galaxies lie on the lower edge of the main sequence. We see that in the simulations, the galaxies quench their inner regions first, while the outskirts quench $1-2~\mathrm{Gyrs}$ later (red line in Figure~\ref{Fig_Quenching_Time_Profile}). 

In T15b we have used an empirical model to constrain the progression of inside-out quenching motivated by observations. The main idea was to use the measured stellar mass and SFR profiles for the most massive galaxies in that sample ($M_{\star}\sim10^{11}~M_{\odot}$) and compare them with local $z\sim0$ galaxies. By assuming that the galaxies' star formation rates follow the ones of the star-forming main sequence at any given time, we estimated when the galaxy has to stop forming stars at the certain radius so that they do not overshoot the $z=0$ stellar mass profile. We found in T15b that $z\sim2$ galaxies have to cease the star-formation in their centres very soon ($<<1~\mathrm{Gyr}$), while they keep forming stars for a few Gyr in their outskirts. Since this empirical model does not take into account mergers or stellar migration, the obtained timescale for quenching is an upper bound, since dry merging would contribute to the mass profile as well, without increasing the SFR there. 

In Figure~\ref{Fig_Quenching_Time_Profile} we show the result from this empirical model in blue. As one can see, the model agrees with the simulations out to a few $R_{\rm e}$ ($\sim5~\mathrm{kpc}$). At $r\sim10$ kpc, the simulations predict a shallower slope in $t_{\rm quench}$ than what we have predicated with the T15b empirical model. As mentioned above, this difference can possibly be explained by merging.

\subsection{Physics of Quenching}\label{subsec:physics_quenching}

What is the physical origin of the inside-out quenching? This is discussed in detail in our companion papers (T15a and Z15). Specifically, in T15a, we study in detail the flow rates in the central 5 kpc (see Figure 9 and D1 of that study). We find that the blue nugget phase comes rather quickly to an end when the inflow rate into the central regions becomes smaller than the SFR $+$ outflow rate inside that region. The gas depletes in the centre quickly due to the formation of new stars and the associated stellar feedback that derives winds, while there is no gas supply from the disc that has shrunk and at least temporarily disappeared. During this phase, the galaxies reduce its star-formation substantially (especially in the inner regions) and the sSFR profile is rising toward the outskirts as a new, extended gas ring is forming. Newly accreted gas through the halo and into the galaxy can initiate new star-formation, such that the galaxy does not necessarily quench for good.

For full and permanent quenching, we argue in T15a that the disc replenishment time (i.e., timescale for accretion of fresh gas into the galaxy) is longer than the depletion time (sometimes also called the gas consumption time). This is the case when the circumgalactic medium in the host halo becomes hot, typically when the halo mass is above the critical mass for virial shock heating, and at late times when cold streams do not penetrate efficiently \citep{birnboim03, keres05, dekel06, zolotov15}. As mentioned in Section~\ref{subsec:limitations}, full quenching to very low sSFR is not always reached in the current simulations because the feedback may still be underestimated (e.g., the adiabatic phase of supernova feedback is not resolved, and AGN feedback is not yet incorporated). However, 12 out of our 26 galaxies drop to more than 1$\sigma$ below the main sequence ridge by $z=1$, and all are continuously reducing their sSFR for the last several Gyrs, indicating that they are in a long-term quenching process.

\section{Conclusion}\label{sec:Conclusion}

We have used a suite of 26 zoom-in cosmological simulations to investigate the development of the density profiles in galaxies at $z=1-7$ through their characteristic phases of evolution. The galaxies compactify to blue nuggets and then quench inside-out, developing a red-nugget core that is usually surrounded by a star-forming ring. Overall, we find good agreement between observations and simulations for the main features of wet compaction, inside-out quenching, and saturation of dense stellar cores. We summarize our main findings below.

To shed light on the physical process in these high-$z$ galaxies, we have stacked the profiles of the galaxies based on their four evolutionary states with respect to compaction and quenching: the pre-blue nugget, blue nugget, post-blue nugget, and quenching phases, following Z15 and T15a. The emerging picture is that a wet compaction \citep{dekel14_nugget}, caused by gas-rich mergers or smoother gas streams (such as counter-rotating streams and low-angular momentum recycled gas), leads to a high central gas density above $10^8~M_{\odot}~\mathrm{kpc}^{-2}$ and an episode of high central SFR, i.e. a blue nugget. This short phase can be interpreted as an outside-in quenching episode, where the SFR slows down in the shrinking disc, while in the centre the SFR peaks in the blue nugget phase. In this phase, the gas and SFR profiles are cuspy. 

The blue nugget phase comes rather quickly to an end when the balance of the flow rates in the central 1 kpc tips to inflow rate $<$ SFR $+$ outflow rate. The immediate reason for this is that while the SFR $+$ outflow rate are at their maximum, the inflow is suppressed because the disc has largely disappeared (shrunk). In parallel, the bulge has grown, so the remaining disc is stable gravitationally (morphological quenching, see \citealt{martig09, cacciato12, genzel14a, tacchella15_sci}) and does not drive more gas inward. After the gas has depleted in the centre, a star-forming, gas-rich ring forms in some of the galaxies. This is consistent with the 1D axisymmetric thin disc model of \citet{forbes14} based on continuous disc instability with \citet{toomre64} $Q$-parameter of the order of unity. They showed that at every radius galaxies tend to be in a slowly evolving equilibrium state (as predicted in a cosmological setting by \citealt{dekel09b} and in \citealt{cacciato12}) and they explicitly predicted the formation of a central gas ``hole'' with an extended gas ring. 

In the post-blue nugget phase, the sSFR is rising toward the outskirts, which we interpret as inside-out quenching phase. We argue that a galaxy departs from the main sequence and quenches once the disc replenishment time is longer than the core depletion time. This naturally happens at late redshifts, and earlier for more massive galaxies (``downsizing'', T15a). Quenching can be maintained for long term when the replenishment time becomes even longer, e.g., once the circum galactic medium in the host halo becomes hot, typically when the halo mass is above the critical mass for virial shock heating, and at late times when cold streams do not penetrate efficiently \citep{birnboim03, keres05, dekel06, zolotov15}. 

The surface stellar mass density profiles of low-mass galaxies grow self-similarly at $z=1-7$; on average, the stellar mass density increases at all radii with a similar multiplicative pace. This can also be characterised by a flat average sSFR profile. The rate of (total) stellar mass growth for the low-mass galaxies is well described by $M(z)=M(z=1)\times\exp[-1.0(z-1)]$, which is in excellent agreement with the halo growth rate, as measured in simulations and predicted by simple theory, and shows that the sSFR is tightly coupled to the specific halo mass increase \citep{wechsler02, neistein08, dekel13}. 

The stellar mass profiles of high-mass galaxies grow self-similarly only at high redshifts ($z\ga3.5$). At lower redshifts, the high-mass galaxies decouple from the halo mass accretion and grow significantly slower. Comparing the growth of the central region with the outskirts, we find that most of mass increase after $z<3$ happens in the outskirts: the stellar mass within 1 kpc saturates starting at $z\approx2.5$. This is consistent with the strong decline with time of the sSFR within 1 kpc at late times, which drops significantly faster than the total sSFR of the galaxies. We interpret this as inside-out quenching during the post-blue nugget phase, following the outside-in quenching during the pre-blue nugget phase.

The central stellar mass density, a proxy for the bulge component, grows concurrently with the total stellar mass during the star-forming phase, typically at early redshifts and at low masses. At the high-mass end, there is a flattening of the $\Sigma_{M_{\star},1}-M_{\star}$ relation, which is compatible with stellar mass growth in the outskirts for high-mass galaxies at late cosmic epochs. The simulations reproduce the observed $\Sigma_{M_{\star},1}-M_{\star}$ relation both at $z\sim2$ and at $z\sim0$. The mass growth in the outskirts at $1-5~R_{\rm e}$ in the redshift range $z\sim1-4$ arises mostly from in-situ star formation and leads to a systematic gradual growth of stellar half-mass size: the simulated galaxies increase their sizes from sub-kpc scales to $2-3$ kpc from $z\sim4$ to 1, which is consistent with the sizes of early-type galaxies in the local universe. We emphasize that $R_{\rm e}$ is evolving rather slowly with time compared to other quantities (such as the total stellar mass $M_{\star}$): $R_{\rm e}$ grows by only a factor of $\sim4$, while the stellar mass grows by a factor of $\sim100$. 

Taking the comparison between observations and simulations one step further, we have compared the profiles of stellar mass, SFR, and sSFR at $z=2.2$. Two important caveats in this comparison are that both the simulated and the observed samples are not mass- and/or SFR-complete (i.e., the observational sample selected mostly SFGs while the simulations did not apply any explicit selection on SFR) and that the SFR in the simulations and observations are obtained in different ways (counting newly born stars versus H$\alpha$ luminosity). For the surface stellar mass density profiles we find good agreement between the simulations and the observations of T15b, both for low-mass and high-mass galaxies, and both for the log-log profile shape and its zero-point normalization. For the SFR profile, the agreement in the profile shape is good, though the overall normalisation in the simulation is substantially lower by up to 1 dex, due to a certain overestimate of the SFR at very high redshifts in the simulations, as discussed in T15a. At $z=2.2$, where most of the galaxies in our sample are in their post-blue nugget phase, we find for both observations and simulations a rising sSFR profile, i.e., the star-formation activity is reduced by an order of magnitude in the centre. This characteristic of inside-out quenching is stronger in the more massive galaxy sample (T15a). 

We conclude that the major compaction event and the subsequent quenching combine to a chain of events of major importance in the history of massive galaxies. In particular, the high-redshift blue nuggets represent a key phase in galaxy evolution. The evolution of profiles of stellar and gas mass and of SFR tightly correlates with these characteristic phases of galaxy evolution, and can help us interpret them theoretically and identify them observationally. Our suggested phases can be directly observed. In the blue nugget phase, we predict a cuspy gas profile, whereas in the post-blue nugget phase, the gas profile is expected to have a hole, sometimes surrounded by a star-forming gas ring that quenches on a longer timescale. These gas signatures can be traced, for example, by IRAM Plateau de Bure and ALMA, which could resolve galaxies at $z\sim1-3$ on sub-kpc scales.

\section*{Acknowledgements}

The authors are thankful to the anonymous referee for her/his useful comments which improved our original manuscript. We acknowledge stimulating discussions with Guillermo Barro, Andi Burkert, Sandy Faber, John Forbes, Reinhard Genzel, David Koo, Mark Krumholz, Simon Lilly, Benny Trakhtenbrot, Joanna Woo, and Adi Zolotov. Development and most of the analysis have been performed in the astro cluster at HU. The simulations were performed at the National Energy Research Scientific Computing Center (NERSC), Lawrence Berkeley National Laboratory, and at NASA Advanced Supercomputing (NAS) at NASA Ames Reserach Center. This work was supported by MINECO grant AYA2012-32295, by ISF grant 24/12, by the I-CORE Program of the PBC, by ISF grant 1829/12, and by NSF grants AST-1010033 and AST-1405962. We acknowledge support by the Swiss National Science Foundation.

\appendix
\section{Stellar-to-halo mass relation}
\label{App:stellarhalo}

\begin{figure}
\includegraphics[width=\linewidth]{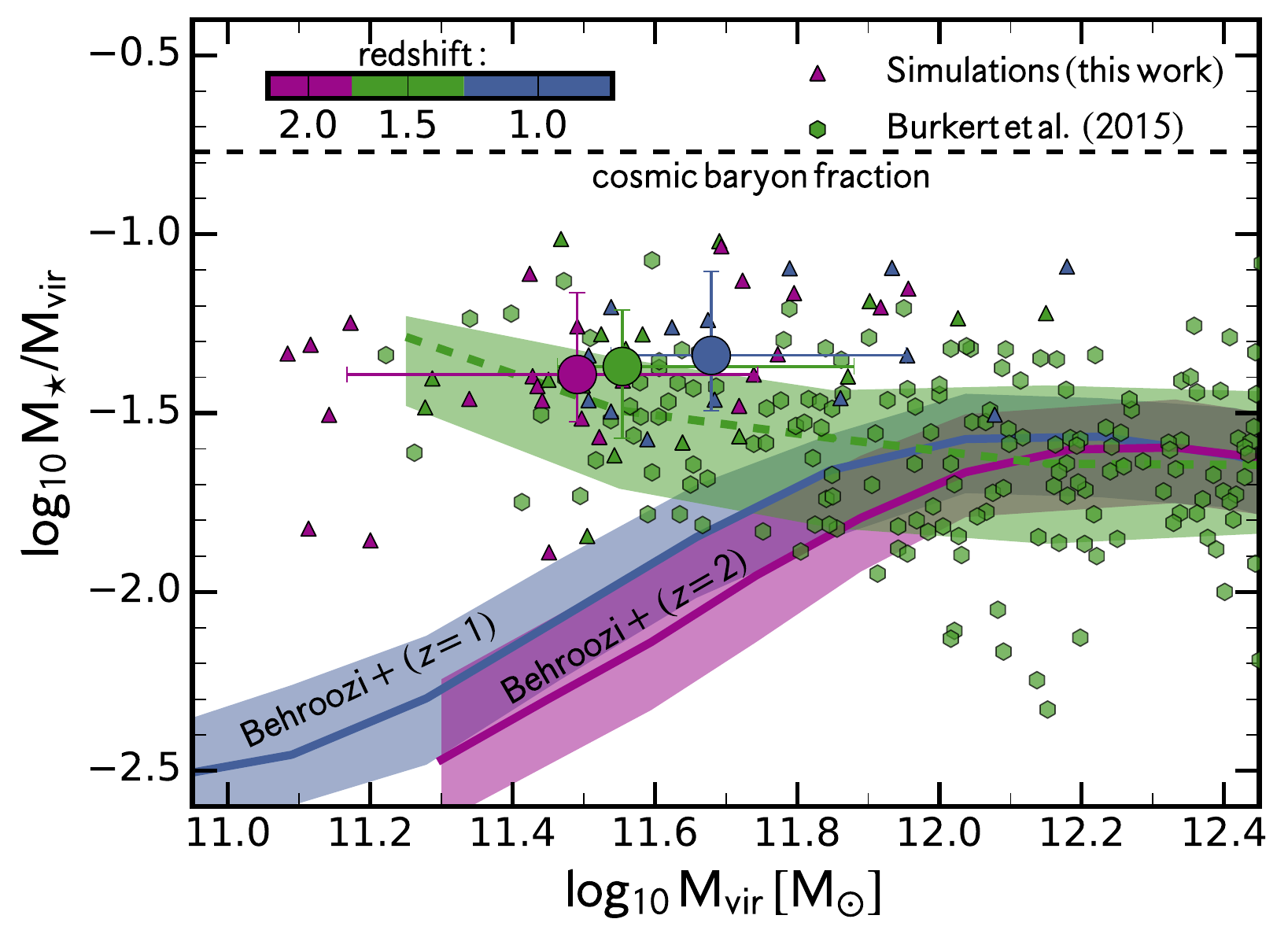} 
\caption{Ratio of stellar mass to halo mass as a function of halo mass. The purple, green and blue point show our simulations at $z=2$, 1.5 and 1.0, respectively. The large circles show the median (at its $1~\sigma$ scatter) at these three redshifts. The horizontal black dashed line shows the cosmic baryon fraction of the Universe. The solid lines show the \citep{behroozi13b} relations. Our simulations lie a factor of $2-6$ above the \citet{behroozi13b} relation at $z\sim1-2$, but are in agreement with recent observations of SFGs at $0.5\la z \la 2.6$ \citep{burkert15}, which are shown as hexagons and their median as dashed line. A similar figure is shown in T15a.}
\label{FigApp:stellarhalo}
\end{figure}

One of the key quantity for simulations to match is the stellar mass made within a given dark-matter halo. In this section, we compare the stellar-to-halo mass ($M_{\star}-M_{\rm vir}$) relation of the simulations with results from abundance matching \citep{conroy09, moster10, moster13, behroozi10, behroozi13b} and observed kinematics \citep{burkert15}. A similar analysis is presented in our companion paper \citep{tacchella15_MS}.

Figure~\ref{FigApp:stellarhalo} shows the ratio of stellar mass to halo mass as a function of halo mass for our simulations and the estimates from observational data \citep{behroozi13b, burkert15}. At $z=2$, we find for our simulated galaxies a median halo mass of $\log_{10}~M_{\rm vir}=11.5\pm0.3$ and a stellar to halo mass ratio of $\log_{10}~M_{\star}/M_{\rm vir}=-1.4\pm0.2$. Via abundance matching, \citet{behroozi13b} found $\log_{10}~M_{\star}/M_{\rm vir}=-2.3\pm0.2$ at $\log_{10}~M_{\rm vir}=11.5$. However, \citet{burkert15} found $\log_{10}~M_{\star}/M_{\rm vir}=-1.6\pm0.3$ at $\log_{10}~M_{\rm vir}=11.5$, which is consistent with our simulations. 

Toward lower redshifts, the overall agreement between the simulations and observations improves. At $z=1$, the simulated galaxies have a median halo mass of $\log_{10}~M_{\rm vir}=11.7\pm0.3$ and a stellar to halo mass ratio of $\log_{10}~M_{\star}/M_{\rm vir}=-1.3\pm0.2$, which is again in agreement with the observational estimates of \citet{burkert15}, and slightly higher than the estimate of \citet{behroozi13b}, who found $\log_{10}~M_{\star}/M_{\rm vir}=-1.8\pm0.2$.

We conclude that our simulations produce stellar to halo mass ratios that are in the ballpark of the values estimated from observations, and within the observational uncertainties, thus, indicating that our feedback model is adequate.

\section{Profiles for all Galaxies}
\label{App:Profiles}

\begin{figure*}
\includegraphics[width=\textwidth]{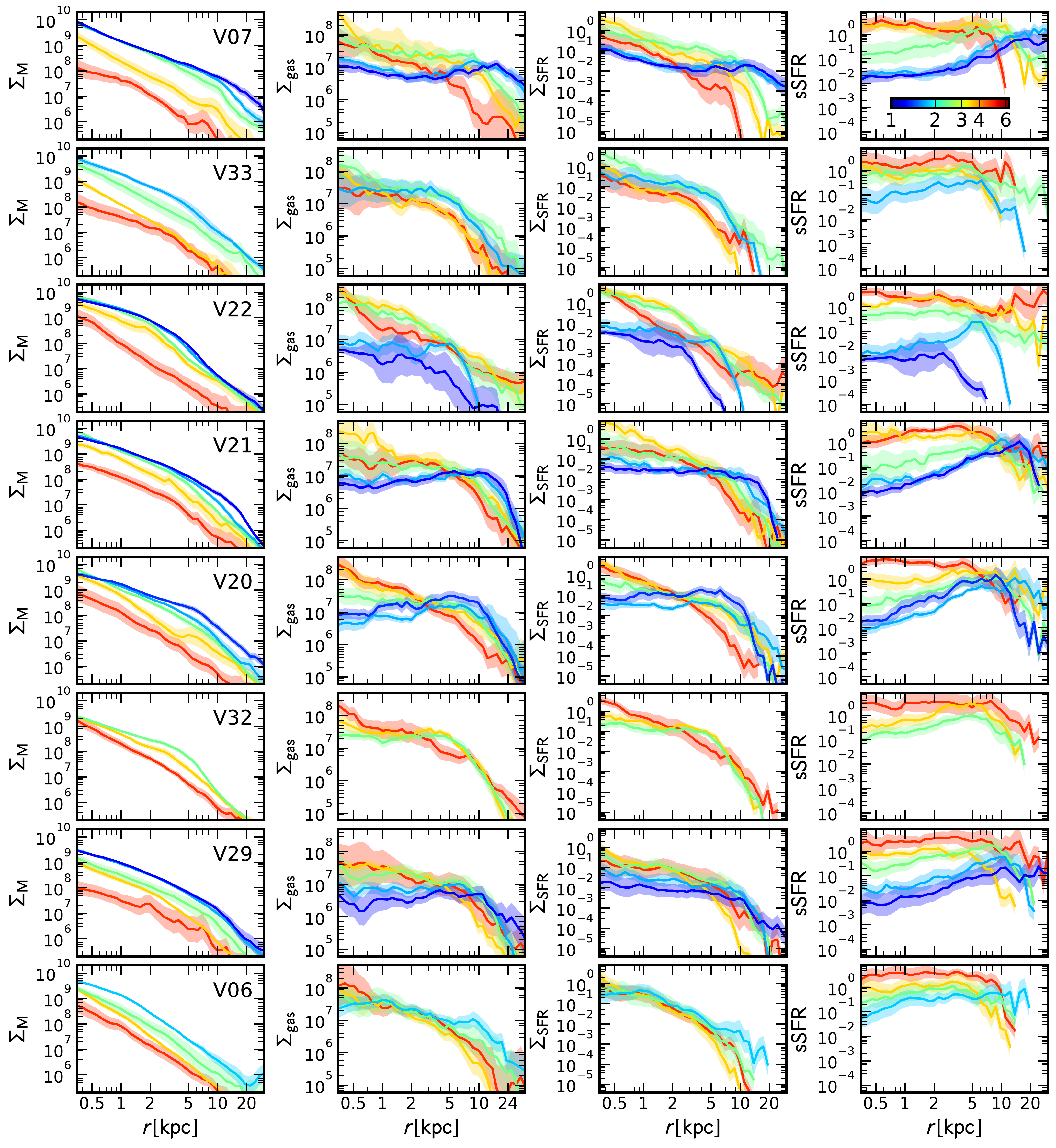} 
\caption{Complete list of stellar mass (left), gas mass (middle-left), SFR (middle-right), and sSFR (right) profiles for all galaxies. We display the profiles at different redshifts bins ($z=6.0, 4.0, 3.0, 2.0, 1.5, 1.0$) with different colours. The shaded regions correspond to the $1~\sigma$ scatter of the profiles in a certain redshift bin. } 
\label{FigA1_Profile}
\end{figure*}

\begin{figure*}
\includegraphics[width=\textwidth]{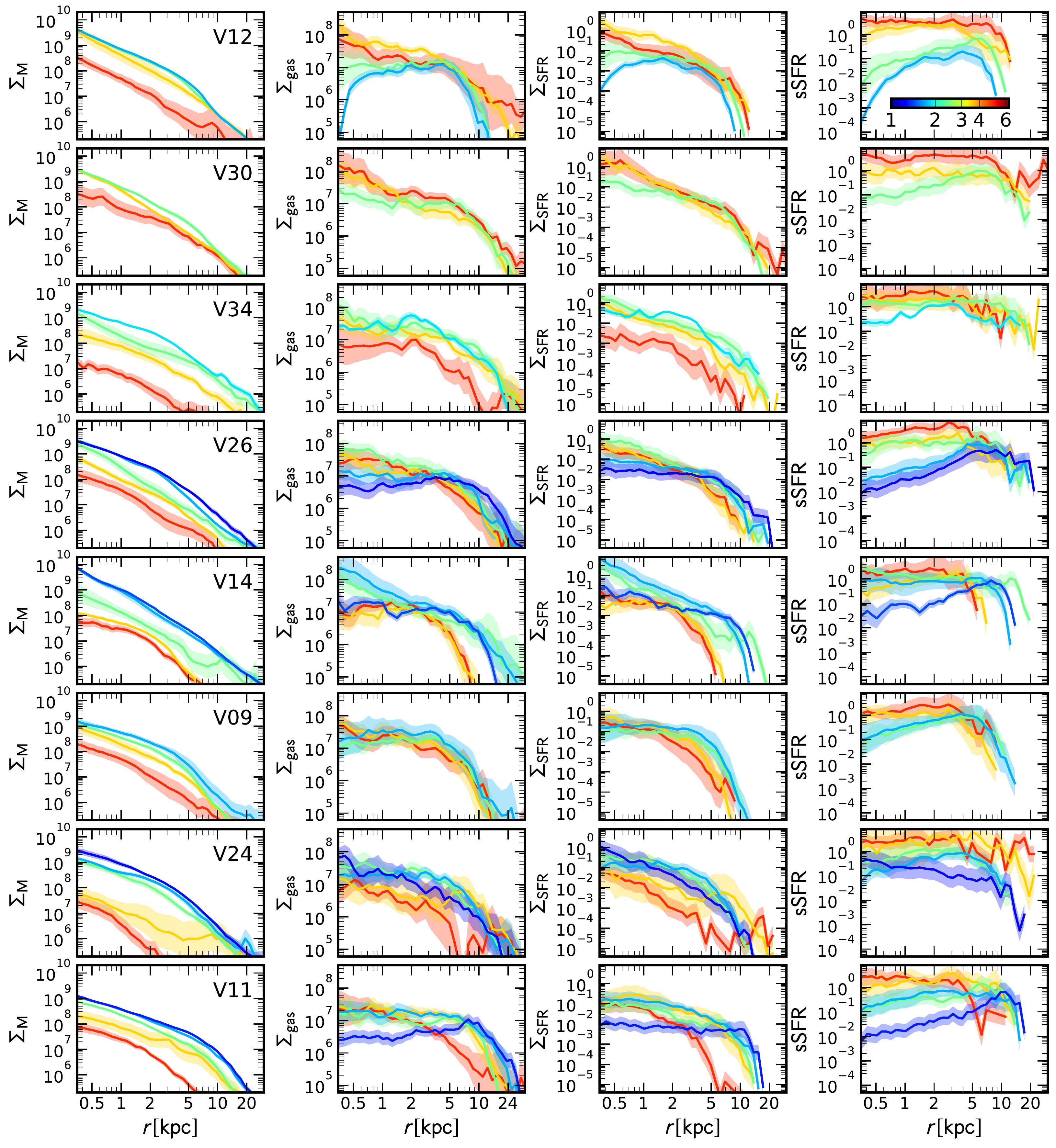} 
\caption{Same as Figure~\ref{FigA1_Profile}. } 
\label{FigA2_Profile}
\end{figure*}

\begin{figure*}
\includegraphics[width=\textwidth]{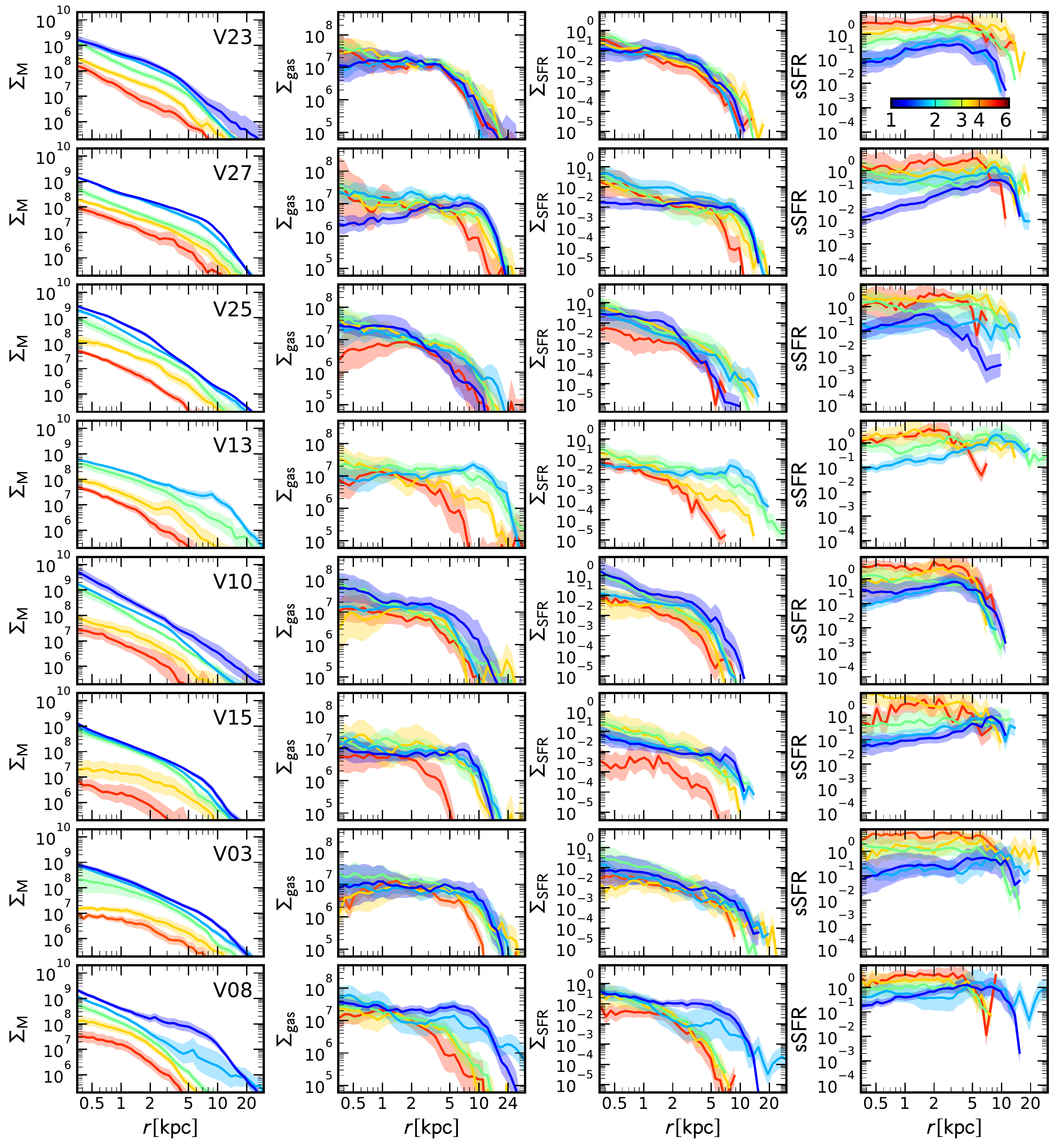} 
\caption{Same as Figure~\ref{FigA1_Profile}. } 
\label{FigA3_Profile}
\end{figure*}

\begin{figure*}
\includegraphics[width=\textwidth]{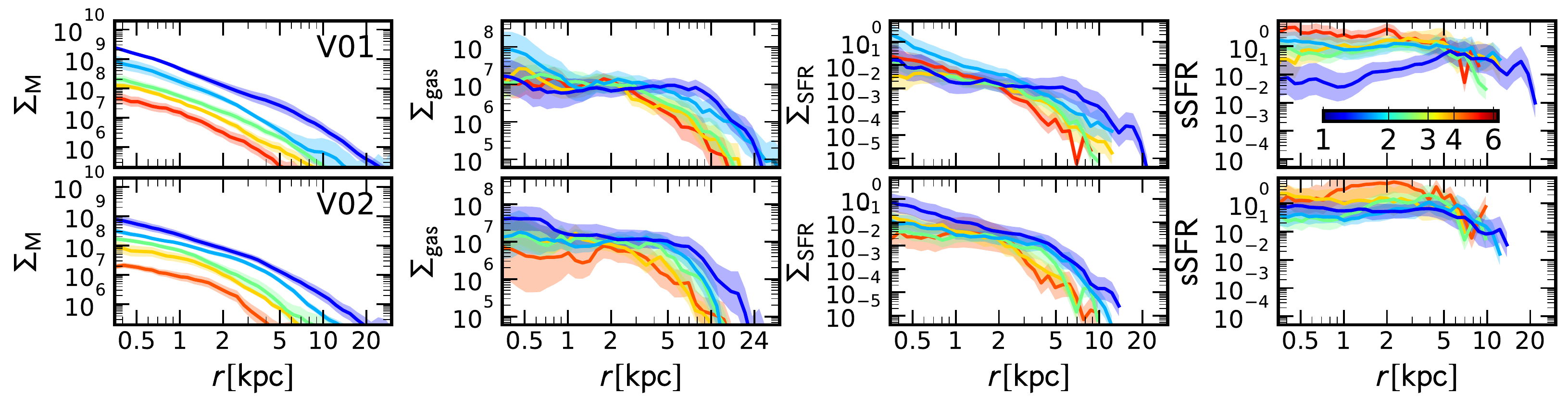} 
\caption{Same as Figure~\ref{FigA1_Profile}. } 
\label{FigA4_Profile}
\end{figure*}

In Figures~\ref{FigA1_Profile}-\ref{FigA4_Profile} we plot the stellar mass, gas mass (total and cold gas), SFR, and sSFR profiles for all the 26 galaxies in our sample. The galaxies are order by decreasing total stellar mass at $z=2.0$. 13 galaxies (V07, V09, V11, V12, V20, V21, V22, V23, V26, V27, V29, V30, V34) show the clear signature of inside-out quenching, with a convergence of the stellar mass density in the inner regions. Four galaxies (V06, V08, V13, V32) show only a weak signature of inside-out quenching, whereas nine galaxies (V01, V02, V03, V15, V10, V24, V25, V33, V34) show no such signature. 

\bsp

\label{lastpage}


\begin{thebibliography}{}
\makeatletter
\relax
\def\mn@urlcharsother{\let\do\@makeother \do\$\do\&\do\#\do\^\do\_\do\%\do\~}
\def\mn@doi{\begingroup\mn@urlcharsother \@ifnextchar [ {\mn@doi@}
  {\mn@doi@[]}}
\def\mn@doi@[#1]#2{\def\@tempa{#1}\ifx\@tempa\@empty \href
  {http://dx.doi.org/#2} {doi:#2}\else \href {http://dx.doi.org/#2} {#1}\fi
  \endgroup}
\def\mn@eprint#1#2{\mn@eprint@#1:#2::\@nil}
\def\mn@eprint@arXiv#1{\href {http://arxiv.org/abs/#1} {{\tt arXiv:#1}}}
\def\mn@eprint@dblp#1{\href {http://dblp.uni-trier.de/rec/bibtex/#1.xml}
  {dblp:#1}}
\def\mn@eprint@#1:#2:#3:#4\@nil{\def\@tempa {#1}\def\@tempb {#2}\def\@tempc
  {#3}\ifx \@tempc \@empty \let \@tempc \@tempb \let \@tempb \@tempa \fi \ifx
  \@tempb \@empty \def\@tempb {arXiv}\fi \@ifundefined
  {mn@eprint@\@tempb}{\@tempb:\@tempc}{\expandafter \expandafter \csname
  mn@eprint@\@tempb\endcsname \expandafter{\@tempc}}}

\bibitem[\protect\citeauthoryear{{Agertz}, {Teyssier}  \& {Moore}}{{Agertz}
  et~al.}{2009}]{agertz09}
{Agertz} O.,  {Teyssier} R.,   {Moore} B.,  2009, \mn@doi [\mnras]
  {10.1111/j.1745-3933.2009.00685.x}, \href
  {http://adsabs.harvard.edu/abs/2009MNRAS.397L..64A} {397, L64}

\bibitem[\protect\citeauthoryear{{Baldry}, {Balogh}, {Bower}, {Glazebrook},
  {Nichol}, {Bamford}  \& {Budavari}}{{Baldry} et~al.}{2006}]{baldry06}
{Baldry} I.~K.,  {Balogh} M.~L.,  {Bower} R.~G.,  {Glazebrook} K.,  {Nichol}
  R.~C.,  {Bamford} S.~P.,   {Budavari} T.,  2006, \mn@doi [\mnras]
  {10.1111/j.1365-2966.2006.11081.x}, \href
  {http://adsabs.harvard.edu/abs/2006MNRAS.373..469B} {373, 469}

\bibitem[\protect\citeauthoryear{{Balogh}, {Baldry}, {Nichol}, {Miller},
  {Bower}  \& {Glazebrook}}{{Balogh} et~al.}{2004}]{balogh04_bimodality}
{Balogh} M.~L.,  {Baldry} I.~K.,  {Nichol} R.,  {Miller} C.,  {Bower} R.,
  {Glazebrook} K.,  2004, \mn@doi [\apjl] {10.1086/426079}, \href
  {http://adsabs.harvard.edu/abs/2004ApJ...615L.101B} {615, L101}

\bibitem[\protect\citeauthoryear{{Barnes} \& {Hernquist}}{{Barnes} \&
  {Hernquist}}{1991}]{barnes91}
{Barnes} J.~E.,  {Hernquist} L.~E.,  1991, \mn@doi [\apjl] {10.1086/185978},
  \href {http://adsabs.harvard.edu/abs/1991ApJ...370L..65B} {370, L65}

\bibitem[\protect\citeauthoryear{{Barro} et~al.,}{{Barro}
  et~al.}{2013}]{barro13}
{Barro} G.,  et~al., 2013, \mn@doi [\apj] {10.1088/0004-637X/765/2/104}, \href
  {http://cdsads.u-strasbg.fr/abs/2013ApJ...765..104B} {765, 104}

\bibitem[\protect\citeauthoryear{{Barro} et~al.,}{{Barro}
  et~al.}{2014}]{barro14}
{Barro} G.,  et~al., 2014, \mn@doi [\apj] {10.1088/0004-637X/791/1/52}, \href
  {http://adsabs.harvard.edu/abs/2014ApJ...791...52B} {791, 52}

\bibitem[\protect\citeauthoryear{{Behroozi}, {Conroy}  \&
  {Wechsler}}{{Behroozi} et~al.}{2010}]{behroozi10}
{Behroozi} P.~S.,  {Conroy} C.,   {Wechsler} R.~H.,  2010, \mn@doi [\apj]
  {10.1088/0004-637X/717/1/379}, \href
  {http://adsabs.harvard.edu/abs/2010ApJ...717..379B} {717, 379}

\bibitem[\protect\citeauthoryear{{Behroozi}, {Wechsler}  \&
  {Conroy}}{{Behroozi} et~al.}{2013a}]{behroozi13b}
{Behroozi} P.~S.,  {Wechsler} R.~H.,   {Conroy} C.,  2013a, \mn@doi [\apj]
  {10.1088/0004-637X/770/1/57}, \href
  {http://adsabs.harvard.edu/abs/2013ApJ...770...57B} {770, 57}

\bibitem[\protect\citeauthoryear{{Behroozi}, {Marchesini}, {Wechsler},
  {Muzzin}, {Papovich}  \& {Stefanon}}{{Behroozi} et~al.}{2013b}]{behroozi13c}
{Behroozi} P.~S.,  {Marchesini} D.,  {Wechsler} R.~H.,  {Muzzin} A.,
  {Papovich} C.,   {Stefanon} M.,  2013b, \mn@doi [\apjl]
  {10.1088/2041-8205/777/1/L10}, \href
  {http://adsabs.harvard.edu/abs/2013ApJ...777L..10B} {777, L10}

\bibitem[\protect\citeauthoryear{{Bell} et~al.,}{{Bell} et~al.}{2012}]{bell12}
{Bell} E.~F.,  et~al., 2012, \mn@doi [\apj] {10.1088/0004-637X/753/2/167},
  \href {http://adsabs.harvard.edu/abs/2012ApJ...753..167B} {753, 167}

\bibitem[\protect\citeauthoryear{{Birnboim} \& {Dekel}}{{Birnboim} \&
  {Dekel}}{2003}]{birnboim03}
{Birnboim} Y.,  {Dekel} A.,  2003, \mn@doi [\mnras]
  {10.1046/j.1365-8711.2003.06955.x}, \href
  {http://adsabs.harvard.edu/abs/2003MNRAS.345..349B} {345, 349}

\bibitem[\protect\citeauthoryear{{Birnboim} \& {Dekel}}{{Birnboim} \&
  {Dekel}}{2011}]{birnboim11}
{Birnboim} Y.,  {Dekel} A.,  2011, \mn@doi [\mnras]
  {10.1111/j.1365-2966.2011.18880.x}, \href
  {http://adsabs.harvard.edu/abs/2011MNRAS.415.2566B} {415, 2566}

\bibitem[\protect\citeauthoryear{{Bluck}, {Mendel}, {Ellison}, {Moreno},
  {Simard}, {Patton}  \& {Starkenburg}}{{Bluck} et~al.}{2014}]{bluck14}
{Bluck} A.~F.~L.,  {Mendel} J.~T.,  {Ellison} S.~L.,  {Moreno} J.,  {Simard}
  L.,  {Patton} D.~R.,   {Starkenburg} E.,  2014, \mn@doi [\mnras]
  {10.1093/mnras/stu594}, \href
  {http://adsabs.harvard.edu/abs/2014MNRAS.441..599B} {441, 599}

\bibitem[\protect\citeauthoryear{{Bouch{\'e}} et~al.,}{{Bouch{\'e}}
  et~al.}{2010}]{bouche10}
{Bouch{\'e}} N.,  et~al., 2010, \mn@doi [\apj] {10.1088/0004-637X/718/2/1001},
  \href {http://adsabs.harvard.edu/abs/2010ApJ...718.1001B} {718, 1001}

\bibitem[\protect\citeauthoryear{{Bournaud}, {Elmegreen}  \&
  {Elmegreen}}{{Bournaud} et~al.}{2007}]{bournaud07}
{Bournaud} F.,  {Elmegreen} B.~G.,   {Elmegreen} D.~M.,  2007, \mn@doi [\apj]
  {10.1086/522077}, \href {http://adsabs.harvard.edu/abs/2007ApJ...670..237B}
  {670, 237}

\bibitem[\protect\citeauthoryear{{Bruce} et~al.,}{{Bruce}
  et~al.}{2012}]{bruce12}
{Bruce} V.~A.,  et~al., 2012, \mn@doi [\mnras]
  {10.1111/j.1365-2966.2012.22087.x}, \href
  {http://adsabs.harvard.edu/abs/2012MNRAS.427.1666B} {427, 1666}

\bibitem[\protect\citeauthoryear{{Bryan} \& {Norman}}{{Bryan} \&
  {Norman}}{1998}]{bryan98}
{Bryan} G.~L.,  {Norman} M.~L.,  1998, \mn@doi [\apj] {10.1086/305262}, \href
  {http://adsabs.harvard.edu/abs/1998ApJ...495...80B} {495, 80}

\bibitem[\protect\citeauthoryear{{Bundy} et~al.,}{{Bundy}
  et~al.}{2010}]{bundy10}
{Bundy} K.,  et~al., 2010, \mn@doi [\apj] {10.1088/0004-637X/719/2/1969}, \href
  {http://adsabs.harvard.edu/abs/2010ApJ...719.1969B} {719, 1969}

\bibitem[\protect\citeauthoryear{{Burkert} et~al.,}{{Burkert}
  et~al.}{2015}]{burkert15}
{Burkert} A.,  et~al., 2015, preprint, \href
  {http://adsabs.harvard.edu/abs/2015arXiv151003262B} {} (\mn@eprint {arXiv}
  {1510.03262})

\bibitem[\protect\citeauthoryear{{Cacciato}, {Dekel}  \& {Genel}}{{Cacciato}
  et~al.}{2012}]{cacciato12}
{Cacciato} M.,  {Dekel} A.,   {Genel} S.,  2012, \mn@doi [\mnras]
  {10.1111/j.1365-2966.2011.20359.x}, \href
  {http://adsabs.harvard.edu/abs/2012MNRAS.421..818C} {421, 818}

\bibitem[\protect\citeauthoryear{{Carollo} et~al.,}{{Carollo}
  et~al.}{2013a}]{carollo13a}
{Carollo} C.~M.,  et~al., 2013a, \mn@doi [\apj] {10.1088/0004-637X/773/2/112},
  \href {http://adsabs.harvard.edu/abs/2013ApJ...773..112C} {773, 112}

\bibitem[\protect\citeauthoryear{{Carollo} et~al.,}{{Carollo}
  et~al.}{2013b}]{carollo13}
{Carollo} C.~M.,  et~al., 2013b, \mn@doi [\apj] {10.1088/0004-637X/776/2/71},
  \href {http://adsabs.harvard.edu/abs/2013ApJ...776...71C} {776, 71}

\bibitem[\protect\citeauthoryear{{Carollo} et~al.,}{{Carollo}
  et~al.}{2014}]{carollo14}
{Carollo} C.~M.,  et~al., 2014, preprint, \href
  {http://adsabs.harvard.edu/abs/2014arXiv1402.1172C} {} (\mn@eprint {arXiv}
  {1402.1172})

\bibitem[\protect\citeauthoryear{{Cassata} et~al.,}{{Cassata}
  et~al.}{2013}]{cassata13}
{Cassata} P.,  et~al., 2013, \mn@doi [\apj] {10.1088/0004-637X/775/2/106},
  \href {http://adsabs.harvard.edu/abs/2013ApJ...775..106C} {775, 106}

\bibitem[\protect\citeauthoryear{{Cattaneo} et~al.,}{{Cattaneo}
  et~al.}{2009}]{cattaneo09}
{Cattaneo} A.,  et~al., 2009, \mn@doi [\nat] {10.1038/nature08135}, \href
  {http://adsabs.harvard.edu/abs/2009Natur.460..213C} {460, 213}

\bibitem[\protect\citeauthoryear{{Ceverino} \& {Klypin}}{{Ceverino} \&
  {Klypin}}{2009}]{ceverino09}
{Ceverino} D.,  {Klypin} A.,  2009, \mn@doi [\apj]
  {10.1088/0004-637X/695/1/292}, \href
  {http://adsabs.harvard.edu/abs/2009ApJ...695..292C} {695, 292}

\bibitem[\protect\citeauthoryear{{Ceverino}, {Dekel}  \& {Bournaud}}{{Ceverino}
  et~al.}{2010}]{ceverino10}
{Ceverino} D.,  {Dekel} A.,   {Bournaud} F.,  2010, \mn@doi [\mnras]
  {10.1111/j.1365-2966.2010.16433.x}, \href
  {http://adsabs.harvard.edu/abs/2010MNRAS.404.2151C} {404, 2151}

\bibitem[\protect\citeauthoryear{{Ceverino}, {Dekel}, {Mandelker}, {Bournaud},
  {Burkert}, {Genzel}  \& {Primack}}{{Ceverino} et~al.}{2012}]{ceverino12}
{Ceverino} D.,  {Dekel} A.,  {Mandelker} N.,  {Bournaud} F.,  {Burkert} A.,
  {Genzel} R.,   {Primack} J.,  2012, \mn@doi [\mnras]
  {10.1111/j.1365-2966.2011.20296.x}, \href
  {http://adsabs.harvard.edu/abs/2012MNRAS.420.3490C} {420, 3490}

\bibitem[\protect\citeauthoryear{{Ceverino}, {Klypin}, {Klimek},
  {Trujillo-Gomez}, {Churchill}, {Primack}  \& {Dekel}}{{Ceverino}
  et~al.}{2014}]{ceverino14_radfeed}
{Ceverino} D.,  {Klypin} A.,  {Klimek} E.~S.,  {Trujillo-Gomez} S.,
  {Churchill} C.~W.,  {Primack} J.,   {Dekel} A.,  2014, \mn@doi [\mnras]
  {10.1093/mnras/stu956}, \href
  {http://adsabs.harvard.edu/abs/2014MNRAS.442.1545C} {442, 1545}

\bibitem[\protect\citeauthoryear{{Ceverino}, {Dekel}, {Tweed}  \&
  {Primack}}{{Ceverino} et~al.}{2015a}]{ceverino15}
{Ceverino} D.,  {Dekel} A.,  {Tweed} D.,   {Primack} J.,  2015a, \mn@doi
  [\mnras] {10.1093/mnras/stu2694}, \href
  {http://adsabs.harvard.edu/abs/2015MNRAS.447.3291C} {447, 3291}

\bibitem[\protect\citeauthoryear{{Ceverino}, {Primack}  \& {Dekel}}{{Ceverino}
  et~al.}{2015b}]{ceverino15b}
{Ceverino} D.,  {Primack} J.,   {Dekel} A.,  2015b, \mn@doi [\mnras]
  {10.1093/mnras/stv1603}, \href
  {http://adsabs.harvard.edu/abs/2015MNRAS.453..408C} {453, 408}

\bibitem[\protect\citeauthoryear{{Chabrier}}{{Chabrier}}{2003}]{chabrier03}
{Chabrier} G.,  2003, \mn@doi [\pasp] {10.1086/376392}, \href
  {http://adsabs.harvard.edu/abs/2003PASP..115..763C} {115, 763}

\bibitem[\protect\citeauthoryear{{Cheung} et~al.,}{{Cheung}
  et~al.}{2012}]{cheung12}
{Cheung} E.,  et~al., 2012, \mn@doi [\apj] {10.1088/0004-637X/760/2/131}, \href
  {http://adsabs.harvard.edu/abs/2012ApJ...760..131C} {760, 131}

\bibitem[\protect\citeauthoryear{{Cibinel} et~al.,}{{Cibinel}
  et~al.}{2013a}]{cibinel13a}
{Cibinel} A.,  et~al., 2013a, \mn@doi [\apj] {10.1088/0004-637X/776/2/72},
  \href {http://adsabs.harvard.edu/abs/2013ApJ...776...72C} {776, 72}

\bibitem[\protect\citeauthoryear{{Cibinel} et~al.,}{{Cibinel}
  et~al.}{2013b}]{cibinel13b}
{Cibinel} A.,  et~al., 2013b, \mn@doi [\apj] {10.1088/0004-637X/777/2/116},
  \href {http://adsabs.harvard.edu/abs/2013ApJ...777..116C} {777, 116}

\bibitem[\protect\citeauthoryear{{Cimatti} et~al.,}{{Cimatti}
  et~al.}{2008}]{cimatti08}
{Cimatti} A.,  et~al., 2008, \mn@doi [\aap] {10.1051/0004-6361:20078739}, \href
  {http://adsabs.harvard.edu/abs/2008A26A...482...21C} {482, 21}

\bibitem[\protect\citeauthoryear{{Ciotti} \& {Ostriker}}{{Ciotti} \&
  {Ostriker}}{2007}]{ciotti07}
{Ciotti} L.,  {Ostriker} J.~P.,  2007, \mn@doi [\apj] {10.1086/519833}, \href
  {http://adsabs.harvard.edu/abs/2007ApJ...665.1038C} {665, 1038}

\bibitem[\protect\citeauthoryear{{Conroy} \& {Wechsler}}{{Conroy} \&
  {Wechsler}}{2009}]{conroy09}
{Conroy} C.,  {Wechsler} R.~H.,  2009, \mn@doi [\apj]
  {10.1088/0004-637X/696/1/620}, \href
  {http://adsabs.harvard.edu/abs/2009ApJ...696..620C} {696, 620}

\bibitem[\protect\citeauthoryear{{Croton} et~al.,}{{Croton}
  et~al.}{2006}]{croton06}
{Croton} D.~J.,  et~al., 2006, \mn@doi [\mnras]
  {10.1111/j.1365-2966.2005.09675.x}, \href
  {http://adsabs.harvard.edu/abs/2006MNRAS.365...11C} {365, 11}

\bibitem[\protect\citeauthoryear{{Daddi} et~al.,}{{Daddi}
  et~al.}{2005}]{daddi05}
{Daddi} E.,  et~al., 2005, \mn@doi [\apj] {10.1086/430104}, \href
  {http://adsabs.harvard.edu/abs/2005ApJ...626..680D} {626, 680}

\bibitem[\protect\citeauthoryear{{Danovich}, {Dekel}, {Hahn}  \&
  {Teyssier}}{{Danovich} et~al.}{2012}]{danovich12}
{Danovich} M.,  {Dekel} A.,  {Hahn} O.,   {Teyssier} R.,  2012, \mn@doi
  [\mnras] {10.1111/j.1365-2966.2012.20751.x}, \href
  {http://adsabs.harvard.edu/abs/2012MNRAS.422.1732D} {422, 1732}

\bibitem[\protect\citeauthoryear{{Danovich}, {Dekel}, {Hahn}, {Ceverino}  \&
  {Primack}}{{Danovich} et~al.}{2015}]{danovich15}
{Danovich} M.,  {Dekel} A.,  {Hahn} O.,  {Ceverino} D.,   {Primack} J.,  2015,
  \mn@doi [\mnras] {10.1093/mnras/stv270}, \href
  {http://adsabs.harvard.edu/abs/2015MNRAS.449.2087D} {449, 2087}

\bibitem[\protect\citeauthoryear{{Dekel} \& {Birnboim}}{{Dekel} \&
  {Birnboim}}{2006}]{dekel06}
{Dekel} A.,  {Birnboim} Y.,  2006, \mn@doi [\mnras]
  {10.1111/j.1365-2966.2006.10145.x}, \href
  {http://adsabs.harvard.edu/abs/2006MNRAS.368....2D} {368, 2}

\bibitem[\protect\citeauthoryear{{Dekel} \& {Birnboim}}{{Dekel} \&
  {Birnboim}}{2008}]{dekel08}
{Dekel} A.,  {Birnboim} Y.,  2008, \mn@doi [\mnras]
  {10.1111/j.1365-2966.2007.12569.x}, \href
  {http://adsabs.harvard.edu/abs/2008MNRAS.383..119D} {383, 119}

\bibitem[\protect\citeauthoryear{{Dekel} \& {Burkert}}{{Dekel} \&
  {Burkert}}{2014}]{dekel14_nugget}
{Dekel} A.,  {Burkert} A.,  2014, \mn@doi [\mnras] {10.1093/mnras/stt2331},
  \href {http://adsabs.harvard.edu/abs/2014MNRAS.438.1870D} {438, 1870}

\bibitem[\protect\citeauthoryear{{Dekel} \& {Krumholz}}{{Dekel} \&
  {Krumholz}}{2013}]{dekel13a}
{Dekel} A.,  {Krumholz} M.~R.,  2013, \mn@doi [\mnras] {10.1093/mnras/stt480},
  \href {http://adsabs.harvard.edu/abs/2013MNRAS.432..455D} {432, 455}

\bibitem[\protect\citeauthoryear{{Dekel} \& {Mandelker}}{{Dekel} \&
  {Mandelker}}{2014}]{dekel14_bathtube}
{Dekel} A.,  {Mandelker} N.,  2014, \mn@doi [\mnras] {10.1093/mnras/stu1427},
  \href {http://adsabs.harvard.edu/abs/2014MNRAS.444.2071D} {444, 2071}

\bibitem[\protect\citeauthoryear{{Dekel} \& {Silk}}{{Dekel} \&
  {Silk}}{1986}]{dekel86}
{Dekel} A.,  {Silk} J.,  1986, \mn@doi [\apj] {10.1086/164050}, \href
  {http://adsabs.harvard.edu/abs/1986ApJ...303...39D} {303, 39}

\bibitem[\protect\citeauthoryear{{Dekel} et~al.,}{{Dekel}
  et~al.}{2009a}]{dekel09}
{Dekel} A.,  et~al., 2009a, \mn@doi [\nat] {10.1038/nature07648}, \href
  {http://adsabs.harvard.edu/abs/2009Natur.457..451D} {457, 451}

\bibitem[\protect\citeauthoryear{{Dekel}, {Sari}  \& {Ceverino}}{{Dekel}
  et~al.}{2009b}]{dekel09b}
{Dekel} A.,  {Sari} R.,   {Ceverino} D.,  2009b, \mn@doi [\apj]
  {10.1088/0004-637X/703/1/785}, \href
  {http://adsabs.harvard.edu/abs/2009ApJ...703..785D} {703, 785}

\bibitem[\protect\citeauthoryear{{Dekel}, {Zolotov}, {Tweed}, {Cacciato},
  {Ceverino}  \& {Primack}}{{Dekel} et~al.}{2013}]{dekel13}
{Dekel} A.,  {Zolotov} A.,  {Tweed} D.,  {Cacciato} M.,  {Ceverino} D.,
  {Primack} J.~R.,  2013, \mn@doi [\mnras] {10.1093/mnras/stt1338}, \href
  {http://adsabs.harvard.edu/abs/2013MNRAS.435..999D} {435, 999}

\bibitem[\protect\citeauthoryear{{Di Matteo}, {Springel}  \& {Hernquist}}{{Di
  Matteo} et~al.}{2005}]{di-matteo05}
{Di Matteo} T.,  {Springel} V.,   {Hernquist} L.,  2005, \mn@doi [\nat]
  {10.1038/nature03335}, \href
  {http://adsabs.harvard.edu/abs/2005Natur.433..604D} {433, 604}

\bibitem[\protect\citeauthoryear{{Dressler}}{{Dressler}}{1980}]{dressler80}
{Dressler} A.,  1980, \mn@doi [\apj] {10.1086/157753}, \href
  {http://adsabs.harvard.edu/abs/1980ApJ...236..351D} {236, 351}

\bibitem[\protect\citeauthoryear{{Dullo} \& {Graham}}{{Dullo} \&
  {Graham}}{2013}]{dullo13}
{Dullo} B.~T.,  {Graham} A.~W.,  2013, \mn@doi [\apj]
  {10.1088/0004-637X/768/1/36}, \href
  {http://adsabs.harvard.edu/abs/2013ApJ...768...36D} {768, 36}

\bibitem[\protect\citeauthoryear{{Elmegreen}, {Elmegreen}, {Fernandez}  \&
  {Lemonias}}{{Elmegreen} et~al.}{2009}]{elmegreen09}
{Elmegreen} B.~G.,  {Elmegreen} D.~M.,  {Fernandez} M.~X.,   {Lemonias} J.~J.,
  2009, \mn@doi [\apj] {10.1088/0004-637X/692/1/12}, \href
  {http://adsabs.harvard.edu/abs/2009ApJ...692...12E} {692, 12}

\bibitem[\protect\citeauthoryear{{Fabian}}{{Fabian}}{2012}]{fabian12}
{Fabian} A.~C.,  2012, \mn@doi [\araa] {10.1146/annurev-astro-081811-125521},
  \href {http://adsabs.harvard.edu/abs/2012ARA26A..50..455F} {50, 455}

\bibitem[\protect\citeauthoryear{{Fang}, {Faber}, {Koo}  \& {Dekel}}{{Fang}
  et~al.}{2013}]{fang13}
{Fang} J.~J.,  {Faber} S.~M.,  {Koo} D.~C.,   {Dekel} A.,  2013, \mn@doi [\apj]
  {10.1088/0004-637X/776/1/63}, \href
  {http://adsabs.harvard.edu/abs/2013ApJ...776...63F} {776, 63}

\bibitem[\protect\citeauthoryear{{Feldmann} \& {Mayer}}{{Feldmann} \&
  {Mayer}}{2015}]{feldmann15_quench}
{Feldmann} R.,  {Mayer} L.,  2015, \mn@doi [\mnras] {10.1093/mnras/stu2207},
  \href {http://adsabs.harvard.edu/abs/2015MNRAS.446.1939F} {446, 1939}

\bibitem[\protect\citeauthoryear{{Feldmann}, {Carollo}, {Mayer}, {Renzini},
  {Lake}, {Quinn}, {Stinson}  \& {Yepes}}{{Feldmann} et~al.}{2010}]{feldmann10}
{Feldmann} R.,  {Carollo} C.~M.,  {Mayer} L.,  {Renzini} A.,  {Lake} G.,
  {Quinn} T.,  {Stinson} G.~S.,   {Yepes} G.,  2010, \mn@doi [\apj]
  {10.1088/0004-637X/709/1/218}, \href
  {http://adsabs.harvard.edu/abs/2010ApJ...709..218F} {709, 218}

\bibitem[\protect\citeauthoryear{{Ferland}, {Korista}, {Verner}, {Ferguson},
  {Kingdon}  \& {Verner}}{{Ferland} et~al.}{1998}]{ferland98}
{Ferland} G.~J.,  {Korista} K.~T.,  {Verner} D.~A.,  {Ferguson} J.~W.,
  {Kingdon} J.~B.,   {Verner} E.~M.,  1998, \mn@doi [\pasp] {10.1086/316190},
  \href {http://adsabs.harvard.edu/abs/1998PASP..110..761F} {110, 761}

\bibitem[\protect\citeauthoryear{{Forbes}, {Krumholz}, {Burkert}  \&
  {Dekel}}{{Forbes} et~al.}{2014}]{forbes14}
{Forbes} J.~C.,  {Krumholz} M.~R.,  {Burkert} A.,   {Dekel} A.,  2014, \mn@doi
  [\mnras] {10.1093/mnras/stu1142}, \href
  {http://adsabs.harvard.edu/abs/2014MNRAS.443..168F} {443, 168}

\bibitem[\protect\citeauthoryear{{F{\"o}rster Schreiber}, {Shapley}, {Erb},
  {Genzel}, {Steidel}, {Bouch{\'e}}, {Cresci}  \& {Davies}}{{F{\"o}rster
  Schreiber} et~al.}{2011}]{forster-schreiber11a}
{F{\"o}rster Schreiber} N.~M.,  {Shapley} A.~E.,  {Erb} D.~K.,  {Genzel} R.,
  {Steidel} C.~C.,  {Bouch{\'e}} N.,  {Cresci} G.,   {Davies} R.,  2011,
  \mn@doi [\apj] {10.1088/0004-637X/731/1/65}, \href
  {http://adsabs.harvard.edu/abs/2011ApJ...731...65F} {731, 65}

\bibitem[\protect\citeauthoryear{{F{\"o}rster Schreiber} et~al.,}{{F{\"o}rster
  Schreiber} et~al.}{2014}]{forster-schreiber14}
{F{\"o}rster Schreiber} N.~M.,  et~al., 2014, \mn@doi [\apj]
  {10.1088/0004-637X/787/1/38}, \href
  {http://adsabs.harvard.edu/abs/2014ApJ...787...38F} {787, 38}

\bibitem[\protect\citeauthoryear{{Franx}, {van Dokkum}, {Schreiber}, {Wuyts},
  {Labb{\'e}}  \& {Toft}}{{Franx} et~al.}{2008}]{franx08}
{Franx} M.,  {van Dokkum} P.~G.,  {Schreiber} N.~M.~F.,  {Wuyts} S.,
  {Labb{\'e}} I.,   {Toft} S.,  2008, \mn@doi [\apj] {10.1086/592431}, \href
  {http://adsabs.harvard.edu/abs/2008ApJ...688..770F} {688, 770}

\bibitem[\protect\citeauthoryear{{Genzel} et~al.,}{{Genzel}
  et~al.}{2006}]{genzel06}
{Genzel} R.,  et~al., 2006, \mn@doi [\nat] {10.1038/nature05052}, \href
  {http://adsabs.harvard.edu/abs/2006Natur.442..786G} {442, 786}

\bibitem[\protect\citeauthoryear{{Genzel} et~al.,}{{Genzel}
  et~al.}{2008}]{genzel08}
{Genzel} R.,  et~al., 2008, \mn@doi [\apj] {10.1086/591840}, \href
  {http://adsabs.harvard.edu/abs/2008ApJ...687...59G} {687, 59}

\bibitem[\protect\citeauthoryear{{Genzel} et~al.,}{{Genzel}
  et~al.}{2014}]{genzel14a}
{Genzel} R.,  et~al., 2014, \mn@doi [\apj] {10.1088/0004-637X/785/1/75}, \href
  {http://adsabs.harvard.edu/abs/2014ApJ...785...75G} {785, 75}

\bibitem[\protect\citeauthoryear{{Genzel} et~al.,}{{Genzel}
  et~al.}{2015}]{genzel15}
{Genzel} R.,  et~al., 2015, \mn@doi [\apj] {10.1088/0004-637X/800/1/20}, \href
  {http://adsabs.harvard.edu/abs/2015ApJ...800...20G} {800, 20}

\bibitem[\protect\citeauthoryear{{Haardt} \& {Madau}}{{Haardt} \&
  {Madau}}{1996}]{haardt96}
{Haardt} F.,  {Madau} P.,  1996, \mn@doi [\apj] {10.1086/177035}, \href
  {http://adsabs.harvard.edu/abs/1996ApJ...461...20H} {461, 20}

\bibitem[\protect\citeauthoryear{{Hearin} \& {Watson}}{{Hearin} \&
  {Watson}}{2013}]{hearin13}
{Hearin} A.~P.,  {Watson} D.~F.,  2013, \mn@doi [\mnras]
  {10.1093/mnras/stt1374}, \href
  {http://adsabs.harvard.edu/abs/2013MNRAS.tmp.2143H} {435, 1313}

\bibitem[\protect\citeauthoryear{{Hernquist}}{{Hernquist}}{1990}]{hernquist90}
{Hernquist} L.,  1990, \mn@doi [\apj] {10.1086/168845}, \href
  {http://adsabs.harvard.edu/abs/1990ApJ...356..359H} {356, 359}

\bibitem[\protect\citeauthoryear{{Hopkins}, {Hernquist}, {Cox}, {Di Matteo},
  {Robertson}  \& {Springel}}{{Hopkins} et~al.}{2006}]{hopkins06a}
{Hopkins} P.~F.,  {Hernquist} L.,  {Cox} T.~J.,  {Di Matteo} T.,  {Robertson}
  B.,   {Springel} V.,  2006, \mn@doi [\apjs] {10.1086/499298}, \href
  {http://adsabs.harvard.edu/abs/2006ApJS..163....1H} {163, 1}

\bibitem[\protect\citeauthoryear{{Hopkins}, {Bundy}, {Murray}, {Quataert},
  {Lauer}  \& {Ma}}{{Hopkins} et~al.}{2009}]{hopkins09b}
{Hopkins} P.~F.,  {Bundy} K.,  {Murray} N.,  {Quataert} E.,  {Lauer} T.~R.,
  {Ma} C.-P.,  2009, \mn@doi [\mnras] {10.1111/j.1365-2966.2009.15062.x}, \href
  {http://adsabs.harvard.edu/abs/2009MNRAS.398..898H} {398, 898}

\bibitem[\protect\citeauthoryear{{Hopkins}, {Kere{\v s}}, {Murray}, {Quataert}
  \& {Hernquist}}{{Hopkins} et~al.}{2012}]{hopkins12}
{Hopkins} P.~F.,  {Kere{\v s}} D.,  {Murray} N.,  {Quataert} E.,   {Hernquist}
  L.,  2012, \mn@doi [\mnras] {10.1111/j.1365-2966.2012.21981.x}, \href
  {http://adsabs.harvard.edu/abs/2012MNRAS.427..968H} {427, 968}

\bibitem[\protect\citeauthoryear{{Huang} \& {Kauffmann}}{{Huang} \&
  {Kauffmann}}{2014}]{huang14}
{Huang} M.-L.,  {Kauffmann} G.,  2014, \mn@doi [\mnras]
  {10.1093/mnras/stu1232}, \href
  {http://adsabs.harvard.edu/abs/2014MNRAS.443.1329H} {443, 1329}

\bibitem[\protect\citeauthoryear{{Ilbert} et~al.,}{{Ilbert}
  et~al.}{2013}]{ilbert13}
{Ilbert} O.,  et~al., 2013, \mn@doi [\aap] {10.1051/0004-6361/201321100}, \href
  {http://adsabs.harvard.edu/abs/2013A26A...556A..55I} {556, A55}

\bibitem[\protect\citeauthoryear{{Immeli}, {Samland}, {Gerhard}  \&
  {Westera}}{{Immeli} et~al.}{2004a}]{immeli04a}
{Immeli} A.,  {Samland} M.,  {Gerhard} O.,   {Westera} P.,  2004a, \mn@doi
  [\aap] {10.1051/0004-6361:20034282}, \href
  {http://adsabs.harvard.edu/abs/2004A26A...413..547I} {413, 547}

\bibitem[\protect\citeauthoryear{{Immeli}, {Samland}, {Westera}  \&
  {Gerhard}}{{Immeli} et~al.}{2004b}]{immeli04b}
{Immeli} A.,  {Samland} M.,  {Westera} P.,   {Gerhard} O.,  2004b, \mn@doi
  [\apj] {10.1086/422179}, \href
  {http://adsabs.harvard.edu/abs/2004ApJ...611...20I} {611, 20}

\bibitem[\protect\citeauthoryear{{Kauffmann} et~al.,}{{Kauffmann}
  et~al.}{2003}]{kauffmann03}
{Kauffmann} G.,  et~al., 2003, \mn@doi [\mnras]
  {10.1111/j.1365-2966.2003.07154.x}, \href
  {http://adsabs.harvard.edu/abs/2003MNRAS.346.1055K} {346, 1055}

\bibitem[\protect\citeauthoryear{{Kennicutt}}{{Kennicutt}}{1998}]{kennicutt98}
{Kennicutt} Jr. R.~C.,  1998, \mn@doi [\araa] {10.1146/annurev.astro.36.1.189},
  \href {http://adsabs.harvard.edu/abs/1998ARA26A..36..189K} {36, 189}

\bibitem[\protect\citeauthoryear{{Kere{\v s}}, {Katz}, {Weinberg}  \&
  {Dav{\'e}}}{{Kere{\v s}} et~al.}{2005}]{keres05}
{Kere{\v s}} D.,  {Katz} N.,  {Weinberg} D.~H.,   {Dav{\'e}} R.,  2005, \mn@doi
  [\mnras] {10.1111/j.1365-2966.2005.09451.x}, \href
  {http://adsabs.harvard.edu/abs/2005MNRAS.363....2K} {363, 2}

\bibitem[\protect\citeauthoryear{{Kere{\v s}}, {Katz}, {Fardal}, {Dav{\'e}}  \&
  {Weinberg}}{{Kere{\v s}} et~al.}{2009}]{keres09}
{Kere{\v s}} D.,  {Katz} N.,  {Fardal} M.,  {Dav{\'e}} R.,   {Weinberg} D.~H.,
  2009, \mn@doi [\mnras] {10.1111/j.1365-2966.2009.14541.x}, \href
  {http://adsabs.harvard.edu/abs/2009MNRAS.395..160K} {395, 160}

\bibitem[\protect\citeauthoryear{{Khochfar} \& {Ostriker}}{{Khochfar} \&
  {Ostriker}}{2008}]{khochfar08}
{Khochfar} S.,  {Ostriker} J.~P.,  2008, \mn@doi [\apj] {10.1086/587470}, \href
  {http://adsabs.harvard.edu/abs/2008ApJ...680...54K} {680, 54}

\bibitem[\protect\citeauthoryear{{Kimm} et~al.,}{{Kimm} et~al.}{2009}]{kimm09}
{Kimm} T.,  et~al., 2009, \mn@doi [\mnras] {10.1111/j.1365-2966.2009.14414.x},
  \href {http://adsabs.harvard.edu/abs/2009MNRAS.394.1131K} {394, 1131}

\bibitem[\protect\citeauthoryear{{Knobel} et~al.,}{{Knobel}
  et~al.}{2013}]{knobel13}
{Knobel} C.,  et~al., 2013, \mn@doi [\apj] {10.1088/0004-637X/769/1/24}, \href
  {http://adsabs.harvard.edu/abs/2013ApJ...769...24K} {769, 24}

\bibitem[\protect\citeauthoryear{{Komatsu} et~al.,}{{Komatsu}
  et~al.}{2009}]{komatsu09}
{Komatsu} E.,  et~al., 2009, \mn@doi [\apjs] {10.1088/0067-0049/180/2/330},
  \href {http://adsabs.harvard.edu/abs/2009ApJS..180..330K} {180, 330}

\bibitem[\protect\citeauthoryear{{Kova{\v c}} et~al.,}{{Kova{\v c}}
  et~al.}{2014}]{kovac14}
{Kova{\v c}} K.,  et~al., 2014, \mn@doi [\mnras] {10.1093/mnras/stt2241}, \href
  {http://adsabs.harvard.edu/abs/2014MNRAS.438..717K} {438, 717}

\bibitem[\protect\citeauthoryear{{Kravtsov}}{{Kravtsov}}{2003}]{kravtsov03}
{Kravtsov} A.~V.,  2003, \mn@doi [\apjl] {10.1086/376674}, \href
  {http://adsabs.harvard.edu/abs/2003ApJ...590L...1K} {590, L1}

\bibitem[\protect\citeauthoryear{{Kravtsov}, {Klypin}  \&
  {Khokhlov}}{{Kravtsov} et~al.}{1997}]{kravtsov97}
{Kravtsov} A.~V.,  {Klypin} A.~A.,   {Khokhlov} A.~M.,  1997, \mn@doi [\apjs]
  {10.1086/313015}, \href {http://adsabs.harvard.edu/abs/1997ApJS..111...73K}
  {111, 73}

\bibitem[\protect\citeauthoryear{{Krumholz} \& {Dekel}}{{Krumholz} \&
  {Dekel}}{2010}]{krumholz10}
{Krumholz} M.~R.,  {Dekel} A.,  2010, \mn@doi [\mnras]
  {10.1111/j.1365-2966.2010.16675.x}, \href
  {http://adsabs.harvard.edu/abs/2010MNRAS.406..112K} {406, 112}

\bibitem[\protect\citeauthoryear{{Krumholz} \& {Dekel}}{{Krumholz} \&
  {Dekel}}{2012}]{krumholz12b}
{Krumholz} M.~R.,  {Dekel} A.,  2012, \mn@doi [\apj]
  {10.1088/0004-637X/753/1/16}, \href
  {http://adsabs.harvard.edu/abs/2012ApJ...753...16K} {753, 16}

\bibitem[\protect\citeauthoryear{{Krumholz} \& {Thompson}}{{Krumholz} \&
  {Thompson}}{2012}]{krumholz12}
{Krumholz} M.~R.,  {Thompson} T.~A.,  2012, \mn@doi [\apj]
  {10.1088/0004-637X/760/2/155}, \href
  {http://adsabs.harvard.edu/abs/2012ApJ...760..155K} {760, 155}

\bibitem[\protect\citeauthoryear{{Lang} et~al.,}{{Lang} et~al.}{2014}]{lang14}
{Lang} P.,  et~al., 2014, \mn@doi [\apj] {10.1088/0004-637X/788/1/11}, \href
  {http://adsabs.harvard.edu/abs/2014ApJ...788...11L} {788, 11}

\bibitem[\protect\citeauthoryear{{Lilly}, {Carollo}, {Pipino}, {Renzini}  \&
  {Peng}}{{Lilly} et~al.}{2013}]{lilly13_bathtube}
{Lilly} S.~J.,  {Carollo} C.~M.,  {Pipino} A.,  {Renzini} A.,   {Peng} Y.,
  2013, \mn@doi [\apj] {10.1088/0004-637X/772/2/119}, \href
  {http://adsabs.harvard.edu/abs/2013ApJ...772..119L} {772, 119}

\bibitem[\protect\citeauthoryear{{Lu}, {Mo}, {Lu}, {Katz}, {Weinberg}, {van den
  Bosch}  \& {Yang}}{{Lu} et~al.}{2015}]{lu15}
{Lu} Z.,  {Mo} H.~J.,  {Lu} Y.,  {Katz} N.,  {Weinberg} M.~D.,  {van den Bosch}
  F.~C.,   {Yang} X.,  2015, \mn@doi [\mnras] {10.1093/mnras/stv667}, \href
  {http://adsabs.harvard.edu/abs/2015MNRAS.450.1604L} {450, 1604}

\bibitem[\protect\citeauthoryear{{Magdis} et~al.,}{{Magdis}
  et~al.}{2012}]{magdis12}
{Magdis} G.~E.,  et~al., 2012, \mn@doi [\apj] {10.1088/0004-637X/760/1/6},
  \href {http://adsabs.harvard.edu/abs/2012ApJ...760....6M} {760, 6}

\bibitem[\protect\citeauthoryear{{Mandelker}, {Dekel}, {Ceverino}, {Tweed},
  {Moody}  \& {Primack}}{{Mandelker} et~al.}{2014}]{mandelker14}
{Mandelker} N.,  {Dekel} A.,  {Ceverino} D.,  {Tweed} D.,  {Moody} C.~E.,
  {Primack} J.,  2014, \mn@doi [\mnras] {10.1093/mnras/stu1340}, \href
  {http://adsabs.harvard.edu/abs/2014MNRAS.443.3675M} {443, 3675}

\bibitem[\protect\citeauthoryear{{Martig}, {Bournaud}, {Teyssier}  \&
  {Dekel}}{{Martig} et~al.}{2009}]{martig09}
{Martig} M.,  {Bournaud} F.,  {Teyssier} R.,   {Dekel} A.,  2009, \mn@doi
  [\apj] {10.1088/0004-637X/707/1/250}, \href
  {http://adsabs.harvard.edu/abs/2009ApJ...707..250M} {707, 250}

\bibitem[\protect\citeauthoryear{{Martig} et~al.,}{{Martig}
  et~al.}{2013}]{martig13}
{Martig} M.,  et~al., 2013, \mn@doi [\mnras] {10.1093/mnras/sts594}, \href
  {http://adsabs.harvard.edu/abs/2013MNRAS.432.1914M} {432, 1914}

\bibitem[\protect\citeauthoryear{{McGrath}, {Stockton}, {Canalizo}, {Iye}  \&
  {Maihara}}{{McGrath} et~al.}{2008}]{mcgrath08}
{McGrath} E.~J.,  {Stockton} A.,  {Canalizo} G.,  {Iye} M.,   {Maihara} T.,
  2008, \mn@doi [\apj] {10.1086/589631}, \href
  {http://adsabs.harvard.edu/abs/2008ApJ...682..303M} {682, 303}

\bibitem[\protect\citeauthoryear{{Mendel}, {Simard}, {Ellison}  \&
  {Patton}}{{Mendel} et~al.}{2013}]{mendel13}
{Mendel} J.~T.,  {Simard} L.,  {Ellison} S.~L.,   {Patton} D.~R.,  2013,
  \mn@doi [\mnras] {10.1093/mnras/sts489}, \href
  {http://adsabs.harvard.edu/abs/2013MNRAS.429.2212M} {429, 2212}

\bibitem[\protect\citeauthoryear{{Mihos} \& {Hernquist}}{{Mihos} \&
  {Hernquist}}{1996}]{mihos96}
{Mihos} J.~C.,  {Hernquist} L.,  1996, \mn@doi [\apj] {10.1086/177353}, \href
  {http://adsabs.harvard.edu/abs/1996ApJ...464..641M} {464, 641}

\bibitem[\protect\citeauthoryear{{Moody}, {Guo}, {Mandelker}, {Ceverino},
  {Mozena}, {Koo}, {Dekel}  \& {Primack}}{{Moody} et~al.}{2014}]{moody14}
{Moody} C.~E.,  {Guo} Y.,  {Mandelker} N.,  {Ceverino} D.,  {Mozena} M.,  {Koo}
  D.~C.,  {Dekel} A.,   {Primack} J.,  2014, \mn@doi [\mnras]
  {10.1093/mnras/stu1534}, \href
  {http://adsabs.harvard.edu/abs/2014MNRAS.444.1389M} {444, 1389}

\bibitem[\protect\citeauthoryear{{Morishita}, {Ichikawa}, {Noguchi}, {Akiyama},
  {Patel}, {Kajisawa}  \& {Obata}}{{Morishita} et~al.}{2015}]{morishita15}
{Morishita} T.,  {Ichikawa} T.,  {Noguchi} M.,  {Akiyama} M.,  {Patel} S.~G.,
  {Kajisawa} M.,   {Obata} T.,  2015, \mn@doi [\apj]
  {10.1088/0004-637X/805/1/34}, \href
  {http://adsabs.harvard.edu/abs/2015ApJ...805...34M} {805, 34}

\bibitem[\protect\citeauthoryear{{Moster}, {Somerville}, {Maulbetsch}, {van den
  Bosch}, {Macci{\`o}}, {Naab}  \& {Oser}}{{Moster} et~al.}{2010}]{moster10}
{Moster} B.~P.,  {Somerville} R.~S.,  {Maulbetsch} C.,  {van den Bosch} F.~C.,
  {Macci{\`o}} A.~V.,  {Naab} T.,   {Oser} L.,  2010, \mn@doi [\apj]
  {10.1088/0004-637X/710/2/903}, \href
  {http://adsabs.harvard.edu/abs/2010ApJ...710..903M} {710, 903}

\bibitem[\protect\citeauthoryear{{Moster}, {Naab}  \& {White}}{{Moster}
  et~al.}{2013}]{moster13}
{Moster} B.~P.,  {Naab} T.,   {White} S.~D.~M.,  2013, \mn@doi [\mnras]
  {10.1093/mnras/sts261}, \href
  {http://adsabs.harvard.edu/abs/2013MNRAS.428.3121M} {428, 3121}

\bibitem[\protect\citeauthoryear{{Murray}, {Quataert}  \& {Thompson}}{{Murray}
  et~al.}{2005}]{murray05}
{Murray} N.,  {Quataert} E.,   {Thompson} T.~A.,  2005, \mn@doi [\apj]
  {10.1086/426067}, \href {http://adsabs.harvard.edu/abs/2005ApJ...618..569M}
  {618, 569}

\bibitem[\protect\citeauthoryear{{Murray}, {Quataert}  \& {Thompson}}{{Murray}
  et~al.}{2010}]{murray10}
{Murray} N.,  {Quataert} E.,   {Thompson} T.~A.,  2010, \mn@doi [\apj]
  {10.1088/0004-637X/709/1/191}, \href
  {http://adsabs.harvard.edu/abs/2010ApJ...709..191M} {709, 191}

\bibitem[\protect\citeauthoryear{{Muzzin} et~al.,}{{Muzzin}
  et~al.}{2013}]{muzzin13}
{Muzzin} A.,  et~al., 2013, \mn@doi [\apj] {10.1088/0004-637X/777/1/18}, \href
  {http://adsabs.harvard.edu/abs/2013ApJ...777...18M} {777, 18}

\bibitem[\protect\citeauthoryear{{Naab}, {Johansson}  \& {Ostriker}}{{Naab}
  et~al.}{2009}]{naab09}
{Naab} T.,  {Johansson} P.~H.,   {Ostriker} J.~P.,  2009, \mn@doi [\apjl]
  {10.1088/0004-637X/699/2/L178}, \href
  {http://adsabs.harvard.edu/abs/2009ApJ...699L.178N} {699, L178}

\bibitem[\protect\citeauthoryear{{Neistein} \& {Dekel}}{{Neistein} \&
  {Dekel}}{2008}]{neistein08}
{Neistein} E.,  {Dekel} A.,  2008, \mn@doi [\mnras]
  {10.1111/j.1365-2966.2008.13525.x}, \href
  {http://adsabs.harvard.edu/abs/2008MNRAS.388.1792N} {388, 1792}

\bibitem[\protect\citeauthoryear{{Nelson} et~al.,}{{Nelson}
  et~al.}{2015}]{nelson15_insideout}
{Nelson} E.~J.,  et~al., 2015, preprint, \href
  {http://adsabs.harvard.edu/abs/2015arXiv150703999N} {} (\mn@eprint {arXiv}
  {1507.03999})

\bibitem[\protect\citeauthoryear{{Newman}, {Ellis}, {Bundy}  \&
  {Treu}}{{Newman} et~al.}{2012}]{newman12a}
{Newman} A.~B.,  {Ellis} R.~S.,  {Bundy} K.,   {Treu} T.,  2012, \mn@doi [\apj]
  {10.1088/0004-637X/746/2/162}, \href
  {http://adsabs.harvard.edu/abs/2012ApJ...746..162N} {746, 162}

\bibitem[\protect\citeauthoryear{{Newman} et~al.,}{{Newman}
  et~al.}{2013}]{newman13a}
{Newman} S.~F.,  et~al., 2013, \mn@doi [\apj] {10.1088/0004-637X/767/2/104},
  \href {http://adsabs.harvard.edu/abs/2013ApJ...767..104N} {767, 104}

\bibitem[\protect\citeauthoryear{{Nipoti}, {Treu}, {Auger}  \&
  {Bolton}}{{Nipoti} et~al.}{2009}]{nipoti09}
{Nipoti} C.,  {Treu} T.,  {Auger} M.~W.,   {Bolton} A.~S.,  2009, \mn@doi
  [\apjl] {10.1088/0004-637X/706/1/L86}, \href
  {http://adsabs.harvard.edu/abs/2009ApJ...706L..86N} {706, L86}

\bibitem[\protect\citeauthoryear{{Noguchi}}{{Noguchi}}{1998}]{noguchi98}
{Noguchi} M.,  1998, \mn@doi [\nat] {10.1038/32596}, \href
  {http://adsabs.harvard.edu/abs/1998Natur.392..253N} {392, 253}

\bibitem[\protect\citeauthoryear{{Ocvirk}, {Pichon}  \& {Teyssier}}{{Ocvirk}
  et~al.}{2008}]{ocvirk08}
{Ocvirk} P.,  {Pichon} C.,   {Teyssier} R.,  2008, \mn@doi [\mnras]
  {10.1111/j.1365-2966.2008.13763.x}, \href
  {http://adsabs.harvard.edu/abs/2008MNRAS.390.1326O} {390, 1326}

\bibitem[\protect\citeauthoryear{{Omand}, {Balogh}  \& {Poggianti}}{{Omand}
  et~al.}{2014}]{omand14}
{Omand} C.~M.~B.,  {Balogh} M.~L.,   {Poggianti} B.~M.,  2014, \mn@doi [\mnras]
  {10.1093/mnras/stu331}, \href
  {http://adsabs.harvard.edu/abs/2014MNRAS.tmp..530O} {}

\bibitem[\protect\citeauthoryear{{Oser}, {Naab}, {Ostriker}  \&
  {Johansson}}{{Oser} et~al.}{2012}]{oser12}
{Oser} L.,  {Naab} T.,  {Ostriker} J.~P.,   {Johansson} P.~H.,  2012, \mn@doi
  [\apj] {10.1088/0004-637X/744/1/63}, \href
  {http://adsabs.harvard.edu/abs/2012ApJ...744...63O} {744, 63}

\bibitem[\protect\citeauthoryear{{Patel} et~al.,}{{Patel}
  et~al.}{2013}]{patel13}
{Patel} S.~G.,  et~al., 2013, \mn@doi [\apj] {10.1088/0004-637X/766/1/15},
  \href {http://adsabs.harvard.edu/abs/2013ApJ...766...15P} {766, 15}

\bibitem[\protect\citeauthoryear{{Peng} et~al.,}{{Peng}
  et~al.}{2010}]{peng10_Cont}
{Peng} Y.-j.,  et~al., 2010, \mn@doi [\apj] {10.1088/0004-637X/721/1/193},
  \href {http://adsabs.harvard.edu/abs/2010ApJ...721..193P} {721, 193}

\bibitem[\protect\citeauthoryear{{Poggianti} et~al.,}{{Poggianti}
  et~al.}{2013}]{poggianti13}
{Poggianti} B.~M.,  et~al., 2013, \mn@doi [\apj] {10.1088/0004-637X/762/2/77},
  \href {http://adsabs.harvard.edu/abs/2013ApJ...762...77P} {762, 77}

\bibitem[\protect\citeauthoryear{{Rees} \& {Ostriker}}{{Rees} \&
  {Ostriker}}{1977}]{rees77}
{Rees} M.~J.,  {Ostriker} J.~P.,  1977, \mnras, \href
  {http://adsabs.harvard.edu/abs/1977MNRAS.179..541R} {179, 541}

\bibitem[\protect\citeauthoryear{{Robaina}, {Hoyle}, {Gallazzi}, {Jim{\'e}nez},
  {van der Wel}  \& {Verde}}{{Robaina} et~al.}{2012}]{robaina12}
{Robaina} A.~R.,  {Hoyle} B.,  {Gallazzi} A.,  {Jim{\'e}nez} R.,  {van der Wel}
  A.,   {Verde} L.,  2012, \mn@doi [\mnras] {10.1111/j.1365-2966.2012.21804.x},
  \href {http://adsabs.harvard.edu/abs/2012MNRAS.427.3006R} {427, 3006}

\bibitem[\protect\citeauthoryear{{Salim}, {Fang}, {Rich}, {Faber}  \&
  {Thilker}}{{Salim} et~al.}{2012}]{salim12}
{Salim} S.,  {Fang} J.~J.,  {Rich} R.~M.,  {Faber} S.~M.,   {Thilker} D.~A.,
  2012, \mn@doi [\apj] {10.1088/0004-637X/755/2/105}, \href
  {http://adsabs.harvard.edu/abs/2012ApJ...755..105S} {755, 105}

\bibitem[\protect\citeauthoryear{{Saracco}, {Gargiulo}  \&
  {Longhetti}}{{Saracco} et~al.}{2012}]{saracco12}
{Saracco} P.,  {Gargiulo} A.,   {Longhetti} M.,  2012, \mn@doi [\mnras]
  {10.1111/j.1365-2966.2012.20830.x}, \href
  {http://adsabs.harvard.edu/abs/2012MNRAS.422.3107S} {422, 3107}

\bibitem[\protect\citeauthoryear{{Sargent} et~al.,}{{Sargent}
  et~al.}{2014}]{sargent14}
{Sargent} M.~T.,  et~al., 2014, \mn@doi [\apj] {10.1088/0004-637X/793/1/19},
  \href {http://adsabs.harvard.edu/abs/2014ApJ...793...19S} {793, 19}

\bibitem[\protect\citeauthoryear{{Schawinski} et~al.,}{{Schawinski}
  et~al.}{2014}]{schawinski14}
{Schawinski} K.,  et~al., 2014, \mn@doi [\mnras] {10.1093/mnras/stu327}, \href
  {http://adsabs.harvard.edu/abs/2014MNRAS.tmp..527S} {440, 889}

\bibitem[\protect\citeauthoryear{{Scoville} et~al.,}{{Scoville}
  et~al.}{2015}]{scoville15}
{Scoville} N.,  et~al., 2015, preprint, \href
  {http://adsabs.harvard.edu/abs/2015arXiv150502159S} {} (\mn@eprint {arXiv}
  {1505.02159})

\bibitem[\protect\citeauthoryear{{S{\'e}rsic}}{{S{\'e}rsic}}{1968}]{sersic68}
{S{\'e}rsic} J.~L.,  1968, {Atlas de galaxias australes}.
Cordoba, Argentina: Observatorio Astronomico

\bibitem[\protect\citeauthoryear{{Shankar}, {Marulli}, {Bernardi}, {Mei},
  {Meert}  \& {Vikram}}{{Shankar} et~al.}{2013}]{shankar13}
{Shankar} F.,  {Marulli} F.,  {Bernardi} M.,  {Mei} S.,  {Meert} A.,   {Vikram}
  V.,  2013, \mn@doi [\mnras] {10.1093/mnras/sts001}, \href
  {http://adsabs.harvard.edu/abs/2013MNRAS.428..109S} {428, 109}

\bibitem[\protect\citeauthoryear{{Shen}, {Mo}, {White}, {Blanton}, {Kauffmann},
  {Voges}, {Brinkmann}  \& {Csabai}}{{Shen} et~al.}{2003}]{shen03}
{Shen} S.,  {Mo} H.~J.,  {White} S.~D.~M.,  {Blanton} M.~R.,  {Kauffmann} G.,
  {Voges} W.,  {Brinkmann} J.,   {Csabai} I.,  2003, \mn@doi [\mnras]
  {10.1046/j.1365-8711.2003.06740.x}, \href
  {http://adsabs.harvard.edu/abs/2003MNRAS.343..978S} {343, 978}

\bibitem[\protect\citeauthoryear{{Silverman} et~al.,}{{Silverman}
  et~al.}{2015}]{silverman15}
{Silverman} J.~D.,  et~al., 2015, \mn@doi [\apjl]
  {10.1088/2041-8205/812/2/L23}, \href
  {http://adsabs.harvard.edu/abs/2015ApJ...812L..23S} {812, L23}

\bibitem[\protect\citeauthoryear{{Snyder}, {Lotz}, {Moody}, {Peth}, {Freeman},
  {Ceverino}, {Primack}  \& {Dekel}}{{Snyder} et~al.}{2015}]{snyder15_morph}
{Snyder} G.~F.,  {Lotz} J.,  {Moody} C.,  {Peth} M.,  {Freeman} P.,  {Ceverino}
  D.,  {Primack} J.,   {Dekel} A.,  2015, \mn@doi [\mnras]
  {10.1093/mnras/stv1231}, \href
  {http://adsabs.harvard.edu/abs/2015MNRAS.451.4290S} {451, 4290}

\bibitem[\protect\citeauthoryear{{Szomoru}, {Franx}, {Bouwens}, {van Dokkum},
  {Labb{\'e}}, {Illingworth}  \& {Trenti}}{{Szomoru} et~al.}{2011}]{szomoru11}
{Szomoru} D.,  {Franx} M.,  {Bouwens} R.~J.,  {van Dokkum} P.~G.,  {Labb{\'e}}
  I.,  {Illingworth} G.~D.,   {Trenti} M.,  2011, \mn@doi [\apjl]
  {10.1088/2041-8205/735/1/L22}, \href
  {http://adsabs.harvard.edu/abs/2011ApJ...735L..22S} {735, L22}

\bibitem[\protect\citeauthoryear{{Szomoru}, {Franx}  \& {van Dokkum}}{{Szomoru}
  et~al.}{2012}]{szomoru12}
{Szomoru} D.,  {Franx} M.,   {van Dokkum} P.~G.,  2012, \mn@doi [\apj]
  {10.1088/0004-637X/749/2/121}, \href
  {http://adsabs.harvard.edu/abs/2012ApJ...749..121S} {749, 121}

\bibitem[\protect\citeauthoryear{{Szomoru}, {Franx}, {van Dokkum}, {Trenti},
  {Illingworth}, {Labb{\'e}}  \& {Oesch}}{{Szomoru} et~al.}{2013}]{szomoru13}
{Szomoru} D.,  {Franx} M.,  {van Dokkum} P.~G.,  {Trenti} M.,  {Illingworth}
  G.~D.,  {Labb{\'e}} I.,   {Oesch} P.,  2013, \mn@doi [\apj]
  {10.1088/0004-637X/763/2/73}, \href
  {http://adsabs.harvard.edu/abs/2013ApJ...763...73S} {763, 73}

\bibitem[\protect\citeauthoryear{{Tacchella}, {Dekel}, {Carollo}, {Ceverino},
  {DeGraf}, {Lapiner}, {Mandelker}  \& {Primack}}{{Tacchella}
  et~al.}{2015a}]{tacchella15_MS}
{Tacchella} S.,  {Dekel} A.,  {Carollo} C.~M.,  {Ceverino} D.,  {DeGraf} C.,
  {Lapiner} S.,  {Mandelker} N.,   {Primack} J.~R.,  2015a, preprint, \href
  {http://adsabs.harvard.edu/abs/2015arXiv150902529T} {} (\mn@eprint {arXiv}
  {1509.02529})

\bibitem[\protect\citeauthoryear{{Tacchella} et~al.,}{{Tacchella}
  et~al.}{2015b}]{tacchella15_sci}
{Tacchella} S.,  et~al., 2015b, \mn@doi [Science] {10.1126/science.1261094},
  \href {http://adsabs.harvard.edu/abs/2015Sci...348..314T} {348, 314}

\bibitem[\protect\citeauthoryear{{Tacchella} et~al.,}{{Tacchella}
  et~al.}{2015c}]{tacchella15}
{Tacchella} S.,  et~al., 2015c, \mn@doi [\apj] {10.1088/0004-637X/802/2/101},
  \href {http://adsabs.harvard.edu/abs/2015ApJ...802..101T} {802, 101}

\bibitem[\protect\citeauthoryear{{Toomre}}{{Toomre}}{1964}]{toomre64}
{Toomre} A.,  1964, \mn@doi [\apj] {10.1086/147861}, \href
  {http://adsabs.harvard.edu/abs/1964ApJ...139.1217T} {139, 1217}

\bibitem[\protect\citeauthoryear{{Torrey} et~al.,}{{Torrey}
  et~al.}{2015}]{torrey15}
{Torrey} P.,  et~al., 2015, \mn@doi [\mnras] {10.1093/mnras/stv1986}, \href
  {http://adsabs.harvard.edu/abs/2015MNRAS.454.2770T} {454, 2770}

\bibitem[\protect\citeauthoryear{{Trujillo}, {Conselice}, {Bundy}, {Cooper},
  {Eisenhardt}  \& {Ellis}}{{Trujillo} et~al.}{2007}]{trujillo07}
{Trujillo} I.,  {Conselice} C.~J.,  {Bundy} K.,  {Cooper} M.~C.,  {Eisenhardt}
  P.,   {Ellis} R.~S.,  2007, \mn@doi [\mnras]
  {10.1111/j.1365-2966.2007.12388.x}, \href
  {http://adsabs.harvard.edu/abs/2007MNRAS.382..109T} {382, 109}

\bibitem[\protect\citeauthoryear{{Wechsler}, {Bullock}, {Primack}, {Kravtsov}
  \& {Dekel}}{{Wechsler} et~al.}{2002}]{wechsler02}
{Wechsler} R.~H.,  {Bullock} J.~S.,  {Primack} J.~R.,  {Kravtsov} A.~V.,
  {Dekel} A.,  2002, \mn@doi [\apj] {10.1086/338765}, \href
  {http://adsabs.harvard.edu/abs/2002ApJ...568...52W} {568, 52}

\bibitem[\protect\citeauthoryear{{Woo}, {Dekel}, {Faber}  \& {Koo}}{{Woo}
  et~al.}{2015}]{woo15}
{Woo} J.,  {Dekel} A.,  {Faber} S.~M.,   {Koo} D.~C.,  2015, \mn@doi [\mnras]
  {10.1093/mnras/stu2755}, \href
  {http://adsabs.harvard.edu/abs/2015MNRAS.448..237W} {448, 237}

\bibitem[\protect\citeauthoryear{{Wuyts} et~al.,}{{Wuyts}
  et~al.}{2011}]{wuyts11}
{Wuyts} S.,  et~al., 2011, \mn@doi [\apj] {10.1088/0004-637X/742/2/96}, \href
  {http://adsabs.harvard.edu/abs/2011ApJ...742...96W} {742, 96}

\bibitem[\protect\citeauthoryear{{Wuyts} et~al.,}{{Wuyts}
  et~al.}{2012}]{wuyts12}
{Wuyts} S.,  et~al., 2012, \mn@doi [\apj] {10.1088/0004-637X/753/2/114}, \href
  {http://adsabs.harvard.edu/abs/2012ApJ...753..114W} {753, 114}

\bibitem[\protect\citeauthoryear{{Zolotov} et~al.,}{{Zolotov}
  et~al.}{2015}]{zolotov15}
{Zolotov} A.,  et~al., 2015, \mn@doi [\mnras] {10.1093/mnras/stv740}, \href
  {http://adsabs.harvard.edu/abs/2015MNRAS.450.2327Z} {450, 2327}

\bibitem[\protect\citeauthoryear{{van Dokkum} et~al.,}{{van Dokkum}
  et~al.}{2008}]{van-dokkum08}
{van Dokkum} P.~G.,  et~al., 2008, \mn@doi [\apjl] {10.1086/587874}, \href
  {http://adsabs.harvard.edu/abs/2008ApJ...677L...5V} {677, L5}

\bibitem[\protect\citeauthoryear{{van Dokkum} et~al.,}{{van Dokkum}
  et~al.}{2010}]{van-dokkum10}
{van Dokkum} P.~G.,  et~al., 2010, \mn@doi [\apj]
  {10.1088/0004-637X/709/2/1018}, \href
  {http://adsabs.harvard.edu/abs/2010ApJ...709.1018V} {709, 1018}

\bibitem[\protect\citeauthoryear{{van Dokkum} et~al.,}{{van Dokkum}
  et~al.}{2013}]{van-dokkum13}
{van Dokkum} P.~G.,  et~al., 2013, \mn@doi [\apjl]
  {10.1088/2041-8205/771/2/L35}, \href
  {http://adsabs.harvard.edu/abs/2013ApJ...771L..35V} {771, L35}

\bibitem[\protect\citeauthoryear{{van Dokkum} et~al.,}{{van Dokkum}
  et~al.}{2014}]{van-dokkum14_dense_cores}
{van Dokkum} P.~G.,  et~al., 2014, \mn@doi [\apj] {10.1088/0004-637X/791/1/45},
  \href {http://adsabs.harvard.edu/abs/2014ApJ...791...45V} {791, 45}

\bibitem[\protect\citeauthoryear{{van Dokkum} et~al.,}{{van Dokkum}
  et~al.}{2015}]{van-dokkum15}
{van Dokkum} P.~G.,  et~al., 2015, \mn@doi [\apj] {10.1088/0004-637X/813/1/23},
  \href {http://adsabs.harvard.edu/abs/2015ApJ...813...23V} {813, 23}

\makeatother
\end{thebibliography}
\end{document}